\documentclass[12pt,english,floatfix,nofootinbib,superscriptaddress,aps,prd,preprint]{revtex4}
\usepackage[utf8]{inputenc}       
\usepackage[english]{babel}       
\usepackage[utf8]{inputenc}
\usepackage{textcomp}
\usepackage{amsmath,amsthm,amssymb,amsfonts,mathrsfs,amsbsy} 
\usepackage{tensor}               
\usepackage{slashed}  
\usepackage{bbm}
\usepackage{esint}                
\usepackage{cancel}               
\usepackage{graphicx}             
\usepackage{natbib}
\usepackage{float}                
\usepackage{subfig}               
\usepackage[font=small,labelfont=bf]{caption} 
\usepackage{multirow}             
\usepackage{array}                
\usepackage{tikz}                 
\usetikzlibrary{quotes,angles,arrows,decorations.markings} 
\usepackage{makecell} 

\usepackage{lipsum}

\usepackage{xcolor}               
\usepackage{color}                
\usepackage{textcomp}             
\usepackage{units}                

\newcommand{\be}{\begin{equation}}
\newcommand{\ee}{\end{equation}}
\newcommand{\Be}{\begin{eqnarray}}
\newcommand{\Ee}{\end{eqnarray}}

\newcommand{\mincir}{\raise
-3.truept\hbox{\rlap{\hbox{$\sim$}}\raise4.truept\hbox{$<$}\ }}
\newcommand{\magcir}{\raise
-3.truept\hbox{\rlap{\hbox{$\sim$}}\raise4.truept\hbox{$>$}\ }}

\newcolumntype{Y}{>{\centering\arraybackslash}X}
\providecommand{\U}[1]

\usepackage[dvips]{epsfig}        
\usepackage[dvips]{graphicx}      
\usepackage{hyperref}             
\hypersetup{
    colorlinks=true,              
    breaklinks=true,              
    citecolor=blue,               
    linkcolor=[rgb]{0,0.5,0.9},   
    urlcolor=red,                 
    filecolor=green               
}

\newcommand{\ie}{\begin{equation}}
\newcommand{\fe}{\end{equation}}
\newcommand{\se}{\begin{eqnarray}}
\newcommand{\ff}{\end{eqnarray}}

\begin{document}

\title{Gravitational aspects of a new bumblebee black hole}


\author{A. A. Ara\'{u}jo Filho}
\email{dilto@fisica.ufc.br}
\affiliation{Departamento de Física, Universidade Federal da Paraíba, Caixa Postal 5008, 58051--970, João Pessoa, Paraíba,  Brazil.}
\affiliation{Departamento de Física, Universidade Federal de Campina Grande Caixa Postal 10071, 58429-900 Campina Grande, Paraíba, Brazil.}
\affiliation{Center for Theoretical Physics, Khazar University, 41 Mehseti Street, Baku, AZ-1096, Azerbaijan.}


\author{N. Heidari}
\email{heidari.n@gmail.com}

\affiliation{Center for Theoretical Physics, Khazar University, 41 Mehseti Street, Baku, AZ-1096, Azerbaijan.}
\affiliation{School of Physics, Damghan University, Damghan, 3671641167, Iran.}


\author{Iarley P. Lobo}
\email{lobofisica@gmail.com}

\affiliation{Department of Chemistry and Physics, Federal University of Para\'iba, Rodovia BR 079 - km 12, 58397-000 Areia-PB,  Brazil.}
\affiliation{Departamento de Física, Universidade Federal de Campina Grande Caixa Postal 10071, 58429-900 Campina Grande, Paraíba, Brazil.}


\author{V. B. Bezerra}
\email{valdir@fisica.ufpb.br}
\affiliation{Departamento de Física, Universidade Federal da Paraíba, Caixa Postal 5008, 58051--970, João Pessoa, Paraíba,  Brazil.}



\date{\today}

\begin{abstract}

In this paper, we examine the physical consequences of a recently introduced black hole solution in bumblebee gravity \cite{Liu:2025oho,Zhu:2025fiy}. The geometry is first presented and then reformulated through suitable coordinate adjustments, which make its global conical character evident. We then study the propagation of particles by solving the geodesic equations for null and timelike trajectories. The associated critical orbits (or photon spheres) are obtained, and shadow radius are computed and compared with other Lorentz–violating configurations in bumblebee and Kalb--Ramond models, including their charged and cosmological extensions. Massive particle motion is analyzed separately, followed by the construction of the effective potentials for scalar, vector, tensor, and spinor perturbations. These potentials allow the calculations of quasinormal frequencies and the corresponding time–domain evolution. Gravitational lensing phenomena are investigated in the weak and strong deflection regimes, and the light--travel time delay is also evaluated. The study concludes with bounds on the Lorentz--violating parameter based on classical Solar System experiments.

\end{abstract}


\maketitle

\tableofcontents


\section{Introduction }

{The possibility that Lorentz symmetry may not be an exact property of nature has attracted considerable attention in recent years, especially in contexts where quantum--gravity corrections are expected to modify the low--energy geometric structure \cite{colladay1997cpt,kostelecky1989spontaneous,kostelecky2004gravity,kostelecky2011data,kostelecky1999constraints}. Small departures from local Lorentz invariance may emerge through spontaneous symmetry breaking, giving rise to background fields that imprint a preferred direction in spacetime. Within this perspective, the bumblebee construction provides a minimal and consistent effective description, where a vector field develops a vacuum expectation value and generates controlled deformations of standard general relativity \cite{bluhm2005spontaneous,Bluhm:2023kph,Bluhm:2019ato,bluhm2008spontaneous,Maluf:2013nva,Maluf:2014dpa}.

In particular, bumblebee models provide a simple framework in which Lorentz symmetry is not exact but rather emerges as an approximate property of low--energy physics. Their structure is motivated by string--inspired mechanisms for spontaneous Lorentz breaking \cite{kostelecky1989spontaneous,kostelecky1991photon}, vector-tensor extensions of Einstein's theory \cite{jacobson2004einstein}, and general effective field--theory approaches accommodating symmetry-breaking background configurations \cite{kostelecky2004gravity,bluhm2005spontaneous}. In these models, a vector field $B_{\mu}$ is driven to a nonzero vacuum expectation value by a potential of the form $V(B_{\mu}B^{\mu}\mp b^{2})$, thereby selecting a preferred direction in spacetime and triggering spontaneous Lorentz violation \cite{bluhm2005spontaneous,bluhm2008spontaneous}. The resulting spectrum contains Nambu--Goldstone modes that emulate photon--like excitations \cite{bluhm2005spontaneous}, as well as massive modes associated with departures from the fixed--norm condition \cite{bluhm2008spontaneous}.

Bumblebee models have also been formulated in curved backgrounds, leading to a gravitational sector in which the vacuum value of the vector field affects the spacetime geometry \cite{Bertolami:2005bh}. In this setting, black-hole solutions have received particular attention. The geometry introduced in Ref.~\cite{Casana:2017jkc} motivated several subsequent studies, including horizon-related quantum effects \cite{Liu:2024wpa,AraujoFilho:2025hkm}, modifications of gravitational-wave propagation \cite{Liang:2022hxd}, and quasinormal oscillations \cite{Oliveira:2021abg}. Additional developments include configurations with topological-defect-like features \cite{Gullu:2020qzu}, non--commutative corrections \cite{KumarJha:2020ivj}, cosmological-constant extensions \cite{Maluf:2020kgf}, and approximate rotating counterparts \cite{Ding:2019mal,Liu:2019mls}. These backgrounds also supported various analyses of light propagation, including weak and strong lensing studies based on the Gauss--Bonnet method, direct geodesic integration, perturbative rotation, and corrected Newman--Janis constructions \cite{Ovgun:2018ran,Li:2020wvn,asdasdSekhmani:2025zen,asdasd1Deng:2025uvp,Kumar:2025bim}.

Further generalizations of the original bumblebee black hole have also been reported. In the metric--affine formulation, a static solution was obtained in Ref.~\cite{Filho:2022yrk}, followed by an axisymmetric extension in Ref.~\cite{AraujoFilho:2024ykw} and a non--commutative counterpart in Ref.~\cite{AraujoFilho:2025rvn}. Beyond the black-hole sector, the same framework has been employed in cosmological applications \cite{Neves:2022qyb,Jesus:2019nwi}, relativistic stars \cite{Neves:2024ggn}, wormholes and black-bounce geometries \cite{Magalhaes:2025nql,AraujoFilho:2024iox,Magalhaes:2025lti,Ovgun:2018xys,Pereira:2025xnw}, thermodynamic modifications \cite{Gomes:2018oyd,Ovgun:2019ygw,Kanzi:2019gtu}, accretion processes \cite{Yang:2018zef,Yang:2025whw}, and neutrino propagation in metric, metric--affine, and tensorial Lorentz--violating backgrounds \cite{Shi:2025plr,Shi:2025ywa,Shi:2025rfq}. Although these results show the breadth of the framework, the present work is concerned specifically with the gravitational properties of a recently proposed black-hole solution.

Recently, a new black hole configuration in bumblebee gravity was reported in Ref.~\cite{Liu:2025oho}, obtained by allowing the vacuum expectation value of the bumblebee field to contain both temporal and radial components, $b_{\mu}=(b_{t}(r),b_{r}(r),0,0)$. In that construction, the radial metric component takes the form $
g_{rr}=\frac{1+\ell_{1}+\ell_{2}}{1-2M/r}$,
which generalizes the original result of Ref.~\cite{Casana:2017jkc}, where one has $g_{rr}=(1+\ell)/(1-2M/r)$. On the other hand, an independent recent development introduced another static and spherically symmetric solution by imposing the most general vacuum expectation value compatible with this symmetry class \cite{Zhu:2025fiy}. That construction encompasses spacelike, timelike, and lightlike configurations of the vacuum vector, thus covering the full set of admissible causal structures for the bumblebee field. However, the corresponding gravitational signatures were not explored there. The purpose of the present work is precisely to investigate the physical consequences of this solution.

The relevance of this program is reinforced by the fact that black holes have become one of the main arenas for testing gravitational theories beyond general relativity. On the observational side, interferometric detections of gravitational waves by the LIGO and VIRGO collaborations opened a direct window into the strong--gravity regime \cite{LIGOScientific:2016aoc}, while horizon-scale imaging by the Event Horizon Telescope provided direct access to the near horizon structure of supermassive compact objects \cite{Akiyama2022,Akiyama2019}. These developments have turned strong-gravity astrophysics into a setting where modified gravitational scenarios can be directly confronted with observations.

Within this context, two classes of observables have become especially relevant. One of them is associated with the propagation of light near compact objects. The weak--field description based on Schwarzschild-like backgrounds \cite{021} is no longer sufficient in regions where curvature is large, and strong-deflection methods become necessary \cite{019,020,022,mohan2025strong,araujo2025gasdsadravitational}. Black hole shadows, in particular, provide a direct probe of photon trajectories near the compact object. Since the early discussions by Bardeen and collaborators \cite{Cunningham} and the observational proposal by Falcke, Melia, and Agol \cite{Falcke:1999pj}, shadow studies have become an important part of the comparison between general relativity and alternative theories \cite{Afrin:2024khy,Khodadi:2024ubi,Allahyari:2019jqz,Afrin:2021wlj,Nojiri:2024txy,Afrin:2021imp,Nojiri:2024qgx,Bambi:2019tjh,Khodadi:2021gbc,Liu:2024lve,Khodadi:2022pqh,Kumar:2020hgm,Afrin:2022ztr,Afrin:2023uzo,Ghosh:2022kit,Vagnozzi:2019apd,Nojiri:2024nlx,Fu:2021fxn}. Likewise, the strong field lensing formalism developed by Virbhadra and Ellis made it possible to describe photon trajectories that loop around the compact object before reaching the observer \cite{031,virbhadra2000schwarzschild}, and this framework has since been extended to a wide range of alternative geometries, including nontrivial topological structures and other nonstandard backgrounds \cite{033,032,034,heidari2023gravitational,araujo2025antisymmetric,nascimento2024gravitational,chakraborty2017strong,40,Lobo:2020jfl,38.5,38.2,38.1,38.3,38.4,virbhadra2024conservation}.

Another important class of observables is associated with black hole perturbations. After a merger, the remnant object relaxes through exponentially damped oscillations described by quasinormal modes \cite{Konoplya:2019hlu,Konoplya:2013rxa,karmakar2024quasinormal,Konoplya:2007zx,Kokkotas:2010zd,karmakar2022quasinormal,Konoplya:2011qq}. Since these frequencies depend only on the parameters of the underlying geometry, they provide a clean way to characterize departures from general relativity. Their relevance is further emphasized by the connections established with shadow properties \cite{Jusufi:2020dhz} and greybody emission \cite{Konoplya:2024lir,Konoplya:2024vuj}. Although the extraction of individual quasinormal modes from current gravitational wave data is still under discussion \cite{Franchini:2023eda}, this sector remains one of the most promising for testing modified compact object geometries.

In this work, we investigate the gravitational properties of the black-hole solution introduced in Ref.~\cite{Zhu:2025fiy}. After presenting the geometry, we reformulate the metric through suitable coordinate transformations in order to make its asymptotic conical structure explicit. We then study null and timelike geodesics, determine the critical photon orbit and the shadow radius, and compare these results with those of other Lorentz--violating black hole backgrounds. Next, we derive the effective potentials for scalar, vector, tensor, and spinor perturbations, from which the quasinormal spectra and time--domain profiles are obtained. We also examine gravitational lensing in the weak-- and strong--deflection regimes, together with the corresponding time delay of light, and finally derive bounds on the Lorentz--violating parameter from Solar--System tests.}


\section{The new bumblebee black hole }

The new spacetime describing the recently proposed bumblebee black hole solution is given by \cite{Zhu:2025fiy}
\ie
\label{maaaaianametric}
\mathrm{d}s^{2} = - \frac{1}{1+\chi}\left(1 - \frac{2M}{r}         \right)\mathrm{d}t^{2} + \frac{1+\chi}{\left(1 - \frac{2M}{r} \right)} \mathrm{d}r^{2} + r^{2}\mathrm{d}\Omega^{2},
\fe
where $\chi \equiv \alpha \ell$, $\alpha$ denotes the integration constant, and $\ell \equiv \xi b^{2}$ characterizes the Lorentz violation. Here, $\xi$ is a coupling constant, and $b^{2} \equiv b_{\mu} b^{\mu}$ represents the vacuum expectation value of the bumblebee field.  {{Notice that this approach covers the possibility of timelike and spacelike vacuum expectation values and corresponds to a very general configuration, as discussed in \cite{Zhu:2025fiy}.}}

In addition, to explore the properties of the metric tensor presented in Eq. (\ref{maaaaianametric}) more thoroughly, let us introduce some coordinate redefinitions. This step allows for a clearer discussion of specific geometric and physical features. In particular, such an analysis may be relevant when studying entanglement degradation and gravitational lensing. This first aspect lies beyond the scope of the present work, whereas the latter will be examined in detail here by considering both the weak and strong deflection limits.

To rewrite the line element in a form resembling the Schwarzschild geometry, possibly featuring a global--monopole--type structure, the coordinates are redefined by means of an appropriate rescaling $\Tilde{t}=t/\sqrt{1+\chi}$ and $\Tilde{r}=\sqrt{1+\chi}\,r$, accompanied by the redefinitions of the mass parameter $\Tilde{M}=\sqrt{1+\chi}\,M$ and the constant $\Tilde{K}^{2}=1/(1+\chi)$. After applying these transformations, the metric takes the form
\ie
\label{redefinedmetric}
\mathrm{d}s^{2}=-\left(1-\frac{2\Tilde{M}}{\Tilde{r}}\right)\mathrm{d}\Tilde{t}^{\,2}+\frac{1}{1-\frac{2\Tilde{M}}{\Tilde{r}}}\mathrm{d}\Tilde{r}^{\,2}+\Tilde{K}^{2}\Tilde{r}^{2}\mathrm{d}\Omega^{2},
\fe
which maintains the same overall structure as the Schwarzschild line element, differing only by the constant factor $\Tilde{K}^{2}$ that multiplies the angular term. In the asymptotic limit $\Tilde{r}\to\infty$, the spacetime approaches
\ie
\mathrm{d}s^{2}=-\mathrm{d}\Tilde{t}^{\,2}+\mathrm{d}\Tilde{r}^{2}+\Tilde{K}^{2}\Tilde{r}^{\,2}\mathrm{d}\Omega^{2}.
\fe
For small values of $\chi$, one has $\Tilde{K}^{2}=1/(1+\chi)\simeq 1-\chi+\mathcal{O}(\chi^{2})$, allowing the metric to be reformulated as
\ie
\mathrm{d}s^{2}=-\mathrm{d}\Tilde{t}^{\,2}+\mathrm{d}\Tilde{r}^{\,2}+ \Big[1-\chi+\mathcal{O}(\chi^{2})\Big]\Tilde{r}^{\,2}\mathrm{d}\Omega^{2}.
\fe

The obtained metric represents a spacetime that becomes asymptotically conical at large distances, featuring a solid--angle deficit given by $\delta=\chi/(1+\chi)\simeq\chi$ for small $\chi$. This conical behavior is analogous to that found in a global--monopole spacetime, where the total solid angle is smaller than $4\pi$; however, in the present case, the deficit originates from the Lorentz--violating parameter $\chi$. When the motion is confined to the equatorial plane, $\theta=\pi/2$, the line element simplifies to
\ie
\mathrm{d}s^{2}=-\mathrm{d}\Tilde{t}^{\,2}+\mathrm{d}\Tilde{r}^{2}+\Tilde{K}^{2}\Tilde{r}^{\,2}\mathrm{d}\phi^{2}.
\fe
After introducing a new angular coordinate $\Tilde{\phi}=\Tilde{K}\phi=\phi/\sqrt{1+\chi}$, the metric takes the form
\ie
\mathrm{d}s^{2}=-\mathrm{d}\Tilde{t}^{\,2}+\mathrm{d}\Tilde{r}^{2}+\Tilde{r}^{\,2}\mathrm{d}\Tilde{\phi}^{2},
\fe
which clearly corresponds to a locally flat geometry. Nevertheless, the spacetime as a whole possesses a conical deficit, much like the structure produced by a cosmic string or a global monopole. Despite this topological feature, the local region of the spacetime retains flatness, so it may still be treated as locally asymptotically flat. { Notice that information on the $\chi$-parameter is not removed from the geometry by this redefinition. It is absorbed in the deficit angle and contributes to several physical observables, as we will discuss in the next sections. This bears some similarities with the  case of cosmic strings \cite{Hindmarsh:1994re}.}

To explore the gravitational properties of the black hole described by Eq. (\ref{maaaaianametric}), the forthcoming sections examine the motion of both massless and massive particles by solving a system of four coupled differential equations. The analysis then proceeds to determine the critical orbits (photon spheres) and the corresponding shadow radii. Subsequently, the topological aspects of the spacetime are discussed, emphasizing the structure of the topological photon sphere. The study also includes the computation of quasinormal modes and time--domain profiles for scalar, vector, tensor, and spinor perturbations, based on the effective potentials derived for each field. These modes are evaluated using the sixth--order WKB approximation. Moreover, gravitational lensing is analyzed in both the weak and strong deflection regimes. Finally, constraints from solar--system tests are established to place bounds on the Lorentz--violating parameter.

{Before proceeding, we present in Table~\ref{tab:parameters} a summary of the main parameters and symbols used throughout the manuscript in order to improve readability.

\begin{table}[t]
\centering
\begin{tabular}{c l c l}
\hline
Symbol & Description & Symbol & Description \\
\hline
$M$ & Black hole mass & $r$ & Radial coordinate \\
$t$ & Time coordinate & $\theta,\phi$ & Angular coordinates \\
$\chi$ & Lorentz--violating parameter & $\alpha$ & Integration constant \\
$\ell$ & $\ell \equiv \xi b^{2}$ & $\xi$ & Bumblebee coupling constant \\
$b_{\mu}$ & Bumblebee vector field & $b^{2}$ & $b_{\mu}b^{\mu}$ \\
$g_{\mu\nu}$ & Metric tensor & $\delta$ & Solid-angle deficit parameter \\
$\tilde{t}$ & Rescaled time & $\tilde{r}$ & Rescaled radial coordinate \\
$\tilde{M}$ & Rescaled mass & $\tilde{K}^{2}$ & Angular factor constant \\
$E$ & Energy & $L$ & Angular momentum \\
$r_{\mathrm{ph}}$ & Photon sphere radius & $r_{\mathrm{ISCO}}$ & ISCO radius \\
$R_{\mathrm{sh}}$ & Shadow radius & $b_{c}$ & Critical impact parameter \\
$r_{o}$ & Observer position & $r^{*}$ & Tortoise coordinate \\
$V(r,\chi)$ & Null effective potential & $\tilde{V}_{\mathrm{eff}}$ & Massive potential \\
$F_{\mathrm{eff}}$ & Effective force & $V_{s}$ & Scalar potential \\
$V_{v}$ & Vector potential & $V_{t}$ & Tensor potential \\
$V_{\psi}$ & Spinor potential & $\omega$ & Frequency \\
$l$ & Multipole number & $Q$ & Topological charge \\
\hline
\end{tabular}
\caption{Summary of the main parameters and symbols used throughout the manuscript.}
\label{tab:parameters}
\end{table}

}


\section{Geodesics }

In the context of gravitational theories, geodesic motion establishes the bridge between spacetime geometry and the behavior of free particles. The trajectories of such particles reveal the curvature and symmetry properties encoded in the metric. Within bumblebee gravity, this analysis gains particular relevance, since the spontaneous breaking of Lorentz symmetry introduces an additional field that modifies the geometric background itself. Consequently, studying the motion of test particles—both massive and massless—offers an effective means of examining how this Lorentz–violating background alters their paths compared to the predictions of general relativity.

The study of geodesic motion goes beyond merely describing how particles move; it serves as a means to uncover how the geometry—modified by the bumblebee field $B_{\mu}$—affects measurable quantities such as deflection angles, orbital behavior, and photon propagation. Through these trajectories, one can infer the structure of spacetime and assess the influence of the vector field on gravitational phenomena. In general, the motion of a free particle in a curved geometry is governed by the geodesic equation, which can be expressed in covariant form as follows
\begin{equation}
\frac{\mathrm{d}^{2}x^{\mu}}{\mathrm{d}\mathrm{t}^{2}} + \Gamma^\mu_{\nu \lambda} \frac{\mathrm{d}x^{\nu}}{\mathrm{d}\mathrm{t}}\frac{\mathrm{d}x^{\lambda}}{\mathrm{d}\mathrm{t}} = 0.
\label{1g1e1o1d1e1sic1f1u1ll}
\end{equation}
In this context, $\Gamma^\mu_{\nu \lambda}$ represents the Christoffel symbols associated with the metric connection, while $\mathrm{t}$ denotes the affine parameter along the geodesic. The purpose here is to study how the Lorentz--violating parameter $\chi$ affects the trajectories of both photon--like and massive particle modes.

To perform this analysis, the coupled differential equations arising from the geodesic condition in Eq.~\eqref{1g1e1o1d1e1sic1f1u1ll} must be solved. They yield four independent relations, each corresponding to a spacetime coordinate, which are integrated simultaneously using the metric given in Eq.~\eqref{maaaaianametric}. The resulting solutions describe possible departures from general relativity, including changes in orbital stability and deflection behavior, reflecting how the Lorentz--violating geometry alters the motion of test particles, as follows
\begin{align}
&\frac{\mathrm{d}t^{\prime}}{\mathrm{d}\mathrm{t}} =
\frac{2 M r' t'}{2 M r-r^2},\\
&\frac{\mathrm{d}r^{\prime}}{\mathrm{d}\mathrm{t}}  = -\frac{M \left(r'\right)^2}{2 M r-r^2}+ \frac{(2 M-r) \left(M \left(t'\right)^2-r^3 (\chi +1) \left(\left(\theta '\right)^2+\sin ^2(\theta ) \left(\varphi '\right)^2\right)\right)}{r^3 (\chi +1)^2}\\
&\frac{\mathrm{d}\theta^{\prime}}{\mathrm{d}\mathrm{t}} = \sin (\theta ) \cos (\theta ) \left(\varphi '\right)^2-\frac{2 \theta ' r'}{r},\\
&\frac{\mathrm{d}\varphi^{\prime}}{\mathrm{d}\mathrm{t}} = -\frac{2 \varphi ' \left(r'+r \theta ' \cot (\theta )\right)}{r}.
\end{align}
The prime symbol ($'$) is used to denote differentiation along the particle’s path, that is, with respect to the affine parameter that characterizes the evolution of the geodesic.

\begin{figure}
    \centering
     \includegraphics[scale=0.635]{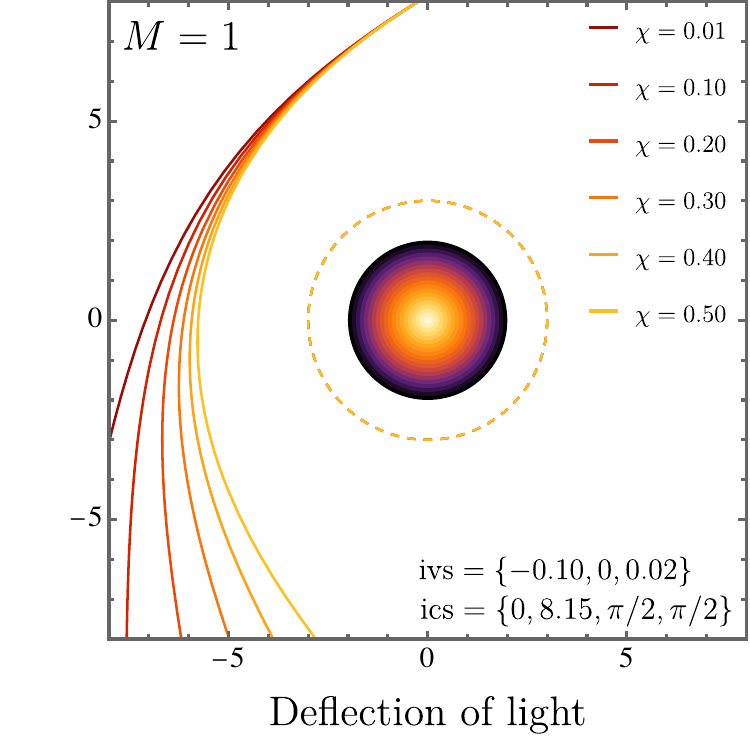}
      \includegraphics[scale=0.635]{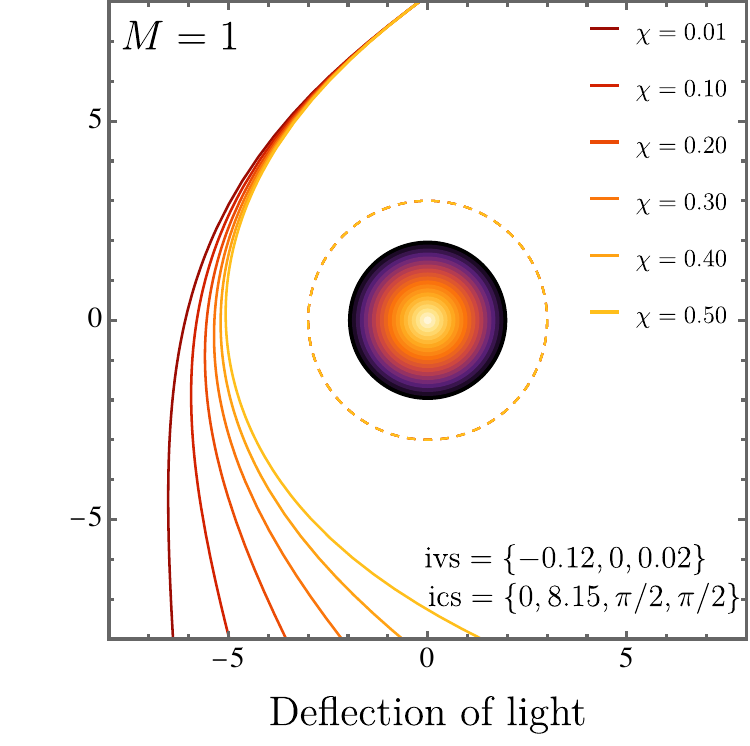}
       \includegraphics[scale=0.635]{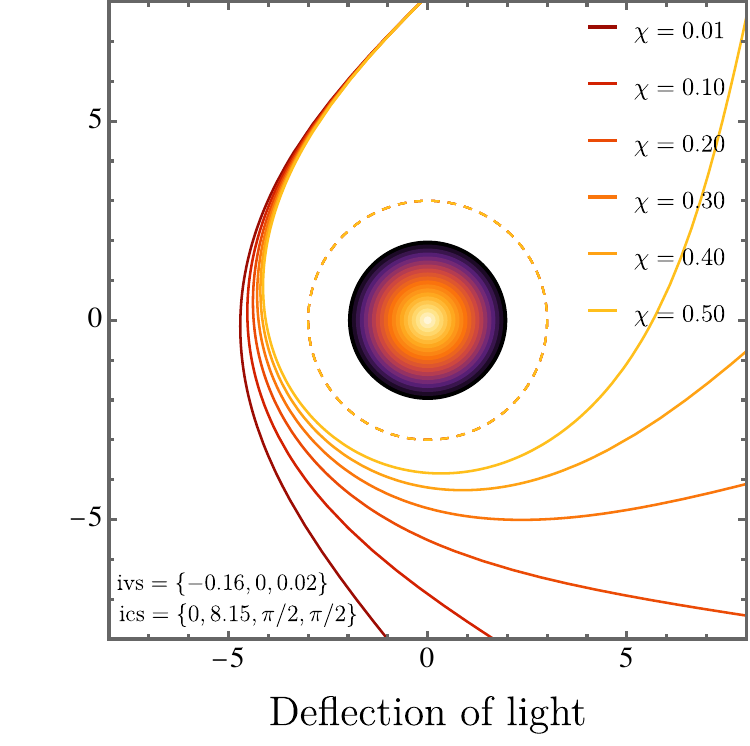}
        \includegraphics[scale=0.635]{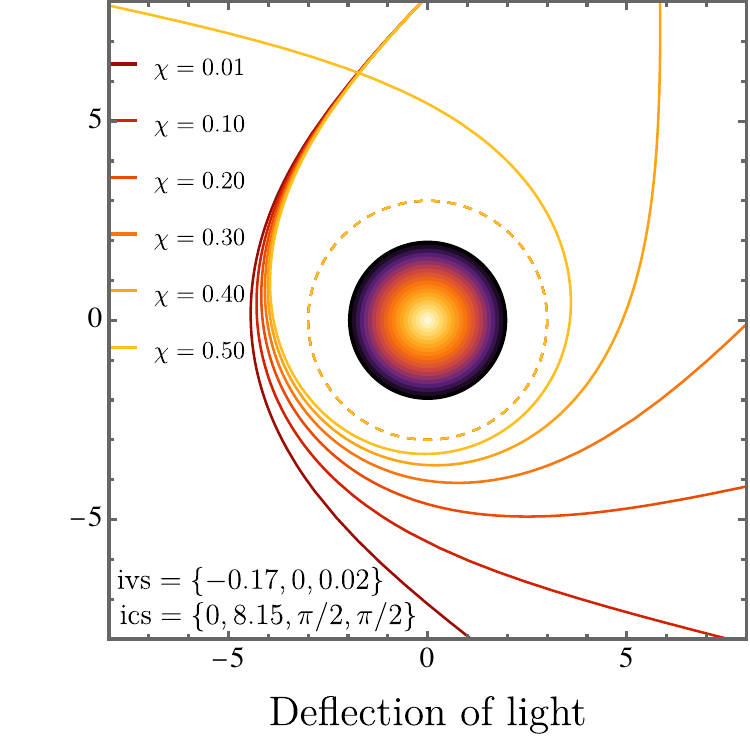}
    \caption{{Light--like geodesics for different values of the Lorentz--violating parameter $\chi$, obtained from the numerical integration of the null geodesic equations for $M=1$. The four panels correspond to the initial values $\mathrm{ivs}=\{-0.10,0,0.02\}$, $\{-0.12,0,0.02\}$, $\{-0.16,0,0.02\}$, and $\{-0.17,0,0.02\}$, respectively, while the same initial conditions $\mathrm{ics}=\{0,8.15,\pi/2,\pi/2\}$ are used in all cases. As $\chi$ increases, the photon trajectories bend more strongly toward the compact object.}}
    \label{geodesicsmassless}
\end{figure}

\begin{figure}
    \centering
     \includegraphics[scale=0.643]{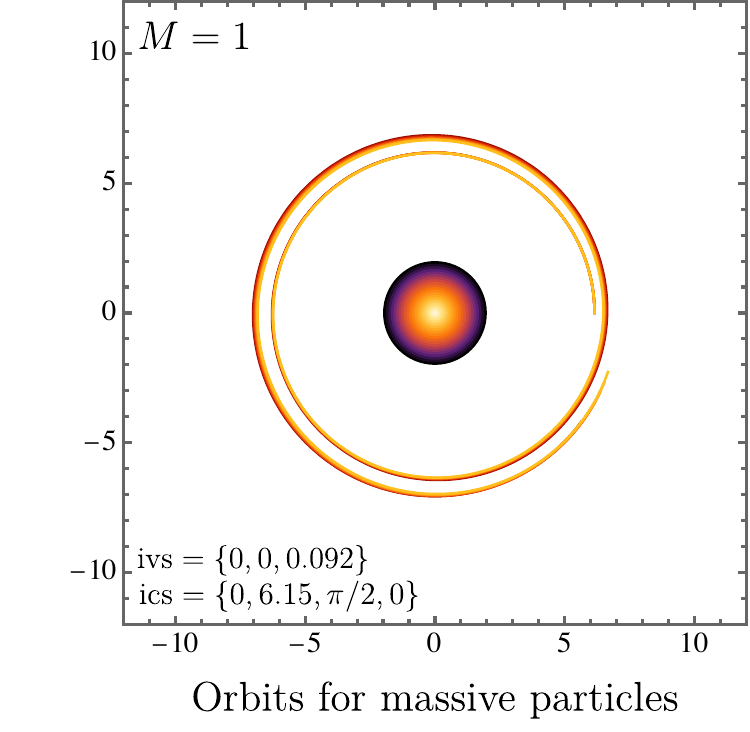}
      \includegraphics[scale=0.643]{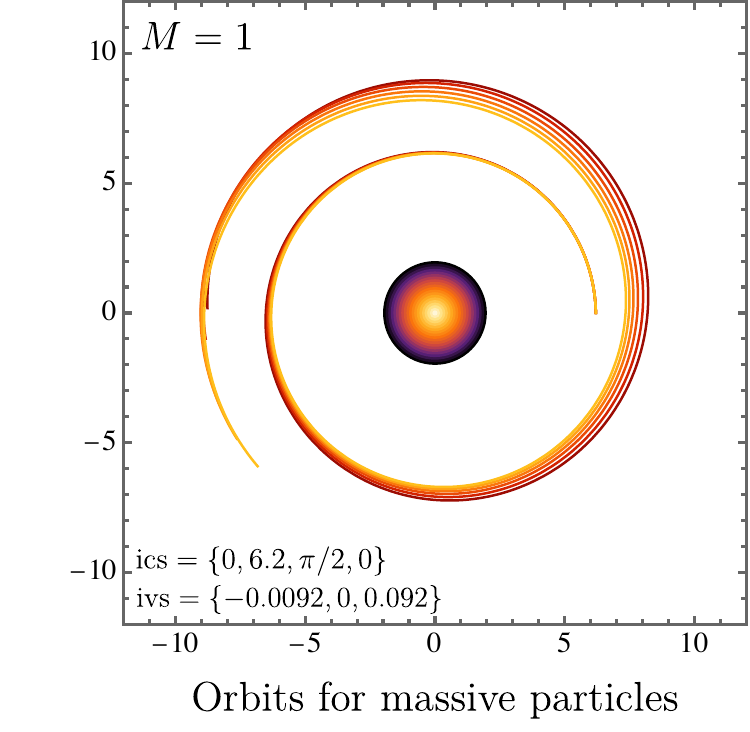}
      \includegraphics[scale=0.643]{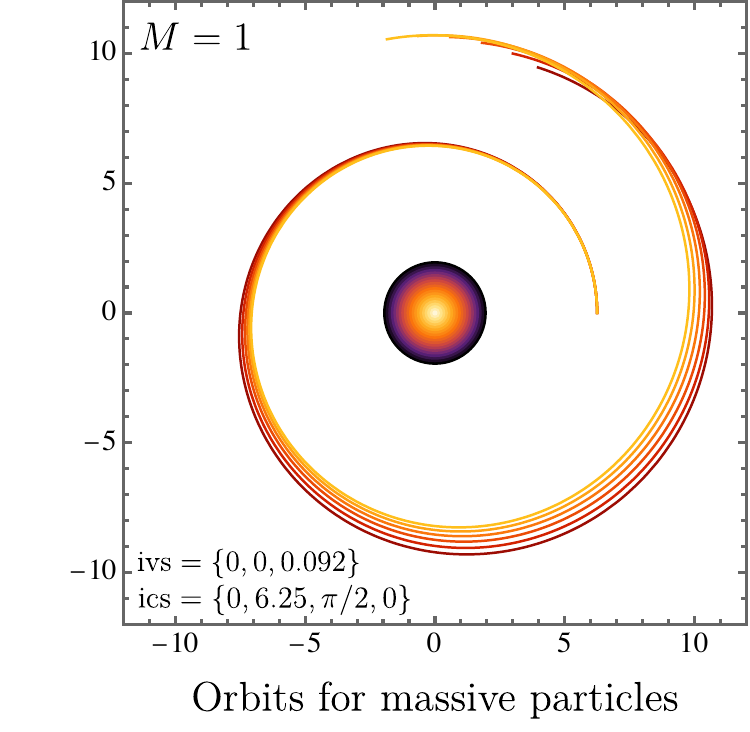}
       \includegraphics[scale=0.643]{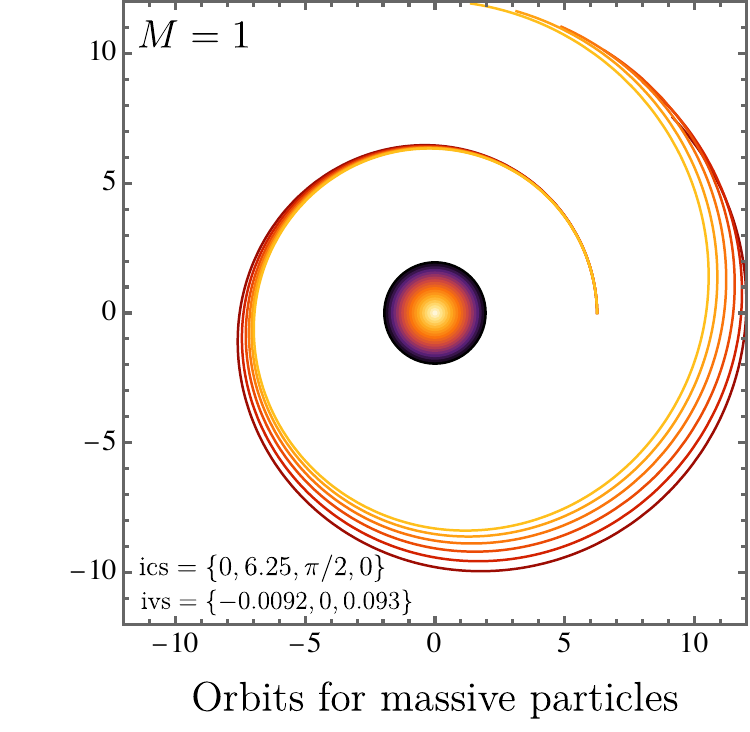}
    \caption{ {Time--like geodesics for different values of the Lorentz--violating parameter $\chi$, obtained from the numerical integration of the timelike geodesic equations for $M=1$. The four panels correspond, respectively, to the pairs $(\mathrm{ivs},\mathrm{ics})=\{0,0,0.092\},\{0,6.15,\pi/2,0\}$, $\{-0.0092,0,0.092\},\{0,6.2,\pi/2,0\}$, $\{0,0,0.092\},\{0,6.25,\pi/2,0\}$, and $\{-0.0092,0,0.093\},\{0,6.25,\pi/2,0\}$. As $\chi$ increases, the particle trajectories bend more strongly toward the compact object.}  }
    \label{geodesicsmassive}
\end{figure}

The system of differential equations was solved numerically, and the corresponding results are displayed in Figs.~\ref{geodesicsmassless} and~\ref{geodesicsmassive}, which illustrate the trajectories of massless (photon–like) and massive particles (time--like ones), respectively. For a fixed set of initial conditions, the numerical solutions show that increasing the parameter $\chi$ produces a contraction effect on both types of trajectories, leading to a noticeable squeezing of the orbits. By verifying the numerical implementation, these outcomes were found to be consistent with the behavior reported for the original bumblebee black hole in Ref.~\cite{Casana:2017jkc}.

A further comparison can be drawn with the recent Kalb--Ramond black hole introduced in Ref.~\cite{Liu:2024oas}. This solution differs from the bumblebee case mainly by the sign of the radial metric component, namely $(1+\ell)$ for the bumblebee configuration and $(1-\ell)$ for the Kalb--Ramond one. As expected, such a change in sign reverses the influence of Lorentz violation on particle motion: whereas the bumblebee parameter $\ell$ causes the trajectories to contract, the Kalb--Ramond counterpart leads to an expansion of the geodesic paths as $\ell$ increases.

\section{Critical orbits and shadows }

For the analysis developed in this section, a general class of spacetime is adopted, whose line element can be expressed in the following general form:
\ie
\mathrm{d}s^{2} = - A(r,\chi) \mathrm{d}t^{2} + \frac{1}{B(r,\chi)} \mathrm{d}r^{2} + C(r,\chi) \mathrm{d}\theta^{2} + D(r,\chi) \mathrm{d}\,\varphi^{2}.
\label{generalll}
\fe

To explore how photons propagate within this background, their motion is examined through the application of the Lagrangian approach, which serves as the foundation for deriving the corresponding equations of motion
\ie
\mathcal{L} =\frac{1}{2} {g_{\mu \nu }}{{\dot x}^\mu }{{\dot x}^\nu } = \epsilon,
\fe
where $\epsilon$ is equal to $0$ and $-1$ for massless and massive particles, respectively. Expressed differently, the formulation can be written as follows
\begin{equation}\label{Eq:L}
2\mathcal{L} = - A(r,\chi){{\dot t}^2} + \frac{1}{B(r,\chi)}{{\dot r}^2} + C(r,\chi){{\dot \theta }^2} + D(r,\chi) {{\dot \varphi }^2}.
\end{equation}
Through the application of the Euler--Lagrange equations, and after restricting the motion to the equatorial plane ($\theta = \pi/2$), two constants naturally emerge from the system’s symmetry: the conserved energy $E$ and angular momentum $L$. Their explicit forms are obtained from the following relations
    \begin{equation}\label{xcaonastaanat}
E = A(r,\chi)\dot t \quad\mathrm{and}\quad L = D(r,\chi)\dot \varphi,
\end{equation}
and taking into account massless particle modes $(\epsilon = 0)$, we get therefore
\begin{equation}\label{xlsaigaahata}
- A(r,\chi){{\dot t}^2} + \frac{1}{B(r,\chi)}{{\dot r}^2} + D(r,\chi){{\dot \varphi }^2} = 0.
\end{equation}
Once the relations in Eq.~(\ref{xcaonastaanat}) are substituted into Eq.~(\ref{xlsaigaahata}) and the algebraic manipulations are completed, the expression reduces to the following form
\begin{equation}
\frac{{{{\dot r}^2}}}{{{{\dot \varphi }^2}}} = {\left(\frac{{\mathrm{d}r}}{{\mathrm{d}\varphi }}\right)^2} = D(r,\chi) B(r,\chi) \left(\frac{{D(r,\chi)}}{{A(r,\chi)}}\frac{{{E^2}}}{{{L^2}}} - 1\right).
\end{equation}
In this manner, we can verify that
\ie
\frac{\mathrm{d}r}{\mathrm{d}\lambda} = \frac{\mathrm{d}r}{\mathrm{d}\varphi} \frac{\mathrm{d}\varphi}{\mathrm{d}\lambda}  = \frac{\mathrm{d}r}{\mathrm{d}\varphi}\frac{L}{D(r,\chi)}, 
\fe
in which
\ie
\Dot{r}^{2} = \left( \frac{\mathrm{d}r}{\mathrm{d}\lambda} \right)^{2} =\left( \frac{\mathrm{d}r}{\mathrm{d}\varphi} \right)^{2} \frac{L^{2}}{D(r,\chi)^{2}}.
\fe
Accordingly, the effective potential $\mathrm{V}(r,\chi)$ takes the following form
\ie
\mathrm{V}(r,\chi) = D(r,\chi) B(r,\chi) \left(\frac{{D(r,\chi)}}{{A(r,\chi)}}\frac{{{E^2}}}{{{L^2}}} - 1\right)\frac{L^{2}}{ D(r,\chi)^{2}}.
\fe

Having set the necessary framework, the next step is to identify the photon sphere. This configuration is obtained by enforcing the condition given below
\begin{equation}
\label{svscsosnsd}
\mathrm{V}(r,\chi)=0, \quad\quad \frac{\mathrm{d} \,{\mathrm{V}(r,\chi)}}{\mathrm{d}r} = 0 .
\end{equation}

Furthermore, defining the critical impact parameter as $b_c = L/E$, the imposed condition yields the following relation
\ie
b_c=\frac{D(r,\chi)}{A(r,\chi)}\Big|_{r=r_{ph}}.
\fe

Proceeding to the evaluation of Eq.~(\ref{svscsosnsd}) using the metric defined in Eq.~(\ref{maaaaianametric}), one obtains a unique physically admissible solution—real and positive—which represents the photonic radius, or equivalently, the critical orbit. The resulting expression for $r_{c}$ is given by
\ie\label{eq:rc}
r_{c} = 3M.
\fe
In other terms, there is no deviation from the Schwarzschild configuration. This outcome equally applies to other bumblebee black hole metrics, whether formulated in the \textit{metric} framework \cite{Casana:2017jkc} or within the \textit{metric--affine} formalism \cite{Filho:2022yrk,Filho:2023etf}. From this point, the discussion can proceed to the shadow radius. Because the spacetime considered here lacks asymptotic flatness, the observer’s position $r_{o}$ must be explicitly included in the computation. Consequently, the shadow radius can be expressed as
\ie
R_{sh} = \sqrt{\frac{A(r_{o})}{A(r_{c})}} r_{c} = r_{c} \sqrt{\frac{2 M r_{c}-r_{c} r_{o}}{2 M r_{o}-r_{c} r_{o}}}.
\fe
It is worth noting that, in the limit $r_{o} \to \infty$, the expression simplifies to
\ie
R_{sh} = 3 \sqrt{3} M,
\fe
where, once more, the Lorentz--violating parameter $\chi$ does not contribute. Therefore, no additional analysis concerning the shadow radius is required, such as those based on the EHT observations of $SgrA^{*}$ and $M87$.

{In particular, the results show that the parameter $\chi$ produces measurable deviations in observables that are not directly associated with the photon sphere or the ISCO, indicating that distinct physical sectors respond differently to the Lorentz--violating deformation. In contrast, as we shall be seeing in the forthcoming sections, $\chi$ enters explicitly in perturbative dynamics, lensing phenomena, time--delay effects, and in the comparison with observational constraints. It is also worth emphasizing that identifying which observables remain unchanged and which are sensitive to $\chi$ is itself a meaningful outcome. This distinction delineates the sectors where Lorentz violation is absent from those where its effects are more directly accessible and, therefore, more suitable for phenomenological investigation.}

{In the following subsection, for clarity and completeness, we compare the results obtained here with those reported in the literature for Lorentz--violating scenarios, including both bumblebee (vector field) and Kalb--Ramond (tensor field) black holes. To present a comprehensive overview, charged and AdS configurations are also taken into account. Furthermore, this analysis revises and corrects a few typographical errors present in earlier published results. }


{\subsection{Comparison among Lorentz-violating vector and tensor field configurations }

As discussed previously, it is instructive to contrast our results with other Lorentz--violating configurations involving both vector and tensor fields—specifically, the bumblebee black holes (including their metric–affine extensions) and the Kalb--Ramond black holes. Except for the AdS geometries, all the cases considered assume the observer to be located at spatial infinity ($r_{o} \to \infty$), ensuring consistency in the comparison of the shadow radii.

It is worth mentioning that some typographical errors appeared in previous studies. In Refs.~\cite{Filho:2023ycx,AraujoFilho:2024ctw}, the expression for the Kalb--Ramond black hole (Model 1) shadow contained a typo; the correct form is $R_{sh} = 3\sqrt{3}(1-\ell)M$, as listed in Table~\ref{photonshadow}. Likewise, Refs.~\cite{Filho:2023etf,AraujoFilho:2025hkm} contain a similar issue for the bumblebee black hole in the \textit{metric–affine} framework—the accurate result is $R_{sh} = 3\sqrt{3}M$, which is also displayed in the table.

The comparative analysis in Table~\ref{photonshadow} includes a wide range of Lorentz–violating scenarios, such as the standard bumblebee case~\cite{Casana:2017jkc}, its AdS extension~\cite{Maluf:2020kgf}, the charged and charged--AdS bumblebee black holes~\cite{Liu:2024axg}, the metric–affine bumblebee formulation~\cite{Filho:2022yrk}, as well as several Kalb--Ramond configurations—Model~1 and Model~2, along with their AdS and charged counterparts~\cite{Yang:2023wtu,Liu:2024oas,Duan:2023gng}.

\begin{table}[!h]
\centering
\caption{\label{photonshadow}
Comparison of the critical orbits and shadow radii for the existing Lorentz–violating configurations associated with bumblebee and Kalb–Ramond black holes. In this context, $\ell$ represents the Lorentz–violating parameter. Also, we have redefined some parameters, such as, $\Xi \equiv \, 3 (\ell+2) M +  \sqrt{(l+2) \left(9 (\ell+2) M^2-16 (\ell+1) Q^2\right)}$, $\Tilde{\gamma} = 3 Q^2-(\ell-1) r_{o} \left(6 (\ell-1) M-\Lambda  r_{o}^3+3 r_{o}\right)$, $\gamma = \frac{1}{1-\ell} -\frac{3 (\ell-1) M^2 \left(\ell-4 \Lambda  Q^2-1\right)}{8 Q^2}+\frac{(\ell-1) M \sqrt{\frac{8 Q^2}{(\ell-1)^3}+9 M^2} \left(\ell+4 \Lambda  Q^2-1\right)}{8 Q^2}+\frac{3 \ell +4 \Lambda  Q^2-3}{6 (\ell-1)^2}$, $\Tilde{\mu} = \sqrt{2 M \sqrt{(l+2) \left(9 (l+2) M^2-16 (l+1) Q^2\right)}+6 (l+2) M^2-8 (l+1) Q^2}$.}
\renewcommand{\arraystretch}{1.25}
\setlength{\tabcolsep}{6pt}
\hspace*{-2cm}\begin{tabular}{l|c|c}
\hline\hline\hline
\textbf{Black holes} & \textbf{Photon sphere} & \textbf{Shadow radii } \\
\hline
This work  &
$3M$ &
$3 \sqrt{3} M$ \\[4pt]
Bumblebee \cite{Casana:2017jkc} &
$3M$ &
$3 \sqrt{3} M$ \\[4pt]
Bumblebee Ads \cite{Maluf:2020kgf} &
$3M$ &
$3 M \sqrt{\frac{\Lambda  (\ell+1) r_{o}^3+6 M-3 r_{o}}{r_{o} \left(9 \Lambda  (\ell+1) M^2-1\right)}}$ \\[4pt]
Charged bumblebee \cite{Liu:2024axg}  &
$ \frac{3 M}{2} + \frac{\sqrt{(\ell+2) \left(9 (\ell+2) M^2-16 (\ell+1) Q^2\right)}}{2 (\ell+2)}$ &
$\frac{\Xi^2}{2 (\ell+2)^{3/2} \Tilde{\mu}}$ \\[4pt]
Charged bumblebee Ads \cite{Liu:2024axg} &
$\frac{3 M}{2} + \frac{\sqrt{(\ell+2) \left(9 (\ell+2) M^2-16 (\ell+1) Q^2\right)}}{2 (\ell+2)}$ &
$\frac{\Xi  \sqrt{\frac{1-\frac{-\frac{6 (\ell+1) Q^2}{\ell+2}+\Lambda  (\ell+1) r_{o}^4+6 M r_{o}}{3 r_{o}^2}}{-\frac{\Lambda  (\ell+1) \Xi ^2}{12 (\ell+2)^2}-\frac{4 (\ell+2) M}{\Xi }+\frac{8 (\ell+1) (\ell+2) Q^2}{\Xi ^2}+1}}}{2 (\ell+2)}$ \\[4pt]
Bumblebee (metric--affine) \cite{Filho:2022yrk} &
$3M$ &
$3\sqrt{3}M$ \\[4pt]
Kalb--Ramond (Model~1) \cite{Yang:2023wtu} &
$3 (1-\ell) M$ &
$3 \sqrt{3} (1-\ell) M$ \\[4pt]
Kalb--Ramond Ads (Model~1) \cite{Yang:2023wtu} &
$3 (1-\ell) M$ &
$3 (1-\ell) M \sqrt{\frac{-6 \ell M+6 M+\Lambda  r_{o}^3-3 r_{o}}{r_{o} \left(9 \Lambda  (\ell-1)^2 M^2-1\right)}}$ \\[4pt]
Kalb--Ramond (Model~2) \cite{Liu:2024oas} &
$3M$ &
$3\sqrt{3}M$ \\[4pt]
Kalb--Ramond Ads (Model~2) \cite{Liu:2024oas} &
$3M$ &
$3 M \sqrt{\frac{\Lambda  (\ell-1) r_{o}^3-6 M+3 r_{o}}{9 \Lambda  (\ell-1) M^2 r_{o}+r_{o}}}$ \\[4pt]
Charged Kalb-Ramond \cite{Duan:2023gng} &
$\frac{3 (1 - \ell) M}{2} \left( 1 + \sqrt{1 - \frac{8 Q^{2}}{9 (1 - \ell)^{3} M^{2}}} \right)$ &
$\frac{3  M}{2} \left( 1 + \sqrt{1 - \frac{8 Q^{2}}{9 M^{2}}} \right)$ \\[4pt]
Charged Kalb-Ramond Ads \cite{Duan:2023gng} &
$\frac{3 (1 - \ell) M}{2} \left( 1 + \sqrt{1 - \frac{8 Q^{2}}{9 (1 - \ell)^{3} M^{2}}} \right)$ &
$\frac{(1-\ell) M}{2 \sqrt{3}} \left(3 \sqrt{\frac{8 Q^2}{9 (\ell-1)^3 M^2}+1}+3\right) \sqrt{\frac{\Tilde{\gamma} }{\gamma  (\ell-1)^2 r_{o}^2}}$ \\[4pt]
\hline
\multicolumn{3}{l}{Note: For all these spacetimes, the observer’s position affects the shadow radius, which is given by $R_{\text{sh}} = \sqrt{\frac{A(r_{o})}{A(r_{c})}} r_{c}$.}\\
\hline\hline\hline
\end{tabular}
\end{table}

}

\subsection{Topological features}

The topological approach has recently emerged as a powerful framework for analyzing the stability and classification of photon spheres \cite{Wei2020,Cunha2020,Sadeghi2024,BahrozBrzo2025,alipour2024weak,Heidari:2025sku}. 
Motivated by these developments, we extend our study to explore the topological characteristics of the photon radius within the present model. 
In this context, the effective potential governing the topological structure of the photon sphere is defined as

\begin{align}\label{eq:Hr}
\mathcal{H}(r, \theta) &=  \sqrt{\frac{A(r,\chi)}{D(r,\chi)}}\\
&=\csc\theta\sqrt{\frac{ (r-2 M)}{r^3 (\chi +1)}}.
\end{align}

\begin{figure}[ht]
    \centering
\includegraphics[width=0.7\linewidth]{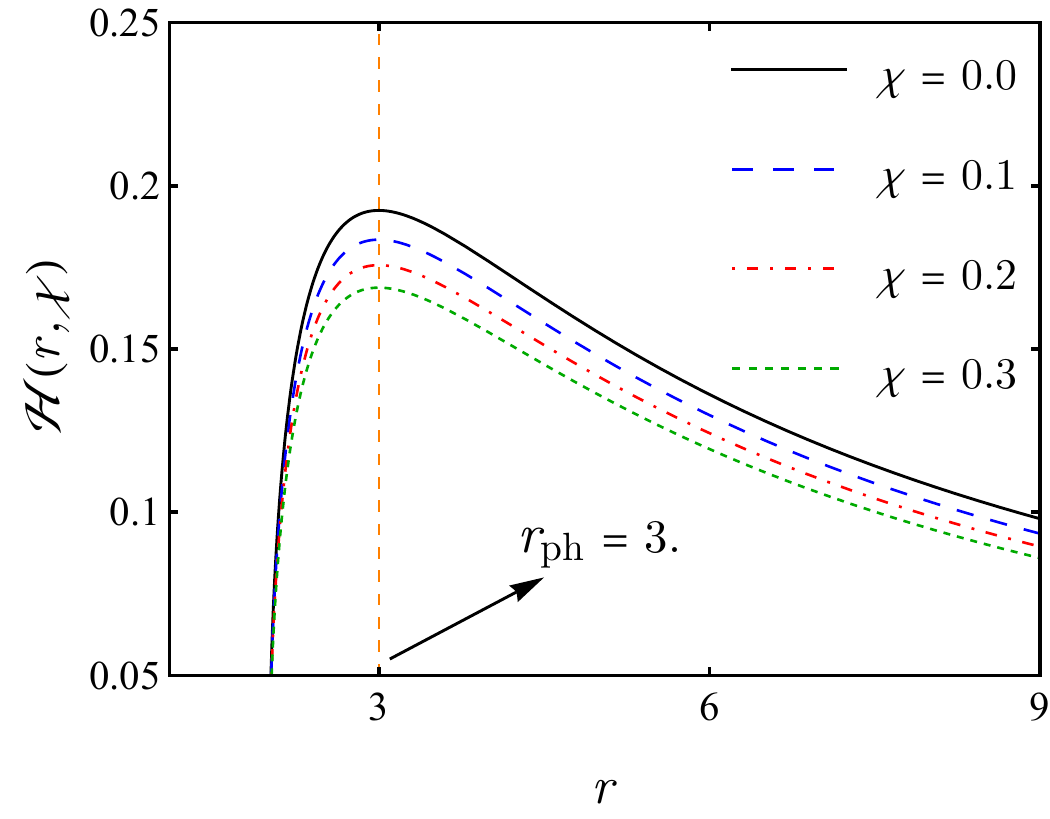}
    \caption{The potential $\mathcal{H}(r,\chi)$ for various values of $\chi$, showing a maximum at the photonic radius $r_{\text{ph}} = 3$. The mass is fixed at $M = 1$.}
    \label{fig:Hr}
\end{figure}

{\color{black}
The radial dependence of the effective potential $\mathcal{H}(r,\chi)$ is illustrated in Fig.~\ref{fig:Hr}. The photon sphere corresponds to the critical points of this potential, determined by the condition $\partial_r \mathcal{H} = 0$. For a unit mass black hole ($M = 1$) and for variations of the bumblebee parameter in the range $\chi = 0$ – $0.3$, the potential exhibits a maximum at $r_{\text{ph}} = 3$, indicating an unstable equilibrium for the photon orbit. This radius coincides with the photon sphere of the Schwarzschild spacetime, implying that the bumblebee framework does not modify the location of the photonic radius. This result is consistent with the previous discussion on the Eq.~\eqref{eq:rc}.
}

{\color{black}
To gain further insight into the topology of the photonic sphere, we introduce a two-dimensional vector field 
\(\boldsymbol{\Phi} = (\Phi_{r}, \Phi_{\theta})\), whose components are defined as
\begin{equation}
\Phi_{r} = \frac{\partial_{r} \mathcal{H}}{\sqrt{g_{rr}}}
           = \sqrt{A(r)}\, \partial_{r}\mathcal{H}, 
\qquad 
\Phi_{\theta} = \frac{\partial_{\theta}\mathcal{H}}{\sqrt{g_{\theta\theta}}}
           = \frac{1}{r}\partial_{\theta}\mathcal{H}.
\label{eq:PhiComponents}
\end{equation}
This vector field allows us to visualize the local behavior of the potential \(\mathcal{H}\) on the \((r,\theta)\) plane and to identify the zero points associated with the photon sphere configuration.

An important topological quantity associated with this field is the winding number \cite{Duane1984}, which characterizes the behavior of \(\boldsymbol{\Phi}\) around a closed contour \(C_i\). It can be defined as
\begin{equation}
\omega_i = \frac{1}{2\pi} \oint_{C_i} d\Omega,
\qquad \Omega = \frac{\Phi_{r}}{\Phi_{\theta}},
\label{eq:winding}
\end{equation}
and the corresponding total topological charge of the system is given by the sum of all winding numbers,
\begin{equation}
Q = \sum_i \omega_i.
\label{eq:charge}
\end{equation}
Applying the definition of potential $\mathcal{H}$ in the Eqs.~\ref{eq:PhiComponents}), the explicit expressions for the components of the vector field are obtained as
\begin{equation}
\Phi_{r} =\frac{\csc \theta  (3 M-r)}{r^3 (\chi +1)},
\label{eq:Phir}
\end{equation}
and
\begin{equation}
\Phi_{\theta} = - \cot \theta \csc \theta  \sqrt{\frac{r-2 M}{r^5 (\chi +1)}}
\label{eq:Phitheta}.
\end{equation}

A zero point of the vector field \(\boldsymbol{\Phi}\) enclosed by a closed curve corresponds to a nontrivial topological charge. Each photon sphere can thus be assigned a definite charge, typically \(+1\) or \(-1\), depending on the orientation of the field around the zero point. The total charge of the system, determined by the chosen closed contour, may take the values \(-1\), \(0\), or \(+1\), depending on whether one, multiple, or no zero points are enclosed. This topological interpretation links the existence and stability of photon spheres to intrinsic geometric properties of the spacetime manifold, as discussed in detail in Ref.~\cite{Wei2020,Sadeghi2024}.

The normalized vector field can be defined as
\begin{equation}
\hat{n}_{j} = \frac{\Phi_{j}}{\|\boldsymbol{\Phi}\|},
\qquad j = (r,\,\theta),
\end{equation}
where $\|\boldsymbol{\Phi}\| = \sqrt{\Phi_{r}^{2} + \Phi_{\theta}^{2}}$ denotes the magnitude of the vector field. This normalization facilitates the visualization of the vector orientation and simplifies the identification of critical points in the photon sphere topology. The vector field structure associated with the photon sphere is depicted in Fig.~\ref{fig:Rphvector}. As shown in the figure, a single critical point appears outside the event horizon, marked by a dot at the photonic radius $r_{\text{ph}} = 3$. This critical point carries a topological charge of $-1$, identifying it as an unstable photon sphere configuration. The direction and convergence pattern of the vector field around this point further confirm its repulsive nature, consistent with the expected instability of null circular orbits in this spacetime.

\begin{figure}[ht]
    \centering
\includegraphics[width=0.7\linewidth]{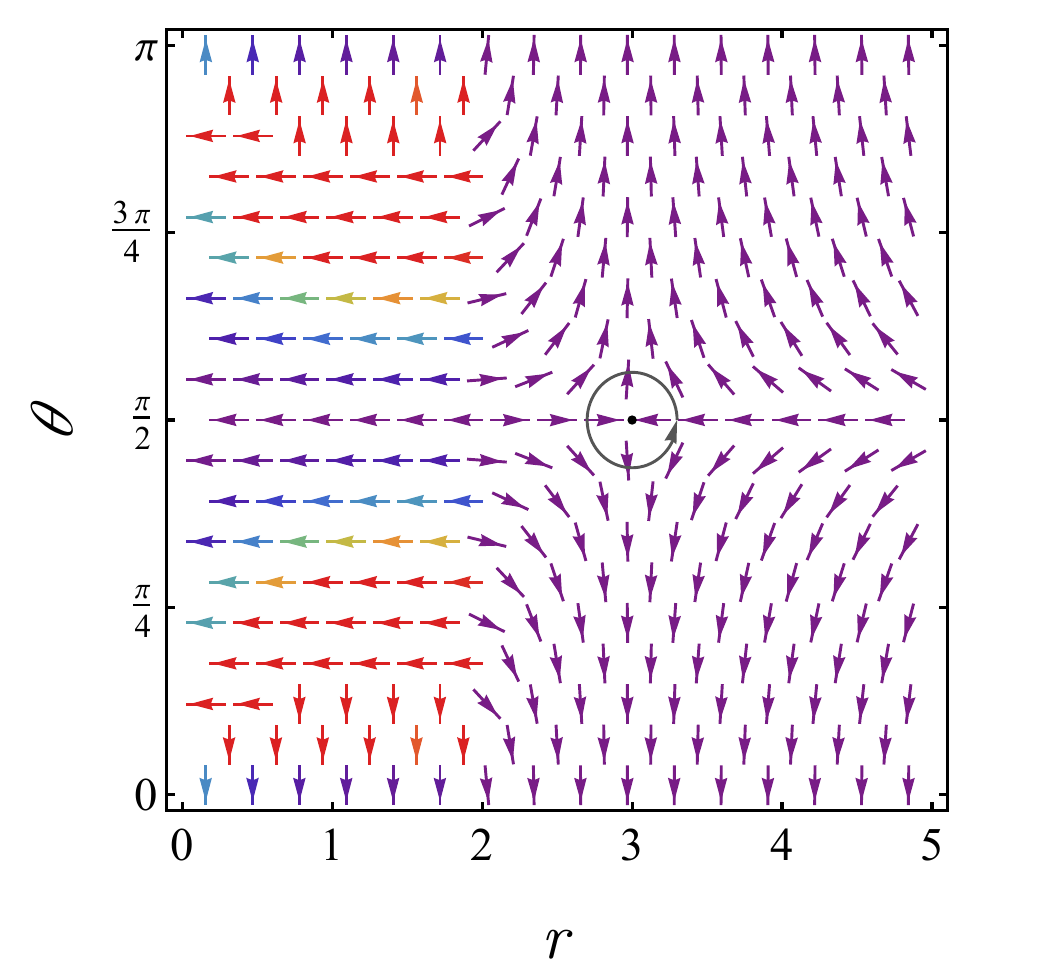}
    \caption{The normalized vector field in the $(r,\theta)$ plane with encircled photon sphere point at $r_{\text{ph}} = 3$. The other parameters are fixed at $M = 1$, $\chi = 0.1$.}
    \label{fig:Rphvector}
\end{figure}


\section{Dynamics of massive particles}
The motion of a neutral massive particle $(\epsilon = -1)$ can be investigated via the Lagrangian in Eq. \eqref{Eq:L}, which leads to the following expression

\begin{equation}\label{xlsaigaahata1}
- A(r,\chi){{\dot t}^2} + \frac{1}{B(r,\chi)}{{\dot r}^2} + D(r,\chi){{\dot \varphi }^2} = -1.
\end{equation}
By substituting the constant of motion in Eq. \eqref{xcaonastaanat}, the radial equation can be described as
\begin{equation}
    {\dot{r}}^2+\tilde{V}_{\text{eff}}=E^2.
\end{equation}
Here $\tilde{V}_{\text{eff}}$ is called the effective potential for the neutral massive particle and is defined by
\begin{align} \label{Veff2}                   \tilde{V}_{\text{eff}}&=A(r,\chi)\left(1+\frac{L^2}{r^2}\right)\\
&=\frac{1}{1+\chi}\left(1 - \frac{2M}{r}         \right).
\end{align}

\subsection{The effective force}
This effective potential plays a crucial role in the motion of the test particle through the effective force acting on it. To further explore the particle dynamics, we compute the effective force, defined as
\begin{equation}
    F_{\text{eff}}=-\frac{\mathrm{d} \tilde{V}_{\text{eff}} }{\mathrm{d} r}.
\end{equation}
By substituting the effective potential from Eq. \eqref{Veff2}, the effective force in the bumblebee framework takes the following form
\begin{equation}
     F_{\text{eff}}=\frac{1}{1
     +\chi }\Big(-\frac{M}{r^2 }-\frac{3 L^2 M}{r^4}+\frac{L^2}{r^3 }\Big).
\end{equation}

The structure of this expression closely resembles the Schwarzschild case, but each term is rescaled by the factor $(1+\chi)^{-1}$, reflecting the influence of Lorentz symmetry breaking through the bumblebee field.
The first contribution, $\frac{1}{1+\chi}\left(\frac{-M}{r^2}\right)$, corresponds to the usual Newtonian gravitational attraction, whose effective strength is weakened for $\chi>0$ and enhanced for $-1<\chi<0$.

The second contribution, $\frac{1}{1+\chi}\left(\frac{L^{2}}{r^{3}}\right)$, represents the outward centrifugal force arising from angular momentum, and is modified in the same manner by the bumblebee coupling. The final term, $\frac{1}{1+\chi}\left(\frac{-3L^{2}M}{r^{4}}\right)$, is the general-relativistic curvature correction that increases the inward force at small radii; it is responsible for relativistic effects such as the shift of circular orbits and the existence of an ISCO. Since all terms scale uniformly with $(\chi+1)^{-1}$, the qualitative structure of the effective force and the corresponding effective potential remains similar to the Schwarzschild case; For further investigation of the impact of the Lorentz-violating parameter on the orbital radius, both $\tilde{V}_{\text{eff}}$ and $F_{\text{eff}}$ are demonstarted in Fig.~\ref{fig:VFeffL} and Fig.~\ref{fig:VFeffchi} for variation of of $\chi$ and $L$, respectively.

In Fig.~\ref{fig:VFeffL}, the upper panel shows the behaviour of the effective potential as a function of the radial coordinate $r$. As the Lorentz-violating parameter increases from $\chi=0$ to $\chi=0.5$, the height of $\tilde{V}_{\text{eff}}$ decreases, whereas the location of its maximum remains unchanged for all values of $\chi$, as indicated by the dashed line. The lower panel presents the corresponding effective force, also plotted against $r$, and demonstrates that for every considered value of $\chi$ the force changes sign at the same radial position, switching from repulsive ($F_{\text{eff}}>0$) to attractive ($F_{\text{eff}}<0$) exactly where predicted by the extremum of the potential. This behavior highlights a fundamental feature of the bumblebee framework: the radial positions of circular orbits are invariant under variations of the Lorentz--violating parameter.

Both the potential maximum and the zero--crossing of the force occur at the same radius for all values of $\chi$, confirming that the condition $\mathrm{d}\Tilde{V}_{\text{eff}}/\mathrm{d}r=0$ depends solely on the mass $M$ and angular momentum $L$, as in standard General Relativity. While $\chi$ modifies the quantitative shape of the effective potential—altering the barrier height and curvature—it leaves the underlying orbital structure unchanged. Consequently, observational signatures based exclusively on circular-orbit radii would remain indistinguishable from their GR counterparts, whereas $\chi$-dependent effects are expected to manifest in more sensitive dynamical quantities.

\begin{figure}
    \centering
     \includegraphics[scale=0.55]{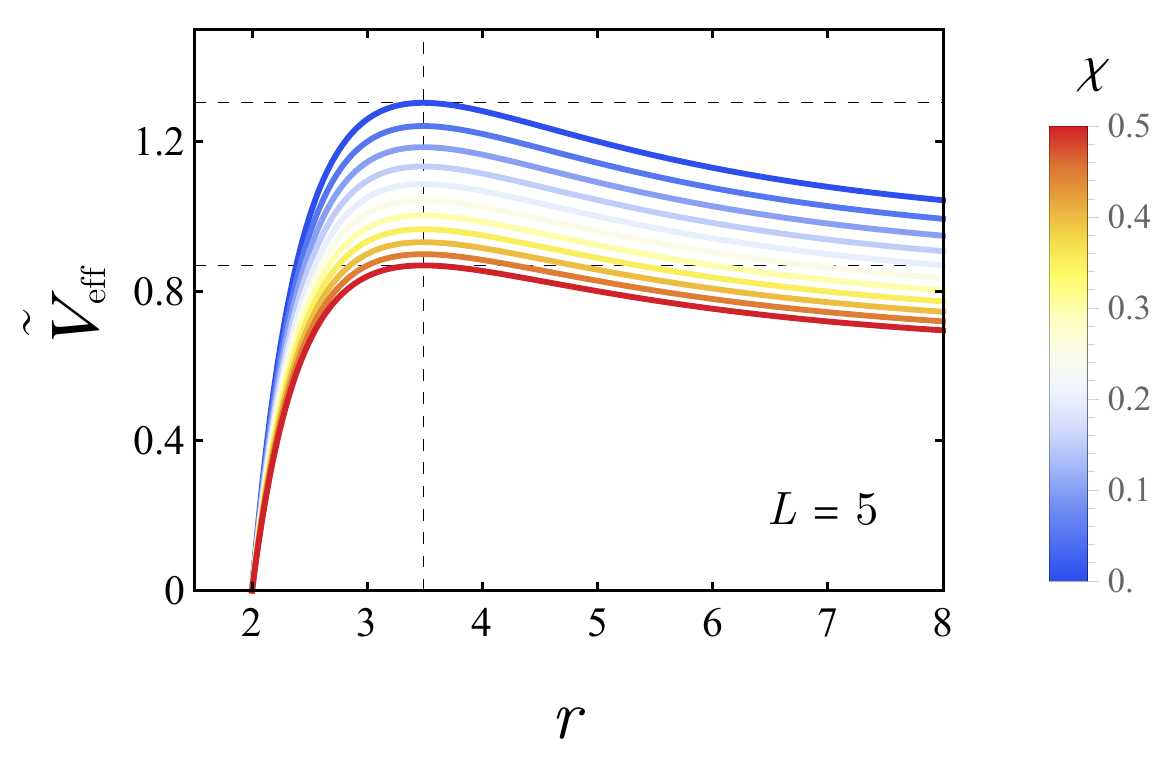}
      \includegraphics[scale=0.55]{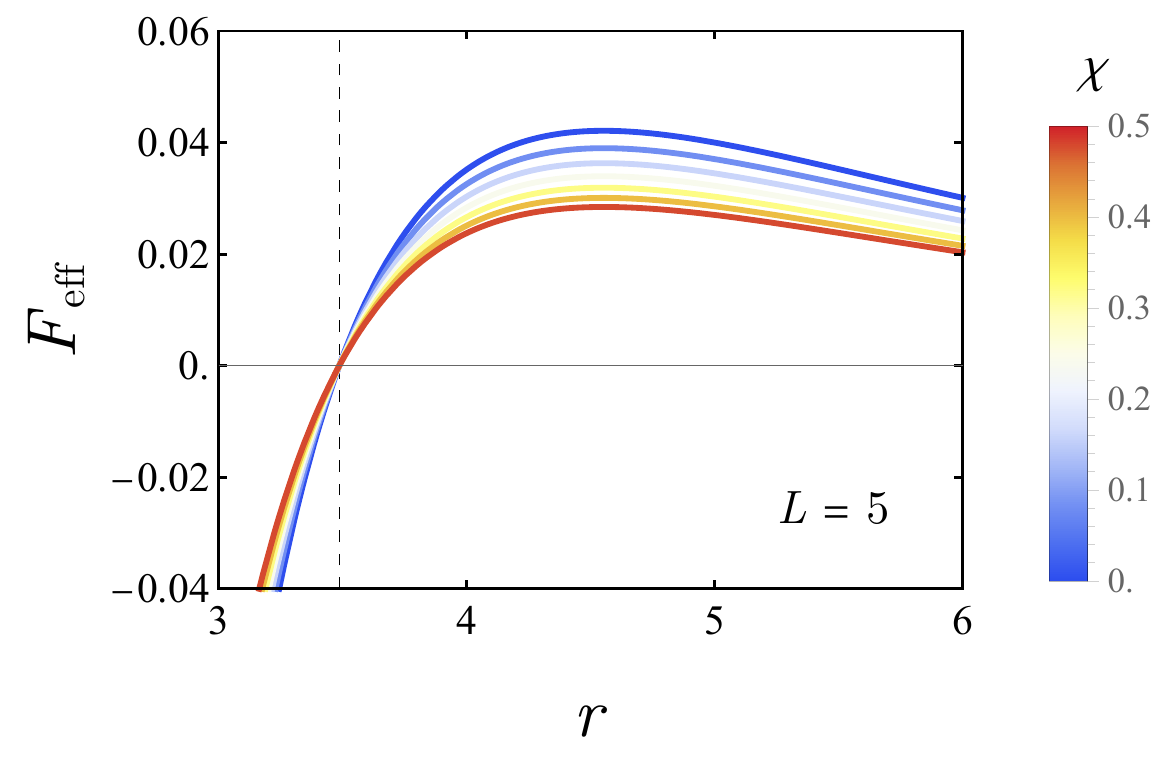}
    \caption{The effective potential and the corresponding effective force on a test particle for variation of the Lorentz--violating parameter $\chi$. The mass and angular momentum are set at $M = 1$ and $L = 5$.}
    \label{fig:VFeffL}
\end{figure}

The effective potential and the effective force on a test particle for different values of the angular momentum $L$ are shown in Fig.~\ref{fig:VFeffchi} and Fig.~\ref{fig:VFeffchi} with $M=1$ and $\chi=0.1$ held fixed. Increasing $L$ raises the height of the potential barrier and shifts the maximum of $\tilde{V}_{\mathrm{eff}}$ inward. The corresponding effective force shows a transition from attraction to repulsion, and the radius at which this sign change occurs moves to smaller values as $L$ decreases, as shown by the dashed lines.

\begin{figure}
    \centering
     \includegraphics[scale=0.55]{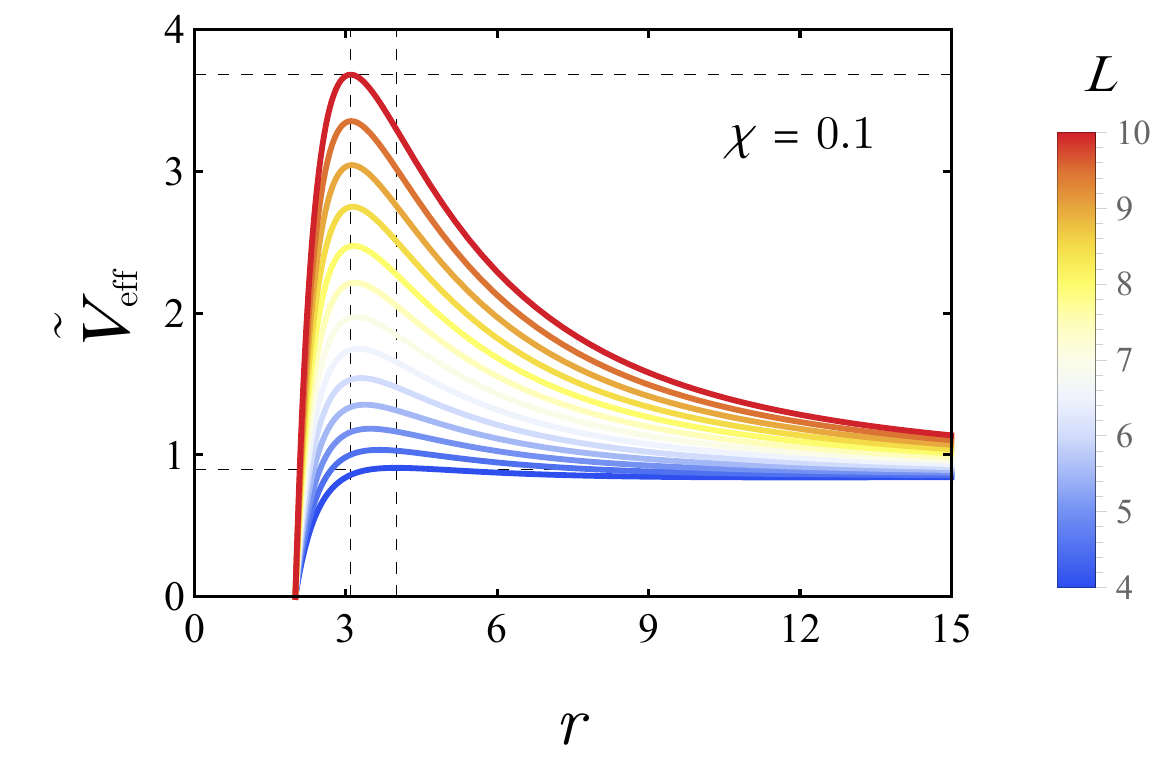}
      \includegraphics[scale=0.55]{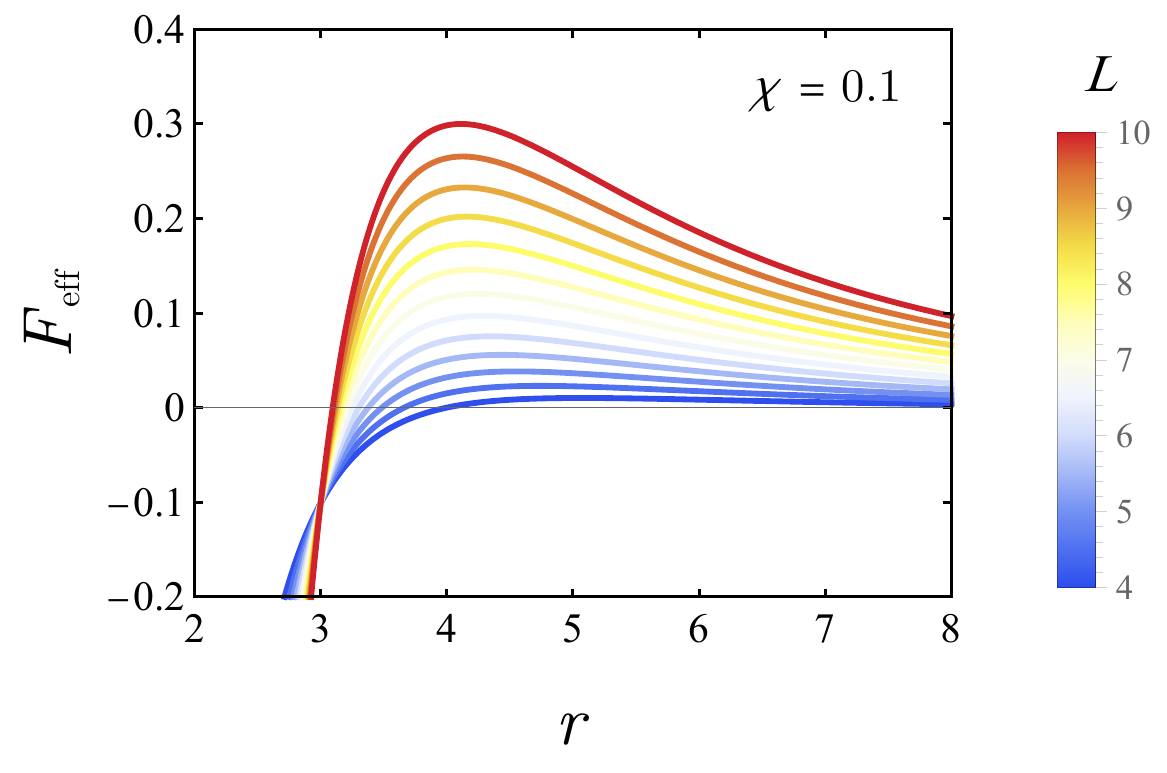}
    \caption{The effective potential and the effective force on a test particle for variation of the angular momentum $L$. The mass and Lorentz--violating parameter are fixed at $M = 1$ and $\chi = 0.1$.}
    \label{fig:VFeffchi}
\end{figure}


\subsection{Innermost stable circular orbits}

Circular orbits are defined by the conditions $\dot{r}=0$ and $\tilde{V}_{\text{eff}}'(r)=0$. Solving the latter yields the angular momentum required for a circular orbit at radius $r$,
\begin{equation}
L^{2}=\frac{M r^{2}}{r-3M},
\end{equation}
which is identical to the Schwarzschild result and notably independent of the Lorentz--violating parameter $\chi$. The innermost stable circular orbit (ISCO) corresponds to the smallest radius at which circular orbits remain stable, determined by the marginal-stability condition $\tilde{V}_{\text{eff}}''(r)=0$. Evaluating the second derivative,
\begin{equation}
\frac{d^{2}\tilde{V}_{\text{eff}}}{dr^{2}}
=\frac{1}{1+\chi}
\left[
-\frac{4M}{r^{3}}
-\frac{24 M L^{2}}{r^{5}}
+\frac{6 L^{2}}{r^{4}}
\right],
\end{equation}
and substituting the circular-orbit relation $L^{2}=M r^{2}/(r-3M)$ gives
\begin{equation}
-2M r^{2}
-12M \left( \frac{M r^{2}}{r-3M} \right)
+3r \left( \frac{M r^{2}}{r-3M} \right)=0.
\end{equation}
Solving this equation yields the ISCO radius
\begin{equation}
r_{\text{ISCO}}=6M,
\end{equation}
which is likewise independent of the Lorentz--violating parameter. The associated specific angular momentum is $L_{\text{ISCO}}=2\sqrt{3}\,M$. In contrast, the specific energy at the ISCO acquires a $\chi$-dependence,
\begin{equation}
E_{\text{ISCO}}=\frac{2\sqrt{2}\,M}{3\sqrt{1+\chi}}.
\end{equation}

These results admit a clear physical interpretation: although the bumblebee parameter $\chi$ modifies the effective gravitational dynamics, it does not affect the geometric location of circular orbits or the ISCO radius itself. Instead, $\chi$ alters the energetics of the motion, implying that observable quantities such as the accretion--disk efficiency or the radiative output may carry signatures of Lorentz violation even though the underlying orbital structure remains unchanged.


\section{Deriving the effective potentials }

A central approach to studying quantum fields in a curved background consists of mapping their dynamics onto an equivalent one--dimensional scattering framework. In the cases of bosons and fermions—governed by the Klein--Gordon and Dirac equations, respectively—this procedure begins with the separation of angular coordinates, which leads to a purely radial equation. The latter can then be rewritten in a Schrödinger--like form involving an effective potential, denoted by $\mathrm{V}_{s,v,t,\psi}$.

In this notation, $\mathrm{V}_{s}$, $\mathrm{V}_{v}$, $\mathrm{V}_{t}$, and $\mathrm{V}_{\psi}$ correspond to scalar, vector, tensor, and spinor perturbations, respectively. The role of this potential is crucial, as it determines the propagation of waves through the background, dictating reflection, transmission, and the resulting physical quantities such as quasinormal spectra, absorption rates, and greybody factors. The next stage consists in deriving the explicit expressions of $\mathrm{V}_{s,v,t,\psi}$ for each spin sector, which will subsequently enable the analysis of quasinormal spectra and the corresponding time--domain evolution.

\subsection{Bosons }

The behavior of bosonic perturbations with spin values $s = 0, 1,$ and $2$ can be effectively reformulated as a one--dimensional wave equation of Schrödinger type.
To investigate these cases, the analysis begins from the covariant field equations corresponding to each spin sector, from which the associated effective potentials are derived. For the scalar configuration ($s=0$), the procedure starts with the Klein–Gordon equation
\ie
\label{klein}
\frac{1}{\sqrt{-g}} \partial_\mu \!\left( \sqrt{-g}\, g^{\mu\nu} \partial_\nu \Psi \right) = 0.
\fe
By introducing a separation of variables ansatz, the scalar field can be decomposed into its temporal, radial, and angular components, allowing the original covariant equation to decouple into independent parts
\ie
\label{ansatz}
    \Psi_{\omega \ell m}(\mathbf{r},t) = \frac{\psi_{\omega \ell}(r)}{r} Y_{\ell m}(\theta, \varphi) e^{-i\omega t},
\fe
and taking into account the so--called tortoise coordinate $r^{*}$, we obtain
\ie
\label{rdsdtdadr}
\mathrm{d}r^{*} = \frac{\mathrm{d}r}{\sqrt{A(r,\chi)B(r,\chi)}},
\fe
the Klein--Gordon equation takes the form of a one--dimensional Schrödinger-type equation for the radial mode, as follows
\ie
\label{waves}
    \left[\frac{\mathrm{d}^2}{\mathrm{d}r^{*2}} + \bigl(\omega^2 - \mathrm{V}_{s}\bigr)\right] \psi_{\omega \ell}(r) = 0.
\fe

In the case of vector perturbations ($s=1$), the dynamics are dictated by the Proca equation,
\ie
\label{proca}
\nabla_\nu F^{\mu\nu} + m^2 A^\mu = 0,
\fe
which, once the angular dependence is separated and the radial components are properly addressed, which yields a single master function governed by a radial equation analogous to Eq.~\eqref{waves}.

For tensor modes ($s=2$), linear perturbations of the Einstein equations give rise to the Regge--Wheeler (axial) and Zerilli (polar) equations, both describing gravitational perturbations through distinct potentials $\mathrm{V}_{t}$. The present treatment considers only the axial sector.

Therefore, in a static and spherically symmetric geometry, the scalar, vector, and tensor perturbations can all be expressed through a common master equation, differing solely in their effective potentials. For the metric defined in Eq.~(\ref{generalll}), the corresponding potentials are given for the scalar case in Refs.~\cite{Heidari:2024bkm,AraujoFilho:2024xhm,AraujoFilho:2024lsi}
\ie
\mathrm{V}_{s}  = A(r,\chi)\left[\frac{{l(l + 1)}}{{{r^2}}} + \frac{1}{{r\sqrt {{A(r,\chi)}{B(r,\chi)^{ - 1}}} }}\frac{\mathrm{d}}{{\mathrm{d}r}}\sqrt {{A(r,\chi)}B(r,\chi)}\right]
\label{effssssss}
\fe
vector \cite{Baruah:2025ifh,Filho:2023abd,Filho:2024ilq}
\ie
\mathrm{V}_{v} =  A(r,\chi)\left[\frac{{l(l + 1)}}{{{r^2}}} \right]
\label{vectorrrrpot}
\fe
and tensor perturbations \cite{AraujoFilho:2024xhm,AraujoFilho:2025hnf,AraujoFilho:2025vgb,Baruah:2025ifh,Chen:2019iuo,Bouhmadi-Lopez:2020oia}
\ie
\mathrm{V}_{t} = A(r,\chi) \left[\frac{2 (B(r,\chi)-1)}{r^2}+\frac{l(l+1)}{r^2}   - \frac{1}{r \sqrt{A(r,\chi)B^{-1}(r,\chi)}} \frac{\partial}{\partial r} \left(  \sqrt{A(r,\chi) B(r,\chi)} \right) \right].
\label{tententen}
\fe
Each effective potential encapsulates the response of the background geometry to disturbances of different spin types. It characterizes how the curvature interacts with the field, governing both the propagation pattern and the stability properties of the resulting modes. From these potentials, one can infer the way curvature alters wave behavior, shapes the scattering process, modifies absorption rates, and defines the spectrum of quasinormal oscillations corresponding to the massless bosonic fields.


\subsection{Fermions }

Apart from the bosonic modes, fermionic perturbations are described by a spin--$\tfrac{1}{2}$ field that obeys the general relativistic Dirac equation, which dictates how spinor particles evolve in a curved background
\ie
\gamma^{\alpha}\!\left( \partial_{\alpha} - \omega_{\alpha} \right)\!\Psi = 0,
\fe
in which $\gamma^\alpha$ are the curved--space gamma matrices, while $\omega_\alpha$ stands for the spin connection associated with the tetrad basis. After implementing the separation of variables, the spinor field splits into angular and radial parts, analogously to what was done for the bosonic sectors. The resulting radial functions, denoted by $\Psi^{\pm}$, obey a decoupled wave equation whose structure leads naturally to the definition of an effective potential
\ie
\left[\frac{\mathrm{d}^2}{\mathrm{d}{r^*}^2} + \bigl(\omega^2 - \mathrm{V}_{\psi}^{\pm}\bigr)\right]\Psi^\pm = 0.
\fe
Here, the symbols $\pm$ label the two independent chiral components of the spinor field. The resulting effective potentials, derived in previous analyses~\cite{albuquerque2023massless, al2024massless, arbey2021hawking, devi2020quasinormal}, take the following form
\ie
\mathrm{V}_{\psi}^{\pm} = \frac{(l + \frac{1}{2})^2}{r^2} A(r,\chi) \pm \left(l + \frac{1}{2}\right) \sqrt{A(r,\chi) B(r,\chi)} \partial_r \left(\frac{\sqrt{A(r,\chi)}}{r}\right).
\label{eq:Veffpm}
\end{equation}
These two potentials constitute a supersymmetric pair, linked through their underlying symmetry structure. In what follows, $\mathrm{V}_{\psi}^{+}$—hereafter denoted simply as $\mathrm{V}_{\psi}$—is adopted as the representative Dirac potential
\ie
\mathrm{V}_{\psi} = \frac{(l + \frac{1}{2})^2}{r^2} A(r,\chi) + \left(l + \frac{1}{2}\right) \sqrt{A(r,\chi) B(r,\chi)} \partial_r \left(\frac{\sqrt{A(r,\chi)}}{r}\right).
\fe

With the effective potentials $\mathrm{V}_s$, $\mathrm{V}_v$, $\mathrm{V}_t$, and $\mathrm{V}_\psi$ now determined, we proceed to the next stage of the investigation. The forthcoming sections are devoted to evaluating the physical quantities derived from these potentials, including the quasinormal modes and the time--domain solution.


\section{Quasinormal modes }

{ In this section, we examine how the bumblebee parameter $\chi$ influences the perturbative dynamics of fields in the black hole background. As will be shown, deviations from both the Schwarzschild solution and the original bumblebee black hole proposed in \cite{Casana:2017jkc} arise, which might become accessible to future observational tests.
}

\subsection{Scalar perturbations }

Since the effective potential for scalar perturbations is defined in Eq.~(\ref{effssssss}), we now substitute into it the metric specified in Eq.~(\ref{maaaaianametric}), yielding
\ie
\mathrm{V}_{s} = \frac{1}{\chi +1}\left(1-\frac{2 M}{r}\right) \left(\frac{l (l+1)}{r^2}+\frac{2 M}{r^3 (\chi +1)}\right).
\fe
It is worth observing that, in the limit $\chi \to 0$, the potential reduces to the standard form corresponding to scalar perturbations in the Schwarzschild spacetime, as one should expect.

Figure~\ref{spot} illustrates the scalar effective potential as a fcuntion of the radial coordinate $r$ for various combinations of $l$ and $\chi$. As $\chi$ increases, the height of the potential barrier decreases, indicating a weaker confining behavior. In a complementary manner, let us examine the effective potential differently; to do so, we express the potential $\mathrm{V}_{s}$ in terms of the tortoise coordinate $r^{*}$. This latter quantity is obtained by substituting Eq.~(\ref{maaaaianametric}) into Eq.~(\ref{rdsdtdadr}), resulting in
\ie
r^{*} = (1+\chi)\left[r + 2M \ln (r - 2M)\right].
\fe

In Fig.~\ref{stortoise} (left panel), $\mathrm{V}_{s}$ is plotted as a function of $r^{*}$ for different multipole indices $l$, with fixed parameters $M=1$ and $\chi=0.1$. The potential profile displays a sine--type shape, which turns out to be suitable for conducting an analysis through the WKB approximation to determine the quasinormal frequencies for instance. As $l$ increases, the barrier becomes higher, consistent with the expected behavior from Schwarzschild and bumblebee black holes. Moreover, as seen from the plot at the right panel, the potential exhibits a single peak (for each $l$ and other parameters), implying the absence of echo signals—an aspect that will be verified later through the time--domain analysis.

\begin{figure}
    \centering
    \includegraphics[scale=0.55]{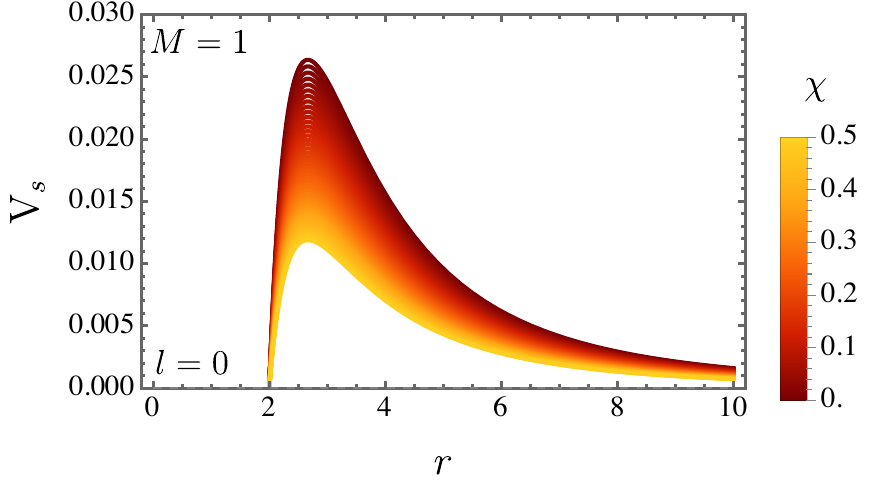}
    \includegraphics[scale=0.55]{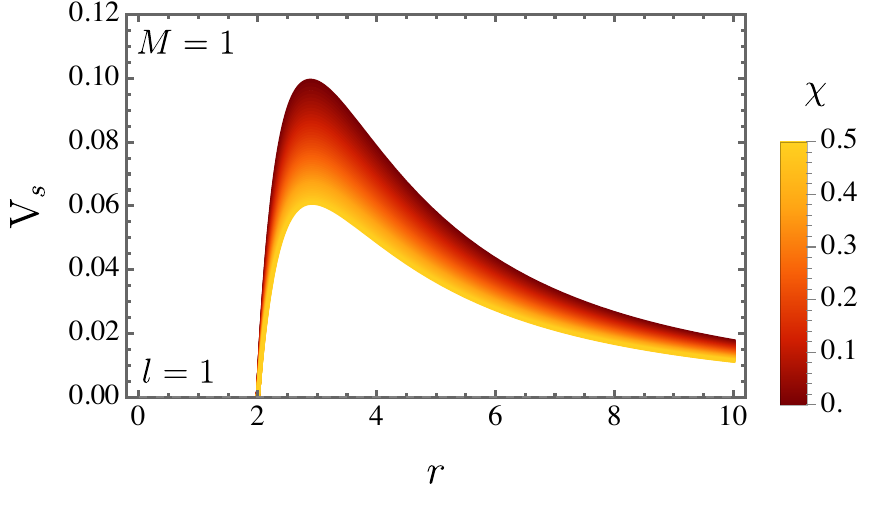}
     \includegraphics[scale=0.55]{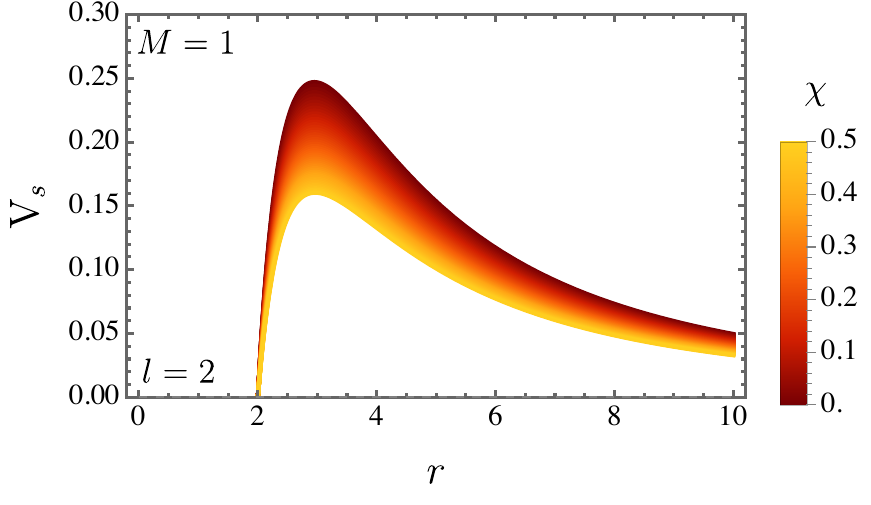}
    \caption{The effective potential for scalar perturbations $\mathrm{V}_{s}$ is shown as a function of the radial coordinate $r$ with $M=1$ and different values of $\chi$: top--left ($l=0$), top--right ($l=1$), and bottom ($l=2$). }
    \label{spot}
\end{figure}

\begin{figure}
    \centering
    \includegraphics[scale=0.5]{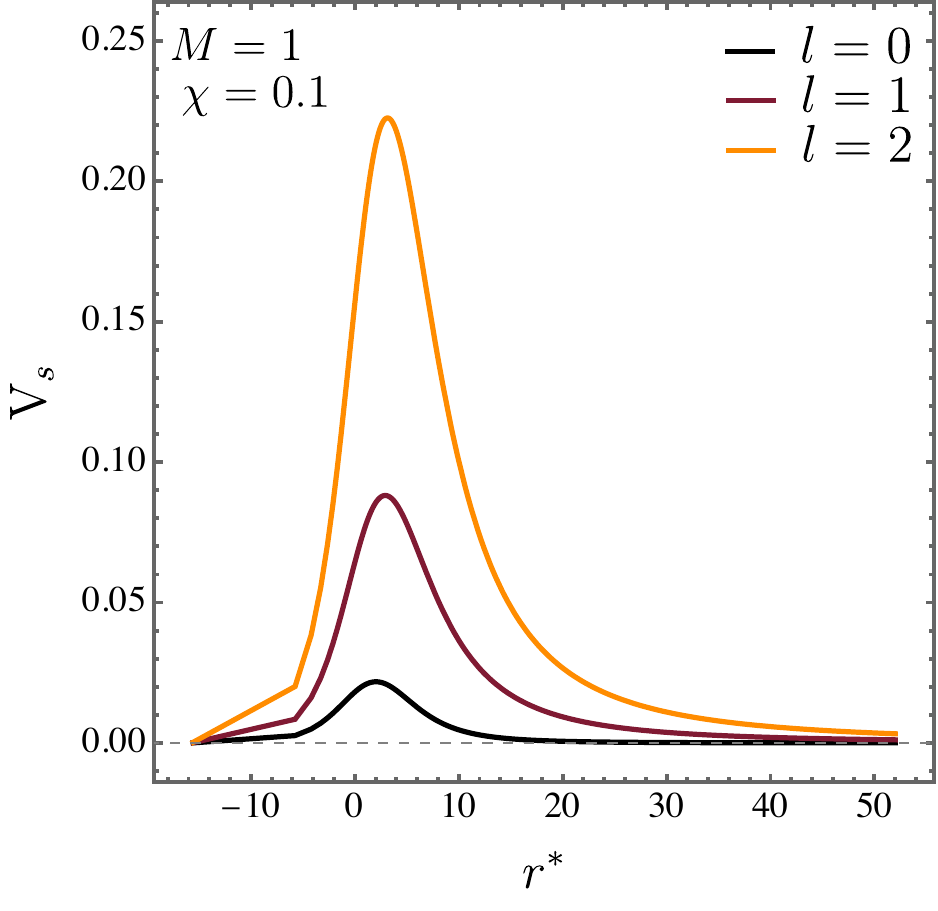}
    \includegraphics[scale=0.625]{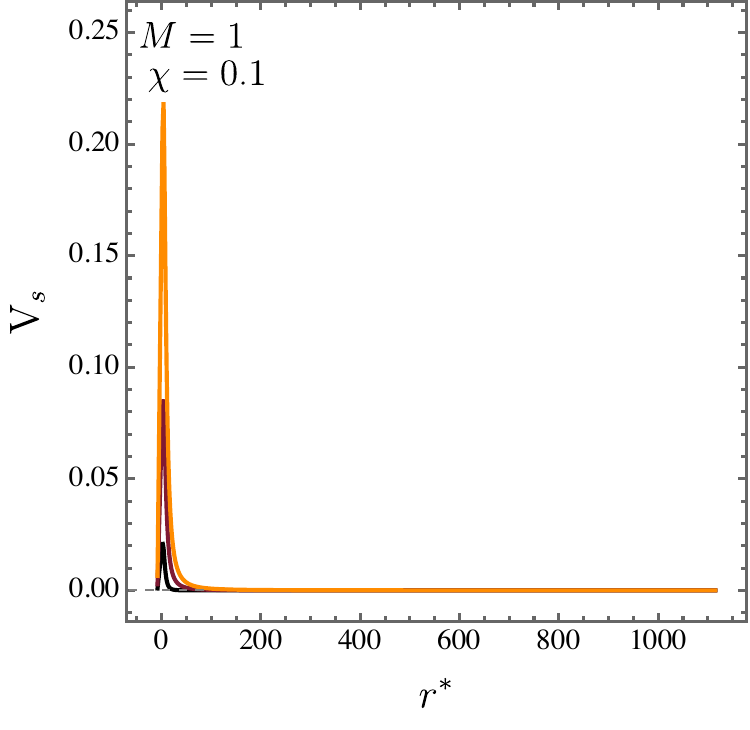}
    \caption{The effective potential $\mathrm{V}_{s}$ for scalar perturbations plotted against the tortoise coordinate $r^{*}$ is exhibited with fixed parameters $M=1$ and $\chi=0.1$ for multipole numbers $l=0,1,2$. In particular, the right panel displays an extended range of $r^{*}$, highlighting that each $l$ value corresponds to a single potential peak. }
    \label{stortoise}
\end{figure}

Since the interpretation of $\mathrm{V}_{s}$ has already been discussed, we now turn to the analysis of the quasinormal modes obtained through the 6th--order WKB approximation. The numerical calculations were performed using the publicly available code from Ref.~\cite{Konoplya:2019hlu}. Tables~\ref{qnmsscalarl0}, \ref{qnmsscalarl1}, and \ref{qnmsscalarl2} list the quasinormal frequencies for $l=0$, $l=1$, and $l=2$, respectively, for different configurations of the Lorentz–violating parameter $\chi$ with fixed $M=1$.

From these results, one observes that increasing $\chi$ systematically lowers both the real and imaginary parts of the frequencies. The reduction in the real part indicates that the oscillations become slower, while the smaller magnitude of the imaginary component implies a weaker damping rate. Consequently, perturbations persist longer as $\chi$ grows, pointing to a gradual stabilization of the system. This behavior is consistent with the softening of the potential barrier discussed earlier. As usual, the higher overtones ($\omega_{1}$, $\omega_{2}$) exhibit faster decay and smaller oscillation frequencies than the fundamental mode, $\omega_{0}$, which dominates the late--time response.

\begin{table}[!h]
\begin{center}
\caption{\label{qnmsscalarl0} Quasinormal frequencies $\omega_{n}$ for scalar perturbations with $l=0$ and fixed mass parameter $M=1$, computed using the 6th--order WKB method for different values of the Lorentz--violating parameter $\chi$.
}
\begin{tabular}{c| c | c | c} 
 \hline\hline\hline 
 \, $M$ \,\,\,\,\,\,  $\chi$  & $\omega_{0}$ & $\omega_{1}$ & $\omega_{2}$  \\ [0.2ex] 
 \hline 
 \,  1.0, \,  0.01  & 0.1093750 - 0.0998163$i$ & 0.0881513 - 0.341104$i$ &  0.189949 - 0.471632$i$ \\
 
\,  1.0, \,  0.1  & 0.1004230 - 0.0916526$i$ & 0.0809328 - 0.313219$i$  & 0.174366 - 0.433146$i$   \\
 
 \, 1.0, \,  0.2  & 0.0920455 - 0.0840229$i$  & 0.0741715 - 0.287182$i$ & 0.159723 - 0.397328$i$ \\
 
\, 1.0, \,  0.3  & 0.0849682 - 0.0775567$i$ & 0.0684714 - 0.265070$i$ & 0.147471 - 0.366679$i$ \\
 
\, 1.0, \,  0.4  & 0.0789021 - 0.0720142$i$ & 0.0635862 - 0.246114$i$ & 0.136976 - 0.340392$i$  \\
   [0.2ex] 
 \hline \hline \hline 
\end{tabular}
\end{center}
\end{table}

\begin{table}[!h]
\begin{center}
\caption{\label{qnmsscalarl1} Quasinormal frequencies $\omega_{n}$ for scalar perturbations with $l=1$ and fixed mass parameter $M=1$, computed using the 6th--order WKB method for different values of the Lorentz--violating parameter $\chi$.}
\begin{tabular}{c| c | c | c} 
 \hline\hline\hline 
 \, $M$ \,\,\,\,\,\,  $\chi$  & $\omega_{0}$ & $\omega_{1}$ & $\omega_{2}$  \\ [0.2ex] 
 \hline 
 \,  1.0, \,  0.01  & 0.291258 - 0.0967802$i$ & 0.263166 - 0.303350$i$ &  0.229927 - 0.536463$i$ \\
 
\,  1.0, \,  0.1  & 0.277531 - 0.0887588$i$ & 0.252269 - 0.277505$i$  & 0.220955 - 0.489923$i$   \\
 
 \, 1.0, \,  0.2  & 0.264315 - 0.0812717$i$  & 0.24168 - 0.253459$i$ & 0.212402 - 0.446611$i$ \\
 
\, 1.0, \,  0.3  & 0.252802 - 0.0749479$i$ & 0.23237 - 0.233215$i$ & 0.204987 - 0.410155$i$ \\
 
\, 1.0, \,  0.4  & 0.242659 - 0.0695363$i$ & 0.224094 - 0.215942$i$ & 0.198456 - 0.379073$i$  \\
   [0.2ex] 
 \hline \hline \hline 
\end{tabular}
\end{center}
\end{table}

\begin{table}[!h]
\begin{center}
\caption{\label{qnmsscalarl2} Quasinormal frequencies $\omega_{n}$ for scalar perturbations with $l=2$ and fixed mass parameter $M=1$, computed using the 6th--order WKB method for different values of the Lorentz--violating parameter $\chi$.}
\begin{tabular}{c| c | c | c} 
 \hline\hline\hline 
 \, $M$ \,\,\,\,\,\,  $\chi$  & $\omega_{0}$ & $\omega_{1}$ & $\omega_{2}$  \\ [0.2ex] 
 \hline 
 \,  1.0, \,  0.01  & 0.481122 - 0.0958029$i$ & 0.461602 - 0.292637$i$ &  0.428548 - 0.503434$i$ \\
 
\,  1.0, \,  0.1  & 0.460084 - 0.0879258$i$ & 0.442774 - 0.268217$i$  & 0.413048 - 0.460499$i$   \\
 
 \, 1.0, \,  0.2  & 0.439657 - 0.0805653$i$  & 0.424353 - 0.245452$i$ & 0.397724 - 0.420587$i$ \\
 
\, 1.0, \,  0.3  & 0.421726 - 0.0743419$i$ & 0.408068 - 0.226245$i$ & 0.384039 - 0.387005$i$ \\
 
\, 1.0, \,  0.4  & 0.40582 - 0.0690109$i$  & 0.393533 - 0.209822$i$ & 0.371710 - 0.358364$i$  \\
   [0.2ex] 
 \hline \hline \hline 
\end{tabular}
\end{center}
\end{table}


\subsection{Vector perturbations }

Let us now carry out an analysis analogous to that performed for the scalar perturbations. Substituting the metric given in Eq.~(\ref{maaaaianametric}) into Eq.~(\ref{vectorrrrpot}), the resulting effective potential for the vector perturbations takes the form
\ie
\mathrm{V}_{v} = \frac{1}{\chi +1}\left(1-\frac{2 M}{r}\right) \left(\frac{l (l+1)}{r^2}\right).
\fe
It is worth noting that, in the limit $\chi \to 0$, the potential reduces to the standard form corresponding to vector perturbations in the Schwarzschild spacetime. Interestingly, while the original bumblebee black hole proposed in Ref.~\cite{Casana:2017jkc} remains unaffected under vector perturbations, the newly obtained solution within the same gravitational framework exhibits nontrivial modifications \cite{Zhu:2025fiy}. This section is therefore devoted to analyzing such effects in detail.

Figure~\ref{vpot} shows the effective potential $\mathrm{V}_{v}$ as a function of the radial coordinate $r$ for different values of $l$ and $\chi$. As $\chi$ increases, the potential barrier becomes progressively lower, indicating a weaker confinement of the perturbative modes, as we could also verify for the scalar perturbations.

Following the same procedure as in the previous subsection, we also plot $\mathrm{V}_{v}$ in terms of the tortoise coordinate $r^{*}$. In Fig.~\ref{vtortoise}, the potential is displayed for $M=1$ and $\chi=0.1$ with multipole indices $l=0,1,2$. The right panel shows a wider range of $r^{*}$, making it evident that each multipole moment corresponds to a single potential peak.

\begin{figure}
    \centering
    \includegraphics[scale=0.55]{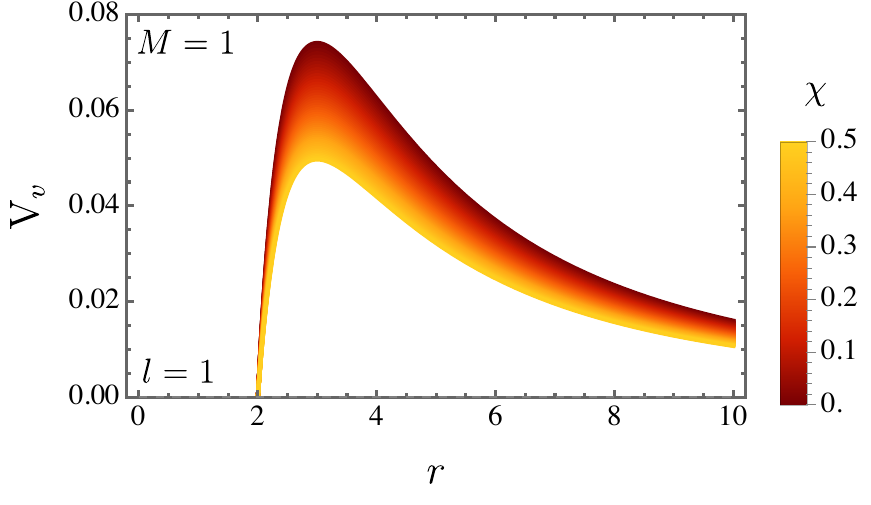}
    \includegraphics[scale=0.55]{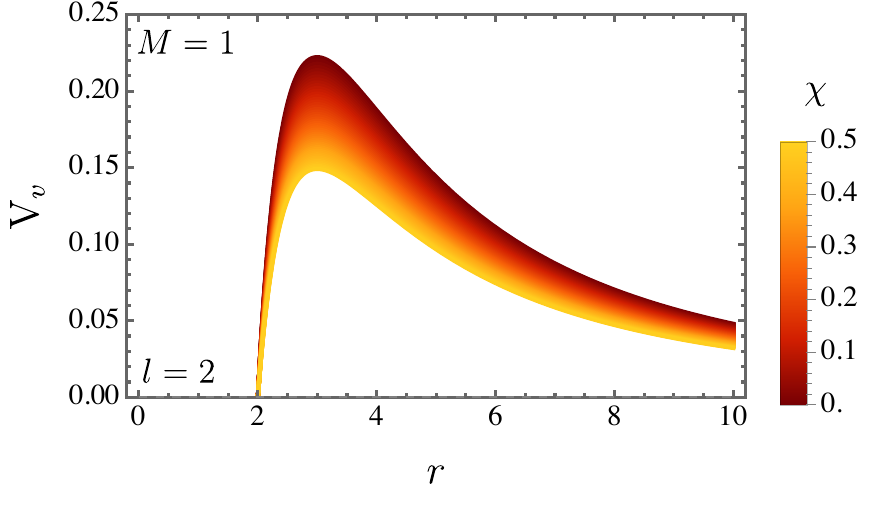}
     \includegraphics[scale=0.55]{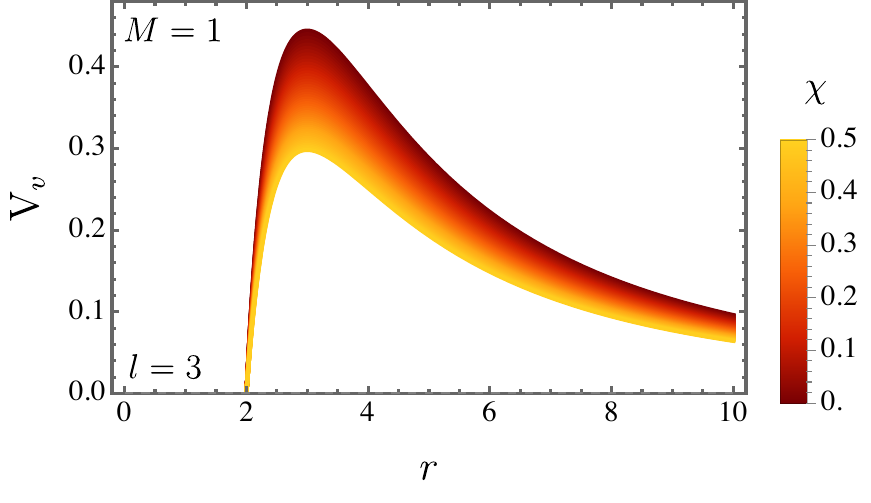}
    \caption{The effective potential for vector perturbations $\mathrm{V}_{v}$ is exhibited as a function of the radial coordinate $r$ with $M=1$ and different values of $\chi$: top--left ($l=1$), top--right ($l=2$), and bottom ($l=3$). }
    \label{vpot}
\end{figure}

\begin{figure}
    \centering
    \includegraphics[scale=0.625]{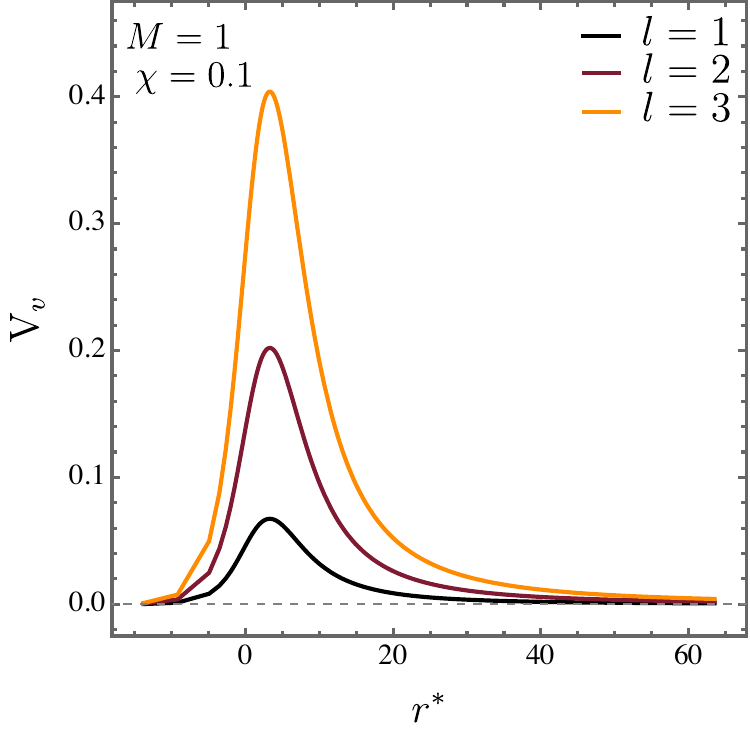}
    \includegraphics[scale=0.625]{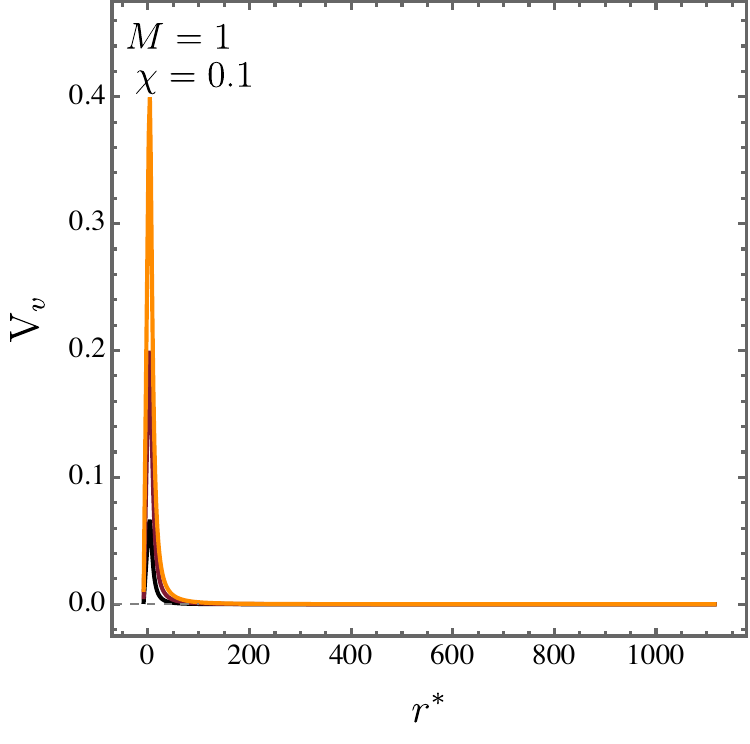}
    \caption{ The effective potential $\mathrm{V}_{v}$ for vector perturbations plotted against the tortoise coordinate $r^{*}$ is exhibited with fixed parameters $M=1$ and $\chi=0.1$ for multipole numbers $l=1,2,3$. In particular, the right panel displays an extended range of $r^{*}$, highlighting that each $l$ value corresponds to a single potential peak. }
    \label{vtortoise}
\end{figure}

Tables~\ref{qnmsvectorl1}, \ref{qnmsvectorl2}, and \ref{qnmsvectorl3} display the quasinormal frequencies for vector perturbations with $l=1$, $l=2$, and $l=3$, respectively, computed using the 6th--order WKB method. The results follow the same qualitative trend as in the scalar sector: increasing the Lorentz–violating parameter $\chi$ reduces both the real and imaginary parts of the frequencies. As a result, the oscillations become slower and the damping weaker, leading to longer–lived perturbations.

When compared with the scalar case, the vector modes exhibit slightly lower oscillation frequencies and smaller damping rates for the same values of $\chi$ and overtone number. The consistent decrease of both real and imaginary parts with increasing $\chi$ indicates that Lorentz violation lowers the effective potential barrier and lengthens the relaxation time of the perturbations. As in the previous cases, the fundamental mode $\omega_{0}$ dominates the late–time behavior, whereas the higher overtones ($\omega_{1}$ and $\omega_{2}$) decay more rapidly.

\begin{table}[!h]
\begin{center}
\caption{\label{qnmsvectorl1} Quasinormal frequencies $\omega_{n}$ for vector perturbations with $l=1$ and fixed mass parameter $M=1$, computed using the 6th--order WKB method for different values of the Lorentz--violating parameter $\chi$.}
\begin{tabular}{c| c | c | c} 
 \hline\hline\hline 
 \, $M$ \,\,\,\,\,\,  $\chi$  & $\omega_{0}$ & $\omega_{1}$ & $\omega_{2}$  \\ [0.2ex] 
 \hline 
 \,  1.0, \,  0.01  & 0.247197 - 0.0917539$i$ & 0.213767 - 0.291163$i$ &  0.247197 - 0.0917539$i$ \\
 
\,  1.0, \,  0.1  & 0.238731 - 0.0845036$i$ & 0.209054 - 0.267010$i$  & 0.171955 - 0.479049$i$   \\
 
 \, 1.0, \,  0.2  & 0.230233 - 0.0776834$i$   & 0.203967 - 0.244465$i$ & 0.169764 - 0.436843$i$ \\
 
\, 1.0, \,  0.3  & 0.222552 - 0.0718819$i$ & 0.199092 - 0.225423$i$ & 0.167481 - 0.401364$i$ \\
 
\, 1.0, \,  0.4  & 0.215571 - 0.0668868$i$ & 0.194453 - 0.209129$i$ & 0.165156 - 0.371137$i$  \\
   [0.2ex] 
 \hline \hline \hline 
\end{tabular}
\end{center}
\end{table}

\begin{table}[!h]
\begin{center}
\caption{\label{qnmsvectorl2} Quasinormal frequencies $\omega_{n}$ for vector perturbations with $l=2$ and fixed mass parameter $M=1$, computed using the 6th--order WKB method for different values of the Lorentz--violating parameter $\chi$.}
\begin{tabular}{c| c | c | c} 
 \hline\hline\hline 
 \, $M$ \,\,\,\,\,\,  $\chi$  & $\omega_{0}$ & $\omega_{1}$ & $\omega_{2}$  \\ [0.2ex] 
 \hline 
 \,  1.0, \,  0.01  & 0.455459 - 0.0940823$i$ & 0.434704 - 0.287831$i$ &  0.399531 - 0.496583$i$ \\
 
\,  1.0, \,  0.1  & 0.437495 - 0.0864737$i$ & 0.419181 - 0.264137$i$  & 0.387707 - 0.454605$i$    \\
 
 \, 1.0, \,  0.2  & 0.419825 - 0.079344$i$  & 0.403707 - 0.242002$i$ & 0.375645 - 0.415540$i$ \\
 
\, 1.0, \,  0.3  & 0.404131 - 0.0733004$i$ & 0.389804 - 0.223289$i$ & 0.364582 - 0.382636$i$  \\
 
\, 1.0, \,  0.4  & 0.390071 - 0.0681122$i$ & 0.377226 - 0.207261$i$ & 0.354399 - 0.354545$i$   \\
   [0.2ex] 
 \hline \hline \hline 
\end{tabular}
\end{center}
\end{table}

\begin{table}[!h]
\begin{center}
\caption{\label{qnmsvectorl3} Quasinormal frequencies $\omega_{n}$ for vector perturbations with $l=3$ and fixed mass parameter $M=1$, computed using the 6th--order WKB method for different values of the Lorentz--violating parameter $\chi$.}
\begin{tabular}{c| c | c | c} 
 \hline\hline\hline 
 \, $M$ \,\,\,\,\,\,  $\chi$  & $\omega_{0}$ & $\omega_{1}$ & $\omega_{2}$  \\ [0.2ex] 
 \hline 
 \,  1.0, \,  0.01  & 0.653735 - 0.0946764$i$ & 0.638795 - 0.286852$i$ &  .611232 - 0.487091$i$ \\
 
\,  1.0, \,  0.1  & 0.653735 - 0.0946764$i$ & 0.614011 - 0.263306$i$  & 0.589559 - 0.446474$i$   \\
 
 \, 1.0, \,  0.2  & 0.601149 - 0.0797653$i$  & 0.589581 - 0.241299$i$ & 0.567954 - 0.408606$i$ \\
 
\, 1.0, \,  0.3  & 0.578114 - 0.0736595$i$ & 0.567843 - 0.222686$i$ & 0.548537 - 0.376653$i$ \\
 
\, 1.0, \,  0.4  & 0.578114 - 0.0736595$i$ & 0.548339 - 0.206739$i$ & 0.530968 - 0.349332$i$  \\
   [0.2ex] 
 \hline \hline \hline 
\end{tabular}
\end{center}
\end{table}


\subsection{Tensor perturbations  }

In this part of the paper, we conclude (in terms) the analysis of the bosonic sector by examining the tensor (axial) perturbations. Proceeding in a manner similar to the scalar and vector cases, we substitute the metric from Eq.~(\ref{maaaaianametric}) into Eq.~(\ref{tententen}), obtaining the corresponding effective potential for the tensor perturbations, which reads
\ie
\mathrm{V}_{t} = \frac{1}{1+\chi} \left(1-\frac{2 M}{r}\right) \left(\frac{l (l+1)}{r^2}-\frac{6 M}{r^3 (\chi +1)}-\frac{2 \chi }{r^2 (\chi +1)}\right).
\fe
As it is straightforward to see, when $\chi \to 0$, the effective potential returns to its usual Schwarzschild form for the corresponding perturbation.

Figure~\ref{tpot} shows the effective potential for tensor perturbations, $\mathrm{V}_{t}$, for several combinations of $l$ and $\chi$. As $\chi$ increases, the potential barrier decreases, indicating a reduction in the confining strength of the perturbations. Furthermore, Figure~\ref{ttortoise} displays $\mathrm{V}_{t}$ as a function of the tortoise coordinate $r^{*}$. Consistent with the scalar and vector cases discussed earlier, the potential exhibits a single peak for each multipole number $l$. Moreover, the overall profile retains a sinusoidal shape, making it suitable for evaluating the quasinormal frequencies using the WKB approximation, as in the previous analyses.

\begin{figure}
    \centering
    \includegraphics[scale=0.55]{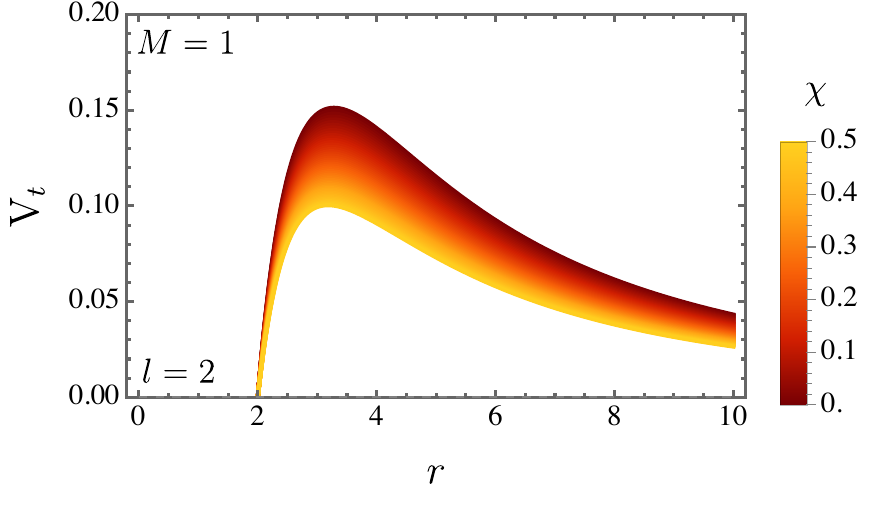}
    \includegraphics[scale=0.55]{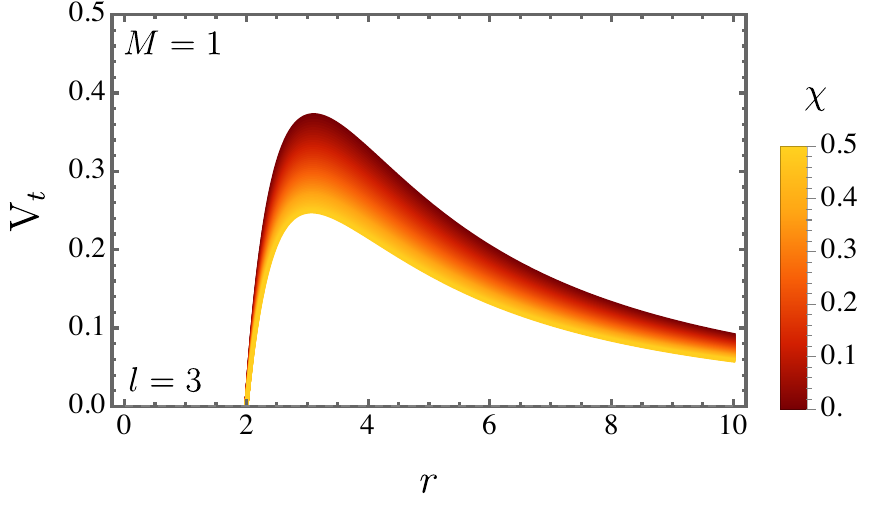}
     \includegraphics[scale=0.55]{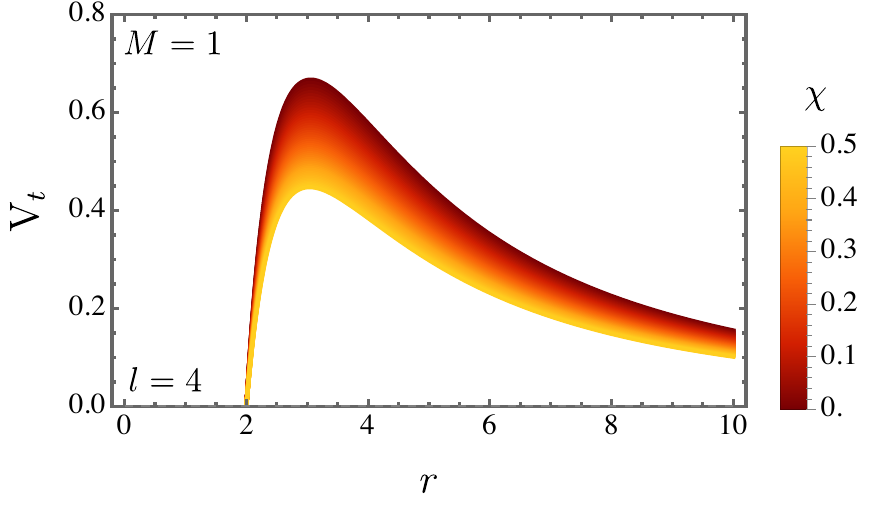}
    \caption{The effective potential for tensor perturbations $\mathrm{V}_{t}$ is exhibited as a function of the radial coordinate $r$ with $M=1$ and different values of $\chi$: top--left ($l=2$), top--right ($l=3$), and bottom ($l=4$). }
    \label{tpot}
\end{figure}

\begin{figure}
    \centering
    \includegraphics[scale=0.625]{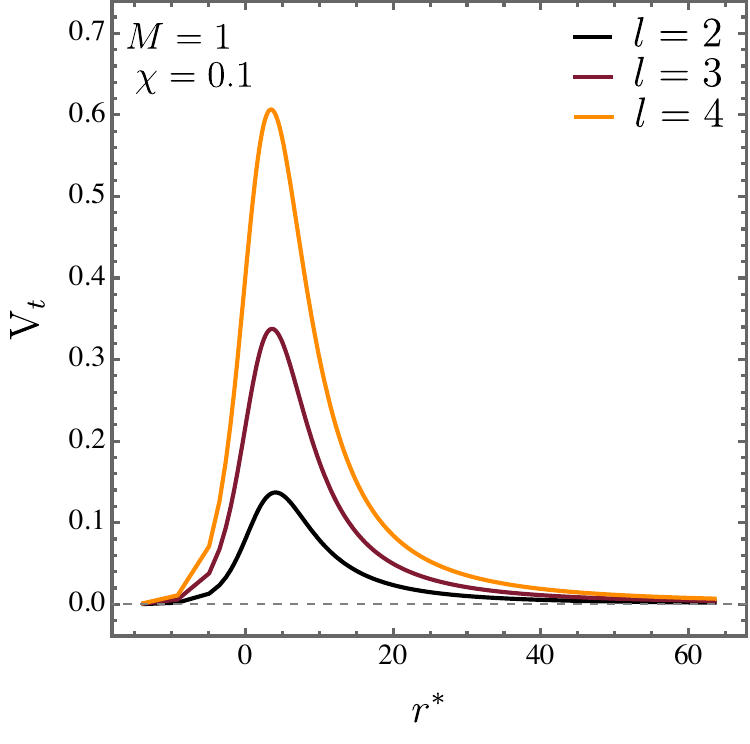}
    \includegraphics[scale=0.625]{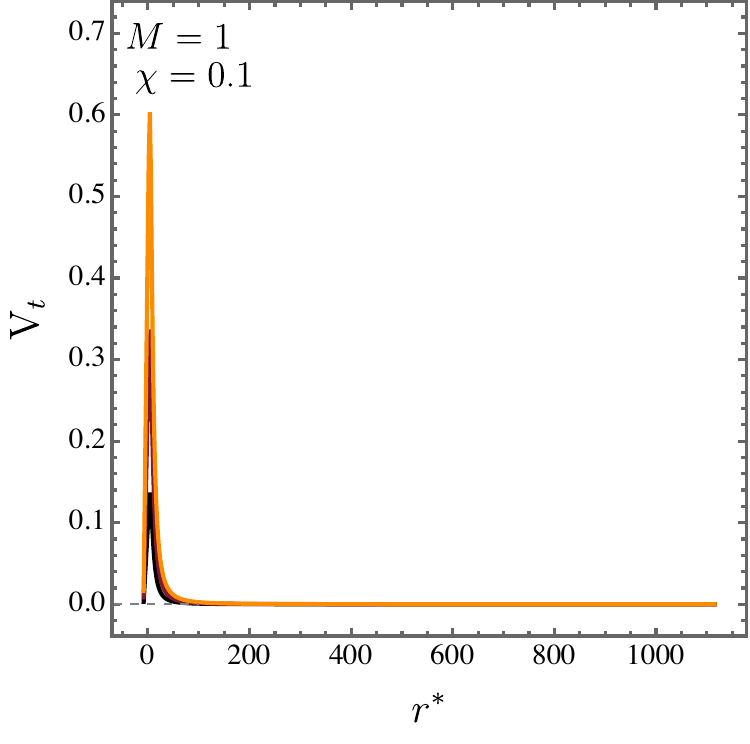}
    \caption{The effective potential $\mathrm{V}_{t}$ for tensor perturbations plotted against the tortoise coordinate $r^{*}$ is exhibited with fixed parameters $M=1$ and $\chi=0.1$ for multipole numbers $l=2,3,4$. In particular, the right panel displays an extended range of $r^{*}$, highlighting that each $l$ value corresponds to a single potential peak. }
    \label{ttortoise}
\end{figure}

Tables~\ref{qnmstensorl2}, \ref{qnmstensorl3}, and \ref{qnmstensorl4} list the quasinormal frequencies associated with tensor (axial) perturbations for $l=2$, $l=3$, and $l=4$, respectively. The overall behavior remains consistent with that observed for the scalar and vector modes: as the Lorentz--violating parameter $\chi$ increases, both the oscillation frequency and the damping rate decrease. This trend indicates that the perturbations become less energetic and persist for longer times as Lorentz violation strengthens.

When compared with the scalar and vector sectors, the tensor modes present lower real frequencies and smaller imaginary parts for the same $\chi$ and overtone number. Therefore, gravitational perturbations oscillate more slowly and decay more gradually, leading to the most sustained ringdown among the three bosonic configurations. This outcome highlights the spin dependence of the effective potentials considered in this work. In all cases, the fundamental mode dominates the late–time response, while the overtones decay faster. Moreover, the monotonic decrease in both the real and imaginary components of the frequencies with increasing $\chi$ across all spin sectors confirms that Lorentz violation lowers the potential barrier and lengthens the relaxation timescale of the perturbations.

\begin{table}[!h]
\begin{center}
\caption{\label{qnmstensorl2} Quasinormal frequencies $\omega_{n}$ for tensor perturbations with $l=2$ and fixed mass parameter $M=1$, computed using the 6th--order WKB method for different values of the Lorentz--violating parameter $\chi$.}
\begin{tabular}{c| c | c | c} 
 \hline\hline\hline 
 \, $M$ \,\,\,\,\,\,  $\chi$  & $\omega_{0}$ & $\omega_{1}$ & $\omega_{2}$  \\ [0.2ex] 
 \hline 
 \,  1.0, \,  0.01  & 0.371834 - 0.0880598$i$ & 0.344952 - 0.270863$i$ &  0.29793 - 0.472789$i$ \\
 
\,  1.0, \,  0.1  & 0.356853 - 0.0812398$i$ & 0.333422 - 0.249451$i$  & 0.292314 - 0.433969$i$   \\
 
 \, 1.0, \,  0.2  & 0.342189 - 0.0748231$i$  & 0.321758 - 0.229384$i$  & 0.28578 - 0.397878$i$ \\
 
\, 1.0, \,  0.3  & 0.329223 - 0.0693586$i$ & 0.311175 - 0.212342$i$  & 0.279264 - 0.367405$i$  \\
 
\, 1.0, \,  0.4  & 0.317648 - 0.0646462$i$ & 0.30154 - 0.197676$i$ & 0.272934 - 0.341294$i$  \\
   [0.2ex] 
 \hline \hline \hline 
\end{tabular}
\end{center}
\end{table}

\begin{table}[!h]
\begin{center}
\caption{\label{qnmstensorl3} Quasinormal frequencies $\omega_{n}$ for tensor perturbations with $l=3$ and fixed mass parameter $M=1$, computed using the 6th--order WKB method for different values of the Lorentz--violating parameter $\chi$.}
\begin{tabular}{c| c | c | c} 
 \hline\hline\hline 
 \, $M$ \,\,\,\,\,\,  $\chi$  & $\omega_{0}$ & $\omega_{1}$ & $\omega_{2}$  \\ [0.2ex] 
 \hline 
 \,  1.0, \,  0.01  & 0.596544 - 0.0918151$i$ & 0.579991 - 0.278568$i$ &  0.549383 - 0.474321$i$ \\
 
\,  1.0, \,  0.1  & 0.57222 - 0.0845345$i$ & 0.557654 - 0.256256$i$  & 0.530574 - 0.435644$i$   \\
 
 \, 1.0, \,  0.2  & 0.548413 - 0.0776926$i$  & 0.535626 - 0.235323$i$  & 0.511729 - 0.399457$i$ \\
 
\, 1.0, \,  0.3  & 0.52736 - 0.0718774$i$ & 0.516016 - 0.217556$i$ & 0.494721 - 0.368823$i$ \\
 
\, 1.0, \,  0.4  & 0.508566 - 0.0668735$i$  & 0.498413 - 0.202288$i$ & 0.479277 - 0.342555$i$  \\
   [0.2ex] 
 \hline \hline \hline 
\end{tabular}
\end{center}
\end{table}

\begin{table}[!h]
\begin{center}
\caption{\label{qnmstensorl4} Quasinormal frequencies $\omega_{n}$ for tensor perturbations with $l=4$ and fixed mass parameter $M=1$, computed using the 6th--order WKB method for different values of the Lorentz--violating parameter $\chi$.}
\begin{tabular}{c| c | c | c} 
 \hline\hline\hline 
 \, $M$ \,\,\,\,\,\,  $\chi$  & $\omega_{0}$ & $\omega_{1}$ & $\omega_{2}$  \\ [0.2ex] 
 \hline 
 \,  1.0, \,  0.01  & 0.805228 - 0.0932507$i$ & 0.792867 - 0.281559$i$ &  0.769274 - 0.475160$i$ \\
 
\,  1.0, \,  0.1  & 0.772104 - 0.0857643$i$ & 0.761225 - 0.258822$i$  & 0.740388 - 0.436369$i$   \\
 
 \, 1.0, \,  0.2  & 0.739706 - 0.0787413$i$  & 0.730155 - 0.237514$i$ & 0.711801 - 0.400080$i$ \\
 
\, 1.0, \,  0.3  & 0.711076 - 0.0727821$i$ & 0.702603 - 0.219449$i$ & 0.686275 - 0.369363$i$  \\
 
\, 1.0, \,  0.4  & 0.685535 - 0.0676619$i$  & 0.677952 - 0.203939$i$ & 0.663301 - 0.343028$i$  \\
   [0.2ex] 
 \hline \hline \hline 
\end{tabular}
\end{center}
\end{table}


\subsection{Spinor perturbations }

Having completed the analysis of the bosonic sector in the previous subsection, we now turn to the fermionic case for completeness. The effective potential governing the spinor perturbations is given by
\ie
\mathrm{V}_{\psi} = \frac{\left(l+\frac{1}{2}\right)^2 \left(1-\frac{2 M}{r}\right)}{r^2 (\chi +1)}+\left(l+\frac{1}{2}\right) \left(\frac{M}{r^3 (\chi +1) \sqrt{\frac{1-\frac{2 M}{r}}{\chi +1}}}-\frac{\sqrt{\frac{1-\frac{2 M}{r}}{\chi +1}}}{r^2}\right) \sqrt{\frac{\left(1-\frac{2 M}{r}\right)^2}{(\chi +1)^2}}.
\fe
It is worth noting that, in the limit $\chi \to 0$, the effective potential reduces to the standard form corresponding to spinor perturbations in the Schwarzschild spacetime, as it is straightforward to check. Figure~\ref{psipot} shows the effective potential for spinorial perturbations for various combinations of $l$ and $\chi$. As $\chi$ increases, the potential barrier becomes lower, indicating a weakening of the confinement of the perturbative modes.

In addition, Fig.~\ref{spinortortoise} shows $\mathrm{V}_{\psi}$ as a function of the tortoise coordinate $r^{*}$ for $M=1$ and $\chi=0.1$, in analogy with the previous analysis for bosonic perturbations. As in those cases, each multipole number $l$ is associated with a single peak. To enable a clearer comparison among all perturbative sectors, we also plot in Fig.~\ref{comptortoise} the effective potentials $\mathrm{V}_{s}$, $\mathrm{V}_{v}$, $\mathrm{V}_{t}$, and $\mathrm{V}_{\psi}$ as functions of $r^{*}$. From this figure, one can observe the ordering $\mathrm{V}_{\psi} > \mathrm{V}_{s} > \mathrm{V}_{v} > \mathrm{V}_{t}$. It is worth noting that the peak of the effective potential primarily determines the real part of the quasinormal frequencies—higher potential barriers generally correspond to higher oscillation frequencies in the WKB approximation. This aspect will be further examined in the section devoted to the time--domain analysis.

\begin{figure}
    \centering
    \includegraphics[scale=0.55]{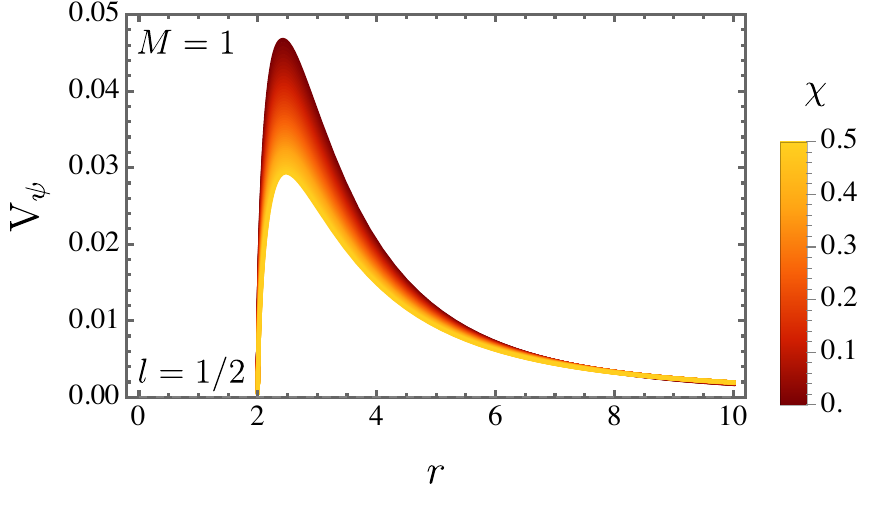}
    \includegraphics[scale=0.55]{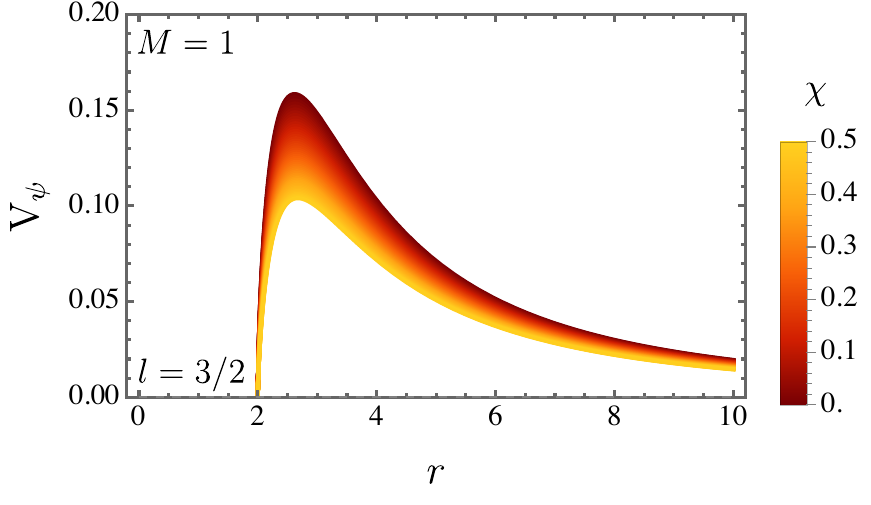}
     \includegraphics[scale=0.55]{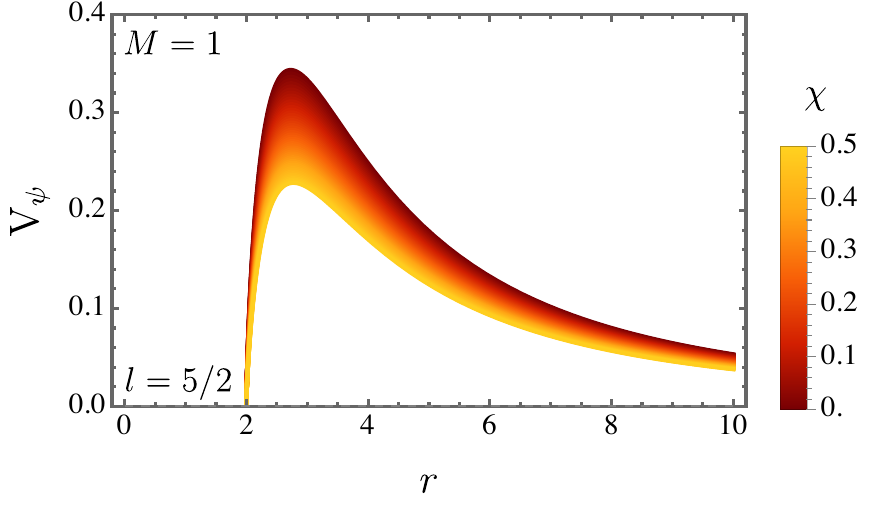}
    \caption{The effective potential for spinor perturbations $\mathrm{V}_{\psi}$ is exhibited as a function of the radial coordinate $r$ with $M=1$ and different values of $\chi$: top--left ($l=1/2$), top--right ($l=3/2$), and bottom ($l=5/2$). }
    \label{psipot}
\end{figure}

\begin{figure}
    \centering
    \includegraphics[scale=0.625]{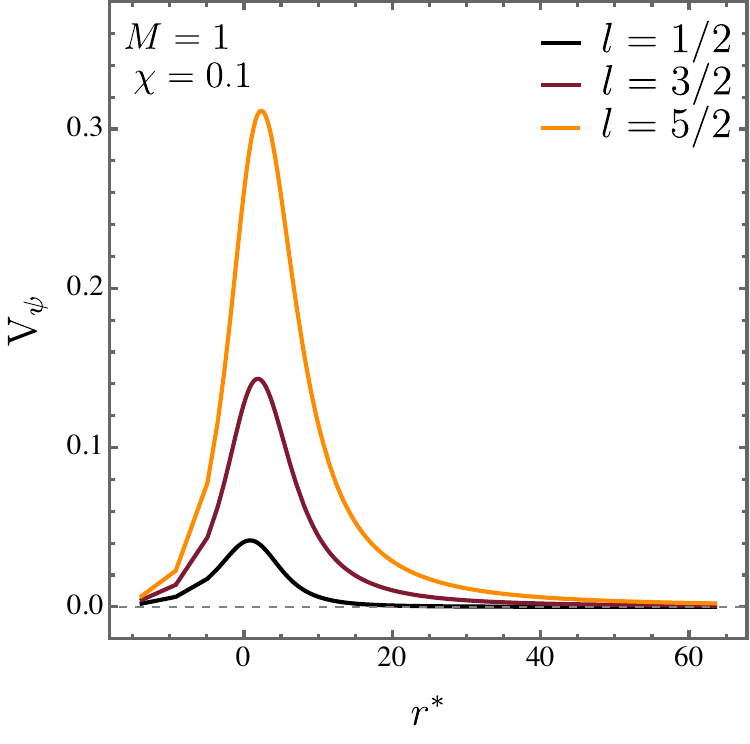}
    \includegraphics[scale=0.625]{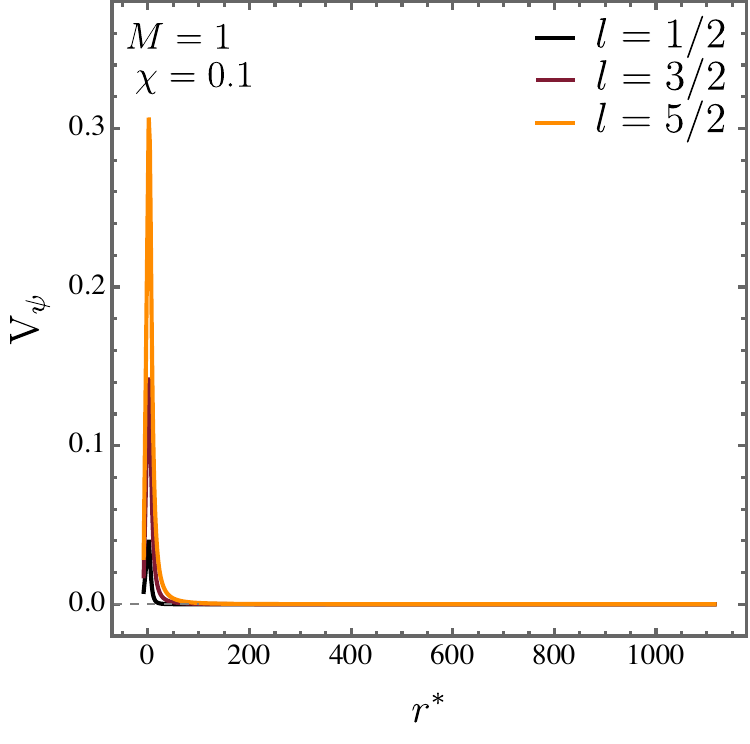}
    \caption{The effective potential $\mathrm{V}_{\psi}$ for spinor perturbations plotted against the tortoise coordinate $r^{*}$ is exhibited with fixed parameters $M=1$ and $\chi=0.1$ for multipole numbers $l=1/2, 3/2, 5/2$. In particular, the right panel displays an extended range of $r^{*}$, highlighting that each $l$ value corresponds to a single potential peak. }
    \label{spinortortoise}
\end{figure}

\begin{figure}
    \centering
    \includegraphics[scale=0.625]{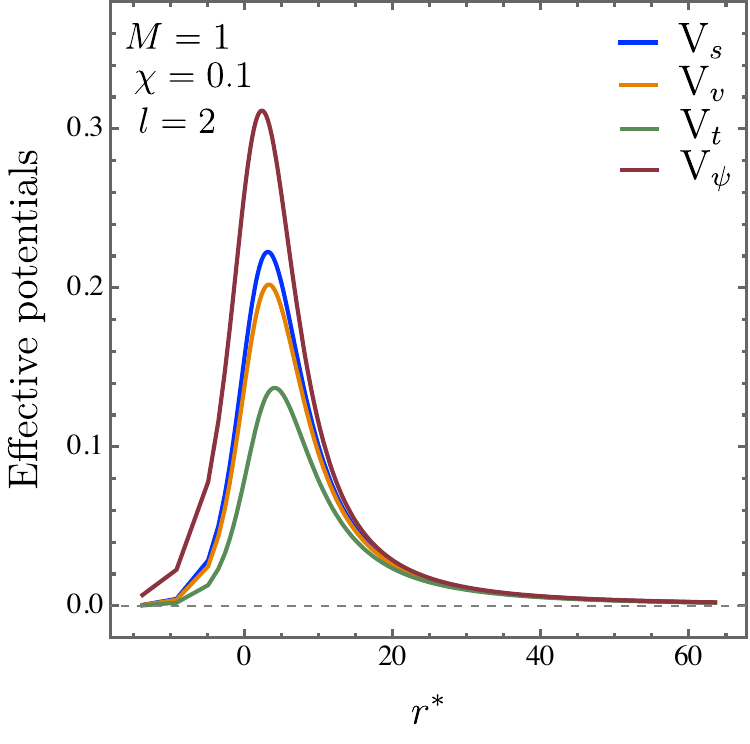}
    \caption{Comparison of the effective potentials obtained for each spin sector, showing the ordering $\mathrm{V}_{\psi} > \mathrm{V}_{s} > \mathrm{V}_{v} > \mathrm{V}_{t}$. For the spinor field, the mode $l = 5/2$ was used, while for all bosonic perturbations the mode $l = 2$ was considered.}
    \label{comptortoise}
\end{figure}

Tables~\ref{qnmsspnior12}, \ref{qnmsspnior32}, and \ref{qnmsspnior52} list the quasinormal frequencies for spinor perturbations with $l=1/2$, $l=3/2$, and $l=5/2$, respectively, obtained through the 3rd--order WKB approximation. The overall behavior mirrors that of the bosonic sector: as the Lorentz--violating parameter $\chi$ increases, both the real and imaginary parts of the frequencies decrease, implying slower oscillations and weaker damping, which lead to longer--lived modes at higher $\chi$.

When compared with the scalar, vector, and tensor sectors, the spinor modes display slightly higher real frequencies and moderately larger imaginary parts (except when compared with the scalar perturbations) for the same values of $\chi$ and overtone number. This behavior indicates that fermionic perturbations decay faster than vector and tensor modes, but more slowly than the scalar ones. A natural question arises: since the fermionic quasinormal modes were obtained through the third--order WKB approximation, would the same pattern persist if the scalar perturbations were computed using the same method? After verification, we found that the same trend indeed holds. Across all spin sectors, the effect of Lorentz violation follows a consistent behavior: it lowers the effective potential barrier, reduces the real part of the frequencies, and weakens the damping as $\chi$ increases. As in the previous analyses, the fundamental mode dominates the late--time evolution, while higher overtones decay much more rapidly. {This coherent behavior distinguishes the effect from scenarios where only specific modes are modified and may serve as a useful signature when combined with multi-mode gravitational-wave observations.}

\begin{table}[!h]
\begin{center}
\caption{\label{qnmsspnior12} Quasinormal frequencies $\omega_{n}$ for spinor perturbations with $l=1/2$ and fixed mass parameter $M=1$, computed via the 3rd--order WKB method for different values of the Lorentz--violating parameter $\chi$.}
\begin{tabular}{c| c | c | c} 
 \hline\hline\hline 
 \, $M$ \,\,\,\,\,\,  $\chi$  & $\omega_{0}$ & $\omega_{1}$ & $\omega_{2}$  \\ [0.2ex] 
 \hline 
 \,  1.0, \,  0.01  & 0.181995 - 0.0939296i&
0.144884 - 0.309934i &
0.125638 - 0.558166i \\
 
\,  1.0, \,  0.1  & 0.175404 - 0.0862788i &
0.142221 - 0.282194i &
0.12284 - 0.505596i \\
 
 \, 1.0, \,  0.2  & 0.168604 - 0.0792391i&
0.138456 - 0.258152i &
0.118226 - 0.464223i \\
 
\, 1.0, \,  0.3  & 0.162496 - 0.0732835i &
0.13493 - 0.237892i &
0.114279 - 0.428954i \\

\, 1.0, \,  0.4  & 0.157095 - 0.0681249i &
0.132143 - 0.219747i &
0.111901 - 0.39494i \\
   [0.2ex] 
 \hline \hline \hline 
\end{tabular}
\end{center}
\end{table}

\begin{table}[!h]
\begin{center}
\caption{\label{qnmsspnior32} Quasinormal frequencies $\omega_{n}$ for spinor perturbations with $l=3/2$ and fixed mass parameter $M=1$, computed via the 3rd--order WKB method for different values of the Lorentz--violating parameter $\chi$.}
\begin{tabular}{c| c | c | c} 
 \hline\hline\hline 
 \, $M$ \,\,\,\,\,\,  $\chi$  & $\omega_{0}$ & $\omega_{1}$ & $\omega_{2}$  \\ [0.2ex] 
 \hline 
 \,  1.0, \,  0.01  & 0.378228 - 0.0954121i &
0.354351 - 0.294257i &
0.317824 - 0.513160i\\ 

\,  1.0, \,  0.1  & 0.36279 - 0.0876054i &
0.341627 - 0.269612i &
0.308509 - 0.468849i \\ 

 \, 1.0, \,  0.2  & 0.34767 - 0.0803025i &
0.328935 - 0.246669i &
0.298948 - 0.427904i \\ 

\, 1.0, \,  0.3  & 0.334298 - 0.0741212i &
0.31757 - 0.227303i &
0.290292 - 0.393406i \\ 

\, 1.0, \,  0.4  & 0.322361 - 0.0688222i &
0.307316 - 0.210740i &
0.282407 - 0.363944i\\
   [0.2ex] 
 \hline \hline \hline 
\end{tabular}
\end{center}
\end{table}

\begin{table}[!h]
\begin{center}
\caption{\label{qnmsspnior52} Quasinormal frequencies $\omega_{n}$ for spinor perturbations with $l=5/2$ and fixed mass parameter $M=1$, computed via the 3rd--order WKB method for different values of the Lorentz--violating parameter $\chi$.}
\begin{tabular}{c| c | c | c} 
 \hline\hline\hline 
 \, $M$ \,\,\,\,\,\,  $\chi$  & $\omega_{0}$ & $\omega_{1}$ & $\omega_{2}$  \\ [0.2ex] 
 \hline 
 \,  1.0, \,  0.01  & 0.571277 - 0.0953527i &
0.554444 - 0.289781i &
0.524362 - 0.494599i \\
 
\,  1.0, \,  0.1  &
0.547658 - 0.087545i &
0.532796 - 0.26578i &
0.505973 - 0.452855i\\
 
 \, 1.0, \,  0.2  & 
0.524567 - 0.0802443i &
0.51148 - 0.243381i &
0.487643 - 0.414009i \\
 
\, 1.0, \,  0.3  & 
0.504171 - 0.0740675i &
0.492532 - 0.224462i &
0.471172 - 0.381283i \\
 
\, 1.0, \,  0.4  & 
0.485982 - 0.0687736i &
0.475544 - 0.208272i &
0.456262 - 0.353342i  \\
   [0.2ex] 
 \hline \hline \hline 
\end{tabular}
\end{center}
\end{table}


\section{Time--domain solution }

A proper examination of the temporal behavior of scalar, vector, and tensor perturbations demands a full dynamical treatment instead of depending exclusively on frequency–domain analyses. Such a formulation allows one to follow the actual evolution of the field and to understand how quasinormal oscillations determine both the damping pattern and the scattering phenomena. Since the corresponding effective potentials usually have a nontrivial structure, a precise numerical algorithm becomes necessary to evolve the perturbations with stability and accuracy. For this purpose, the characteristic integration technique, originally introduced by Gundlach and collaborators~\cite{Gundlach:1993tp}, is adopted.

In line with the numerical frameworks developed in Refs.~\cite{Skvortsova:2024wly,Bolokhov:2024ixe,Guo:2023nkd,Yang:2024rms,Baruah:2023rhd,Gundlach:1993tp,Shao:2023qlt,Lutfuoglu:2025kqp,Santos:2025xbk}, the field equation is recast using double–null coordinates, $u=t-r^{*}$ and $v=t+r^{*}$, which transform the problem into a more convenient form for computation. Expressed in these variables, the wave equation takes the following structure:
\ie
\left(4 \frac{\partial^{2}}{\partial u \, \partial v} + V(u,v)\right) \Tilde{\psi} (u,v) = 0.
\fe

One typically proceeds by transforming the continuous equation into a discrete form, dividing the computational region into a grid of points. Through this finite--difference scheme, the field values at successive grid positions are updated iteratively, enabling the numerical evolution of the wave profile over time
\ie
\Tilde{\psi}(N) = -\Tilde{\psi}(S) + \Tilde{\psi}(W) + \Tilde{\psi}(E) - \frac{h^{2}}{8}V(S)\Big[\Tilde{\psi}(W) + \Tilde{\psi}(E)\Big] + \mathcal{O}(h^{4}).
\fe

The computation starts by constructing a uniform mesh over the $(u,v)$ domain with a chosen step length $h$. Each elementary square of the mesh is labeled by four nodes: $S=(u,v)$ as the reference corner, $W=(u+h,v)$ and $E=(u,v+h)$ as adjacent points, and $N=(u+h,v+h)$ as the point to be determined in the forward direction. The evolution of the field proceeds from the data imposed on the initial null boundaries, $u=u_{0}$ and $v=v_{0}$, which define the starting surfaces of integration. Along the line $u=u_{0}$, the initial perturbation is introduced in the form of a Gaussian profile centered at $v=v_{c}$ with a spread $\sigma$, serving as the seed from which the wave propagation is computed over the entire grid
\ie
\Tilde{\psi}(\Tilde{u} = u_{0},v) = A e^{-(v-v_{0})^{2}}/2\sigma^{2}, \,\,\,\,\,\, \Tilde{\psi}(u,v_{0}) = \Tilde{\psi}_{0}.
\fe

The time evolution is initiated by prescribing boundary data on the line $v=v_{0}$, where the field $\tilde{\psi}(u,v_{0})$ is taken to vanish, ensuring a simple and stable starting configuration. From these initial conditions, the computation proceeds by incrementally updating $\tilde{\psi}$ along successive $v$--values for fixed $u$, following the causal structure imposed by the double--null lattice. To maintain numerical stability and reduce complexity, only massless perturbations are examined throughout this work, with the black hole mass parameter fixed at $M=1$. The initial wave disturbance is modeled as a Gaussian pulse centered at $v=0$ and characterized by a width $\sigma=1$, while the overall amplitude is initially set to zero. The $(u,v)$ plane is uniformly partitioned within the interval $[0,1000]$ using a step size $h=0.1$, a choice that guarantees sufficient precision to capture the full temporal evolution and attenuation of the waveform.

To complement the earlier discussion on quasinormal modes and ensure completeness, the time--domain analysis is extended here to include all spin configurations, as seen in the forthcoming subsections.


\subsection{Scalar perturbations }

This section examines how scalar perturbations evolve over time within the black hole geometry under consideration presented in Eq. (\ref{maaaaianametric}). Figure~\ref{timedomainscalar0} illustrates the numerical evolution of the field $\tilde{\psi}$ for a fixed mass $M=1$, with the Lorentz–violating parameter $\chi$ taking the values $0.1$, $0.2$, $0.3$, and $0.4$. The panels correspond to different angular modes: $l=0$ (upper left), $l=1$ (upper right), and $l=2$ (lower). The obtained waveforms exhibit oscillations whose amplitudes decay exponentially, a distinctive signature of the quasinormal ringing typical of perturbed black hole spacetimes. The time--domain profiles confirm the trend observed from the quasinormal frequency analysis: as $\chi$ increases, the damping rate decreases, leading to more persistent oscillations.

Figure~\ref{timedomainscalar1} further illustrates the attenuation pattern by presenting the evolution of $\ln|\tilde{\psi}|$ for identical values of $\chi$ and $l$. In this logarithmic representation, the exponential damping manifests as straight segments, clearly delineating the quasinormal ringing phase and the subsequent transition into the late--time power--law decay. Consistent with the outcomes obtained from the frequency investigation of the quasinormal modes the previous section, the time--domain profiles reveal that larger values of $\chi$ lead to longer--lived perturbations, indicating that the field decays more gradually as the parameter increases.

Figure~\ref{timedomainscalar2} presents the late--time behavior of $\tilde{\psi}$ on a double--logarithmic scale, using the same panel configuration for consistency. This representation makes the asymptotic regime explicit, showing the smooth transition from the exponentially damped quasinormal ringing to the slower power--law decay that dominates at large times. The emergence of this tail pattern confirms the expected late--time evolution characteristic of black–hole perturbations.

\begin{figure}
    \centering
    \includegraphics[scale=0.51]{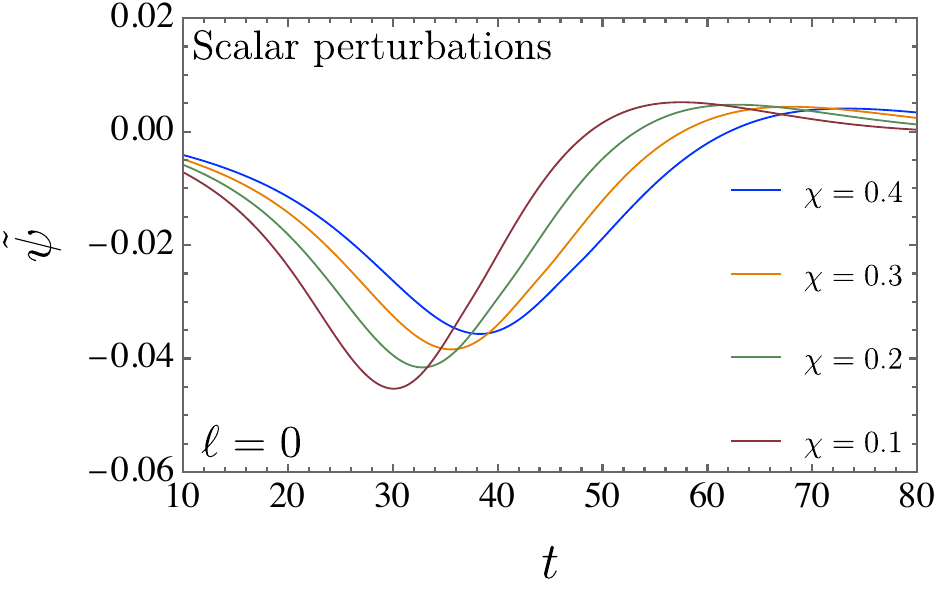}
    \includegraphics[scale=0.51]{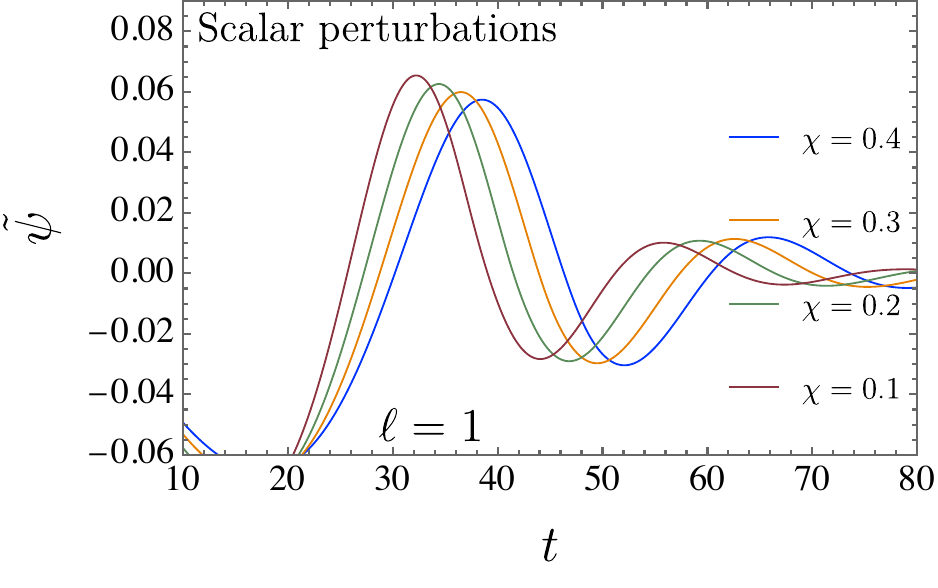}
     \includegraphics[scale=0.51]{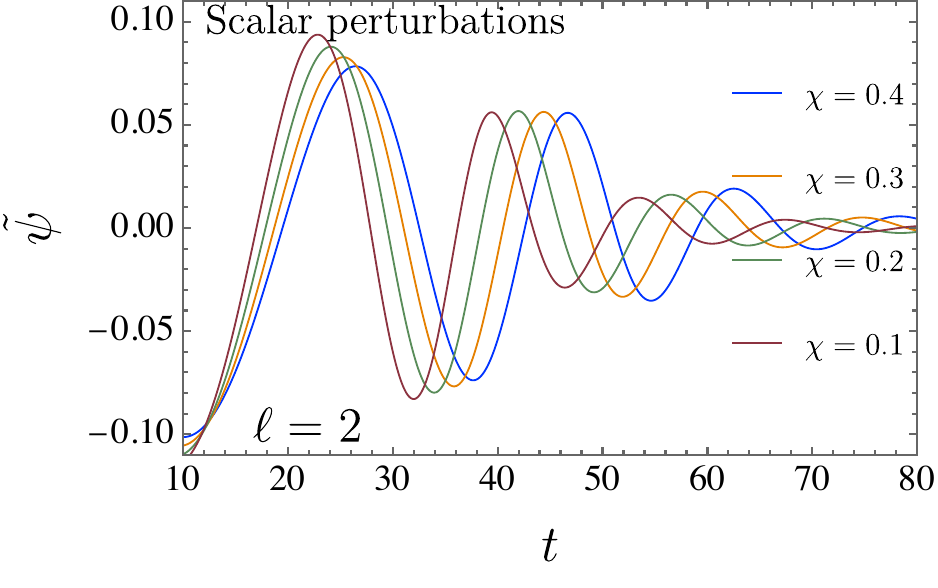}
    \caption{Time evolution of the scalar field $\tilde{\psi}$ for a black hole with fixed mass $M=1$ and different values of the Lorentz–violating parameter $\chi = 0.1, 0.2, 0.3,$ and $0.4$. The panels display the modes $l=0$ (upper left), $l=1$ (upper right), and $l=2$ (lower), illustrating how the waveform changes with increasing $\chi$. }
    \label{timedomainscalar0}
\end{figure}

\begin{figure}
    \centering
    \includegraphics[scale=0.51]{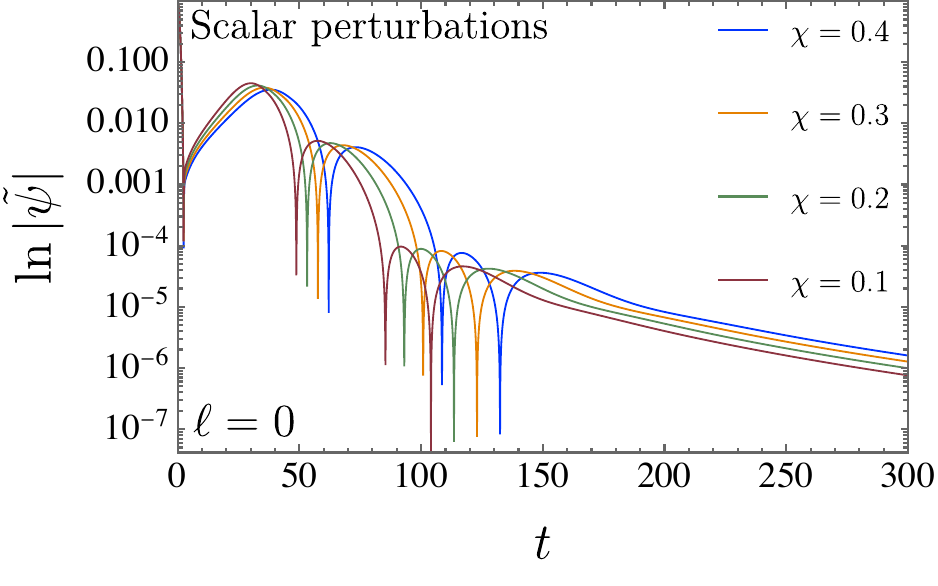}
    \includegraphics[scale=0.51]{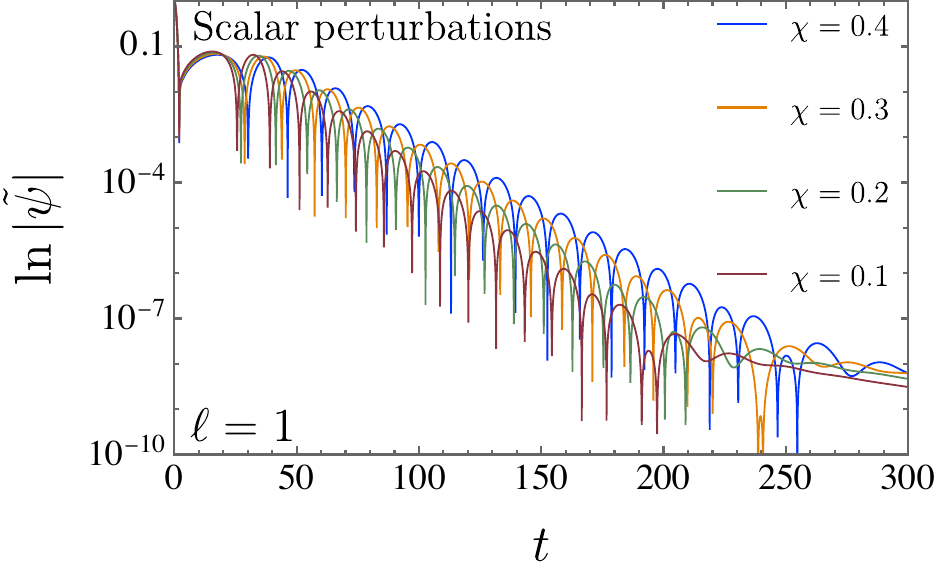}
     \includegraphics[scale=0.51]{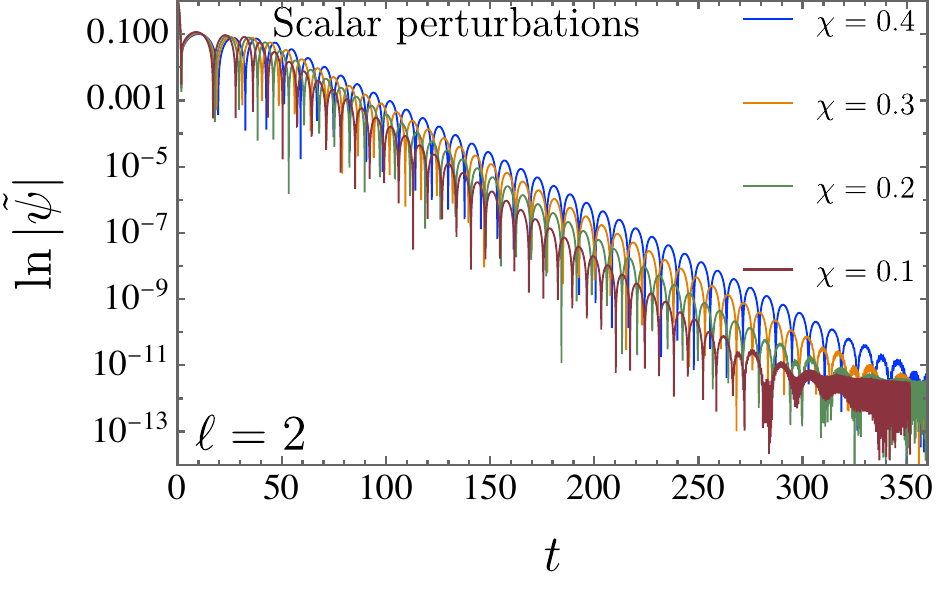}
    \caption{Time evolution of $\ln|\tilde{\psi}|$ for scalar perturbations in a black hole spacetime with fixed mass $M=1$ and Lorentz--violating parameter $\chi = 0.1, 0.2, 0.3,$ and $0.4$. The panels correspond to the angular indices $l=0$ (top left), $l=1$ (top right), and $l=2$ (bottom), showing the damping behavior across different modes.
 }
    \label{timedomainscalar1}
\end{figure}

\begin{figure}
    \centering
    \includegraphics[scale=0.51]{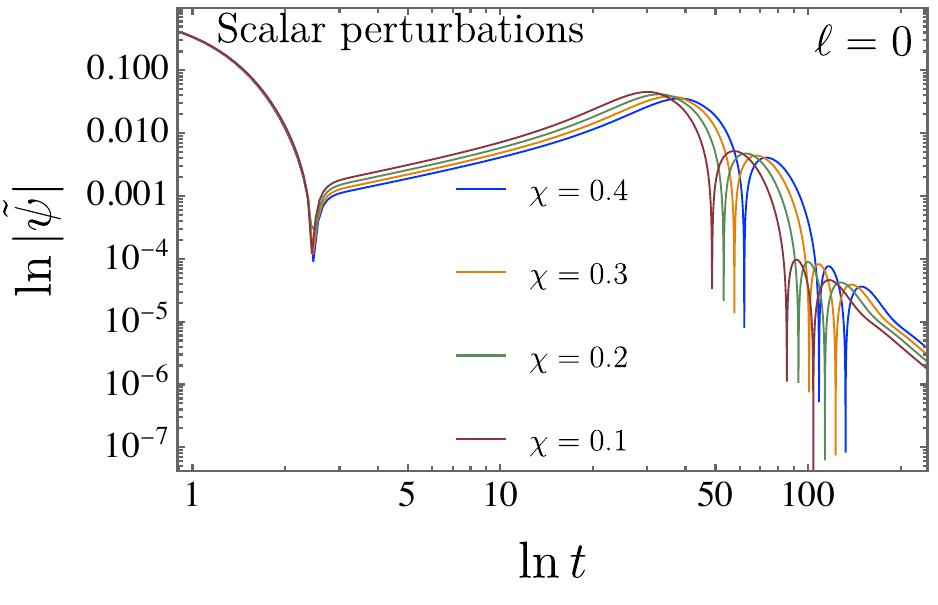}
    \includegraphics[scale=0.51]{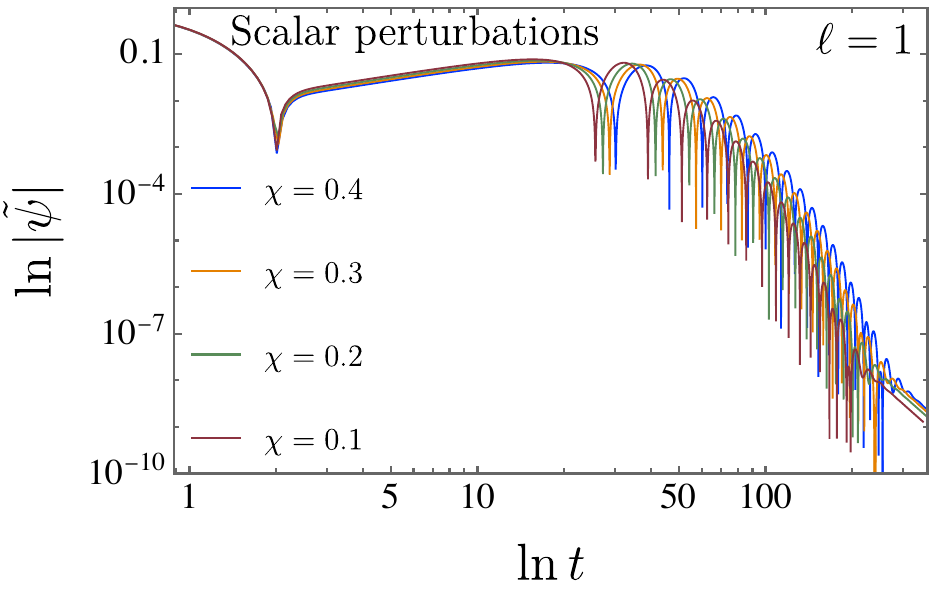}
     \includegraphics[scale=0.51]{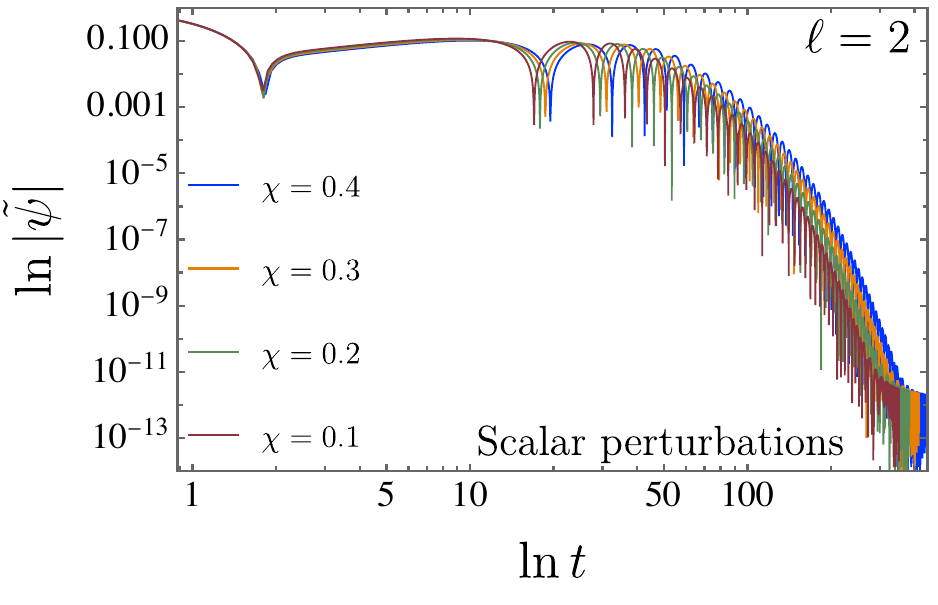}
    \caption{Late--time evolution of the scalar perturbation $\tilde{\psi}$ shown on a double--logarithmic scale, where $\ln|\tilde{\psi}|$ is plotted against $\ln t$ for $M=1$ and $\chi = 0.1, 0.2, 0.3,$ and $0.4$. The subplots correspond to $l=0$ (upper left), $l=1$ (upper right), and $l=2$ (lower), which emphasize the emergence of the power--law decay that characterizes the tail phase of the signal.}
    \label{timedomainscalar2}
\end{figure}


\subsection{Vector perturbations }

The temporal evolution of vector perturbations in the geometry defined by Eq.~(\ref{maaaaianametric}) is analyzed in this section. Figure~\ref{timedomainsvector1} presents the time evolution of the field $\tilde{\psi}$ for a fixed black hole mass $M=1$, with the Lorentz–violating parameter $\chi$ taking the values $0.1$, $0.2$, $0.3$, and $0.4$. The panels correspond to the angular indices $l=1$ (top left), $l=2$ (top right), and $l=3$ (bottom). The waveforms display the characteristic damped oscillations associated with the quasinormal ringing phase. Compared with the scalar case, the vector perturbations oscillate with slightly lower frequencies and decay more slowly, resulting in a longer--lasting signal. As observed previously, increasing $\chi$ weakens the damping, extending the duration of the oscillatory phase.

To highlight the attenuation behavior, Fig.~\ref{timedomainsvector2} plots $\ln|\tilde{\psi}|$ as a function of time for the same parameter set. The linear segments correspond to the exponential damping regime of the quasinormal ringing, which transitions into the late--time power--law tail. These results are consistent with the quasinormal‐mode frequencies obtained from the frequency--domain analysis. Larger values of $\chi$ yield slower decay, and the vector perturbations exhibit a slightly longer relaxation time than the scalar ones.

Finally, Fig.~\ref{timedomainsvector3} displays the late‐time regime on a double‐logarithmic scale, maintaining the same panel arrangement. This representation clearly shows the shift from the exponential ringdown to the asymptotic power‐law tail. Although the qualitative pattern resembles the scalar case, the decay of vector perturbations proceeds more gradually, indicating that they are more persistent due to their weaker damping within the Lorentz--violating background.

\begin{figure}
    \centering
    \includegraphics[scale=0.51]{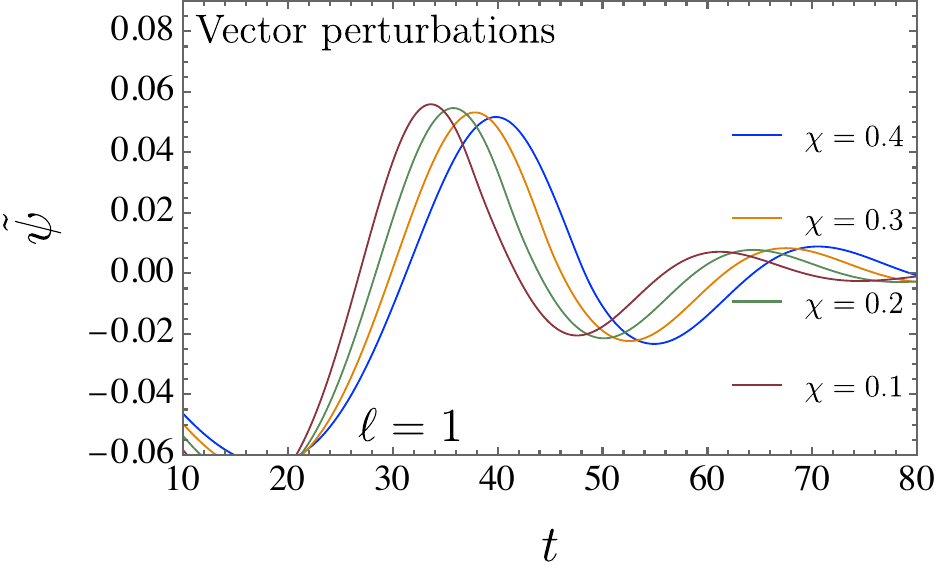}
    \includegraphics[scale=0.51]{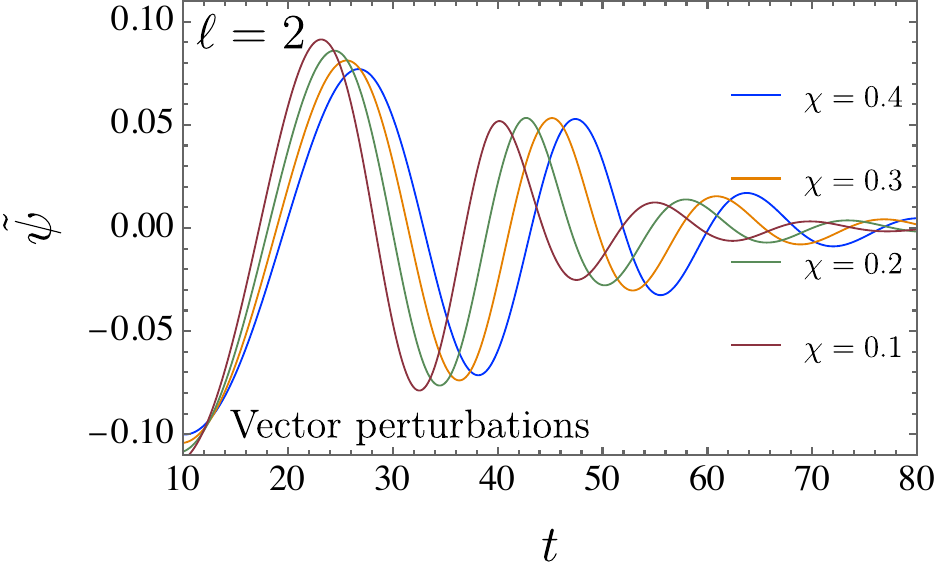}
     \includegraphics[scale=0.51]{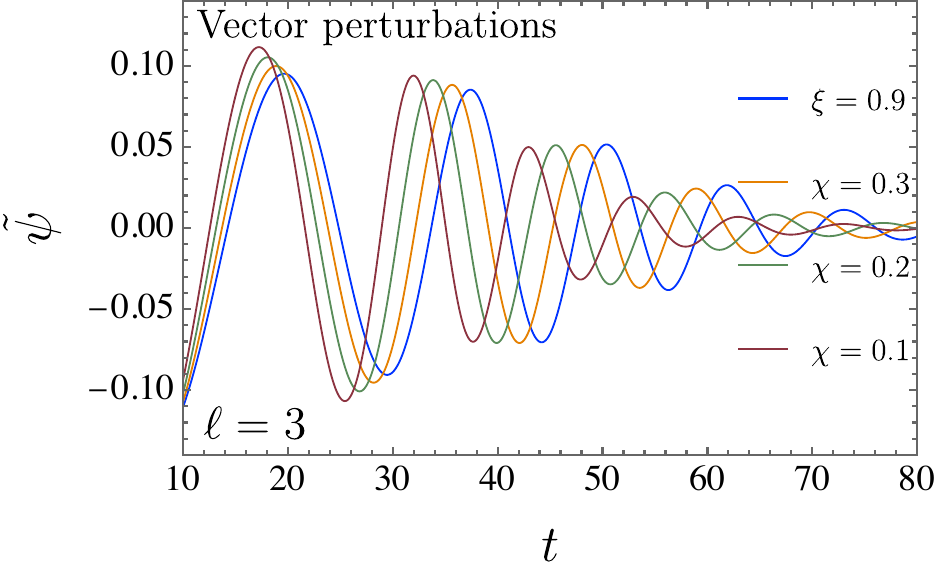}
    \caption{Time evolution of the vector field $\tilde{\psi}$ for a black hole with fixed mass $M=1$ and different values of the Lorentz–violating parameter $\chi = 0.1, 0.2, 0.3,$ and $0.4$. The panels display the modes $l=1$ (upper left), $l=2$ (upper right), and $l=3$ (lower), illustrating how the waveform changes with increasing $\chi$. }
    \label{timedomainsvector1}
\end{figure}

\begin{figure}
    \centering
    \includegraphics[scale=0.51]{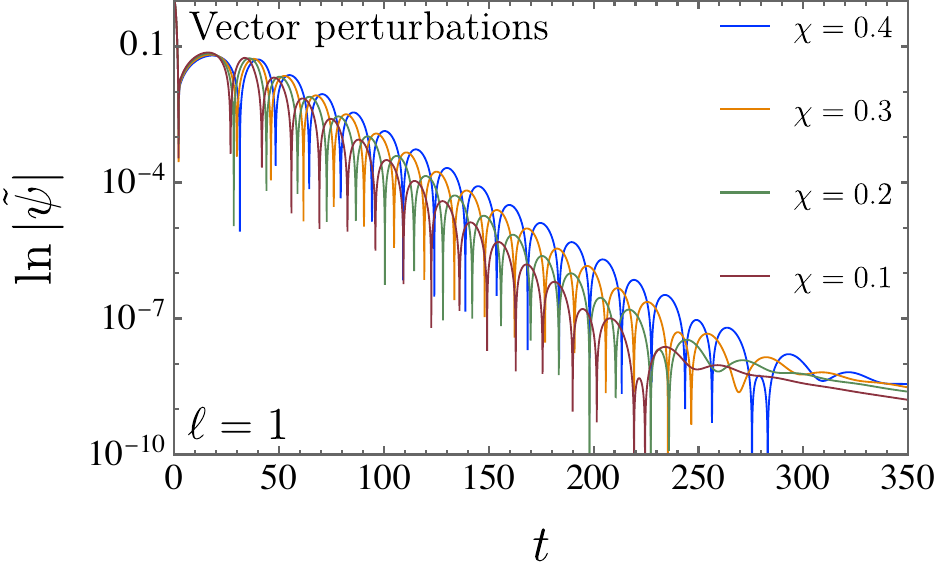}
    \includegraphics[scale=0.51]{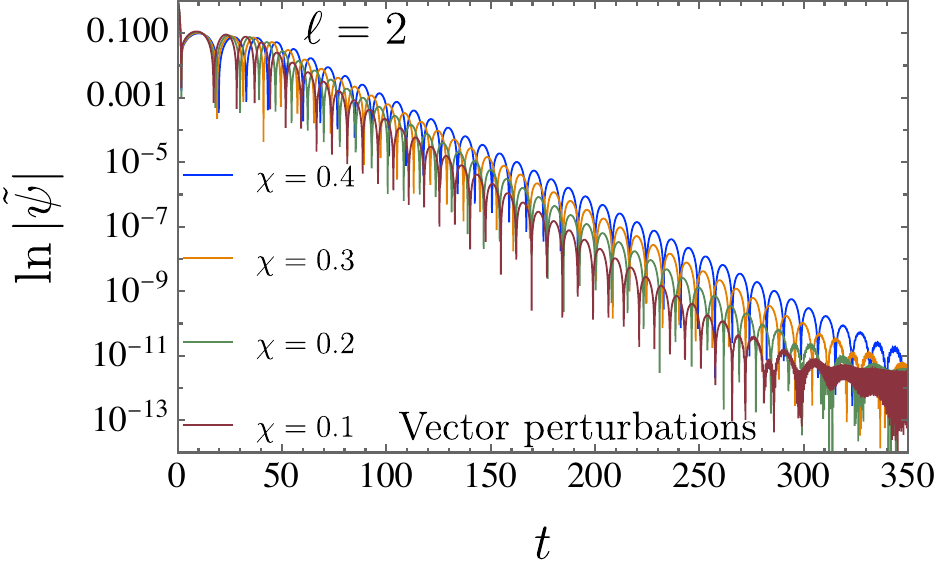}
     \includegraphics[scale=0.51]{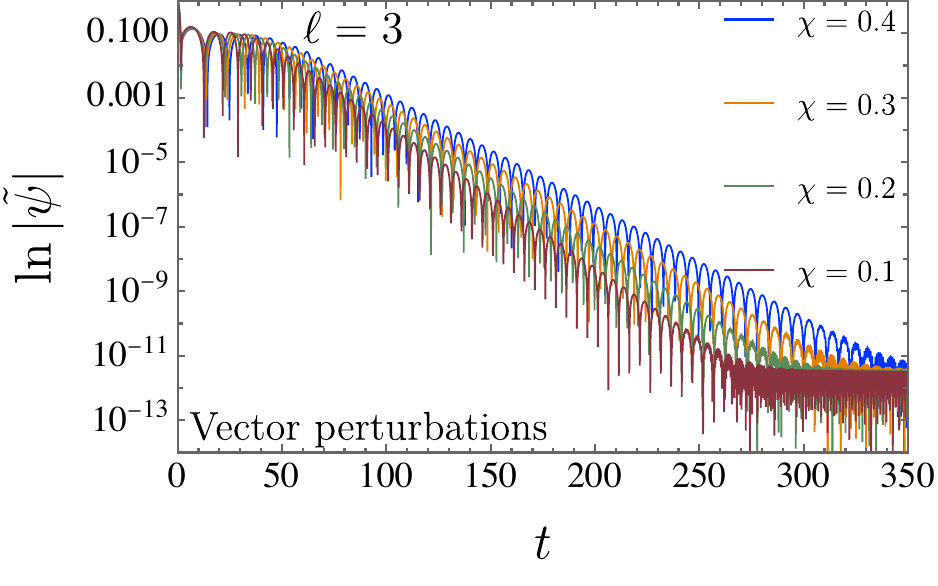}
    \caption{Time evolution of $\ln|\tilde{\psi}|$ for vector perturbations in a black hole spacetime with fixed mass $M=1$ and Lorentz--violating parameter $\chi = 0.1, 0.2, 0.3,$ and $0.4$. The panels correspond to the angular indices $l=1$ (top left), $l=2$ (top right), and $l=3$ (bottom), showing the damping behavior across different modes. }
    \label{timedomainsvector2}
\end{figure}

\begin{figure}
    \centering
    \includegraphics[scale=0.51]{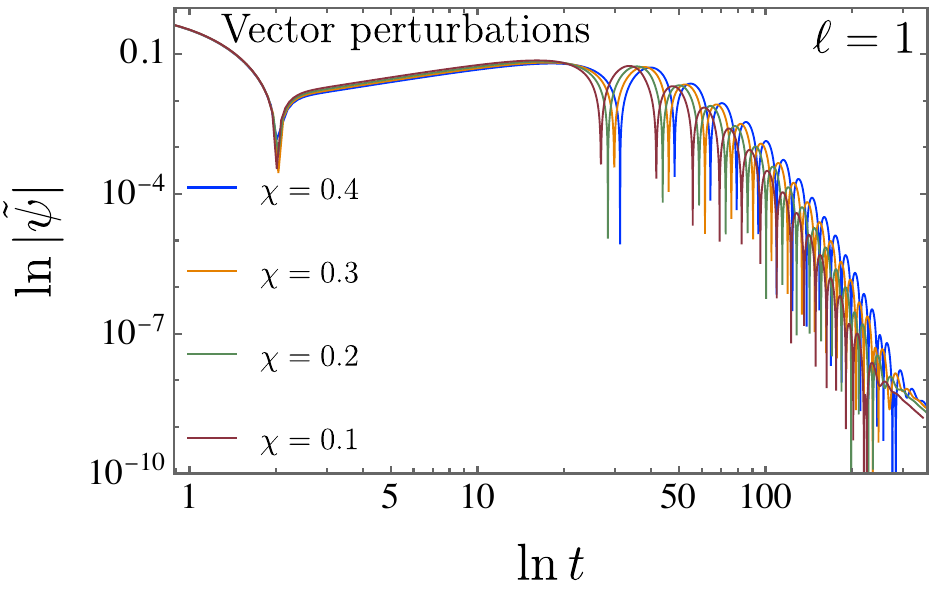}
    \includegraphics[scale=0.51]{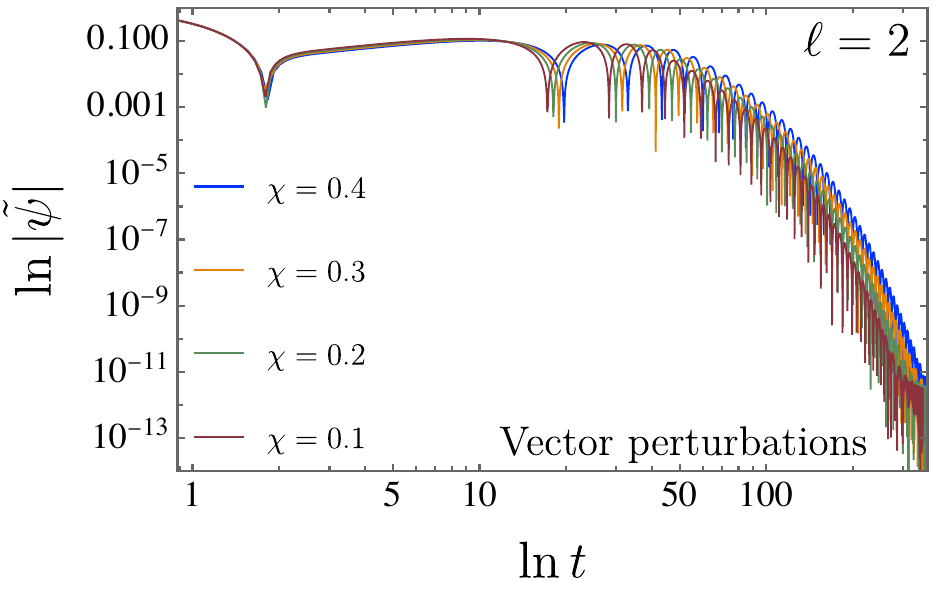}
     \includegraphics[scale=0.51]{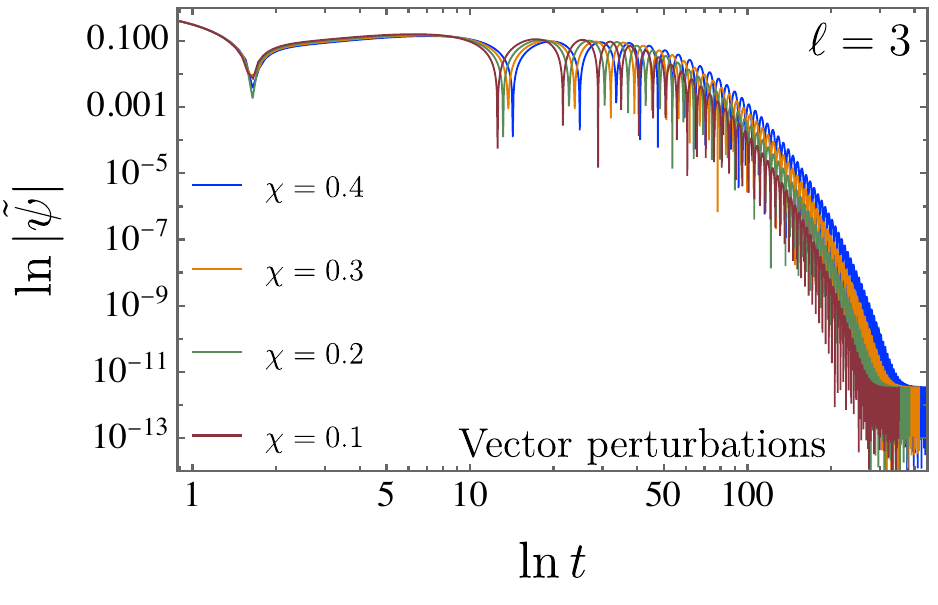}
    \caption{Late--time evolution of the vector perturbation $\tilde{\psi}$ shown on a double--logarithmic scale, where $\ln|\tilde{\psi}|$ is plotted against $\ln t$ for $M=1$ and $\chi = 0.1, 0.2, 0.3,$ and $0.4$. The subplots correspond to $l=1$ (upper left), $l=2$ (upper right), and $l=3$ (lower), which emphasize the emergence of the power--law decay that characterizes the tail phase of the signal. }
    \label{timedomainsvector3}
\end{figure}


\subsection{Tensor perturbations}

The time evolution of tensor perturbations in the spacetime defined by Eq.~(\ref{maaaaianametric}) is now analyzed. Figure~\ref{timedomaintensor2} shows the propagation of the field $\tilde{\psi}$ for a black hole with fixed mass $M=1$ and Lorentz–violating parameter values $\chi = 0.1, 0.2, 0.3,$ and $0.4$. The panels correspond to the angular indices $l=2$ (top left), $l=3$ (top right), and $l=4$ (bottom). The resulting waveforms display the characteristic damped oscillations associated with the quasinormal ringing phase. When compared with the scalar and vector cases, tensor modes oscillate at lower frequencies and decay more slowly, indicating that their damping is weaker and their ringing lasts longer. Nonetheless, the same qualitative behavior is observed: increasing $\chi$ reduces the damping rate and extends the duration of the oscillatory phase.

Figure~\ref{timedomaintensor3} provides a clearer representation of this attenuation by plotting $\ln|\tilde{\psi}|$ against time for the same set of parameters. The nearly linear sections correspond to the exponential decay typical of the quasinormal ringing stage, after which the curves transition to the late–time power–law tail. As in the frequency–domain analysis, a larger $\chi$ results in slower decay. The tensor sector, however, exhibits the slowest damping among all spin configurations, outlasting both the scalar and vector perturbations before reaching equilibrium.

Finally, Fig.~\ref{timedomaintensor4} displays the late–time behavior on a double--logarithmic scale, keeping the same panel arrangement for comparison. This representation clearly shows the transition from the exponential ringdown to the power–law tail that dominates at large times. While the overall structure mirrors that of the scalar and vector cases, the tensor perturbations decay at a notably slower rate, confirming that they are the most persistent of the three spin sectors within the Lorentz--violating background.

\begin{figure}
    \centering
    \includegraphics[scale=0.51]{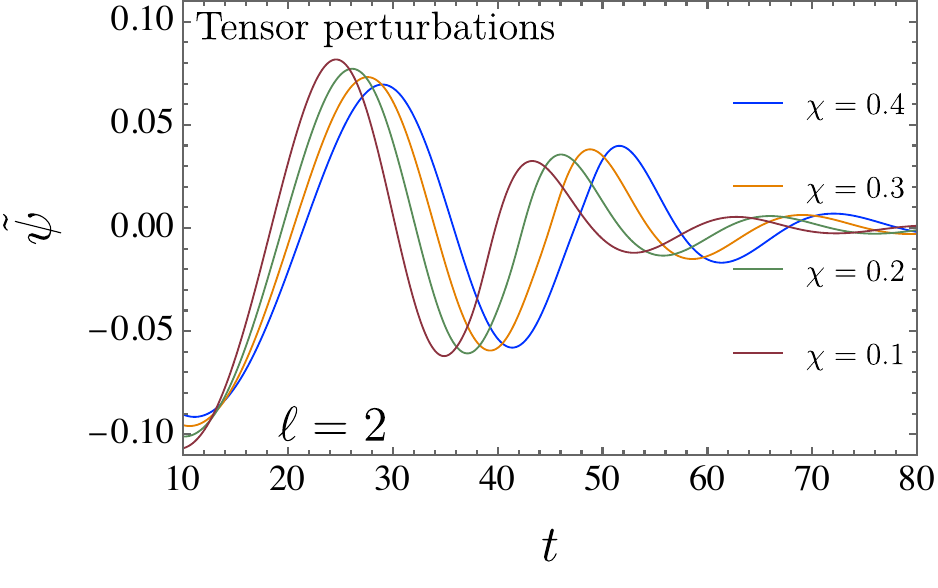}
    \includegraphics[scale=0.51]{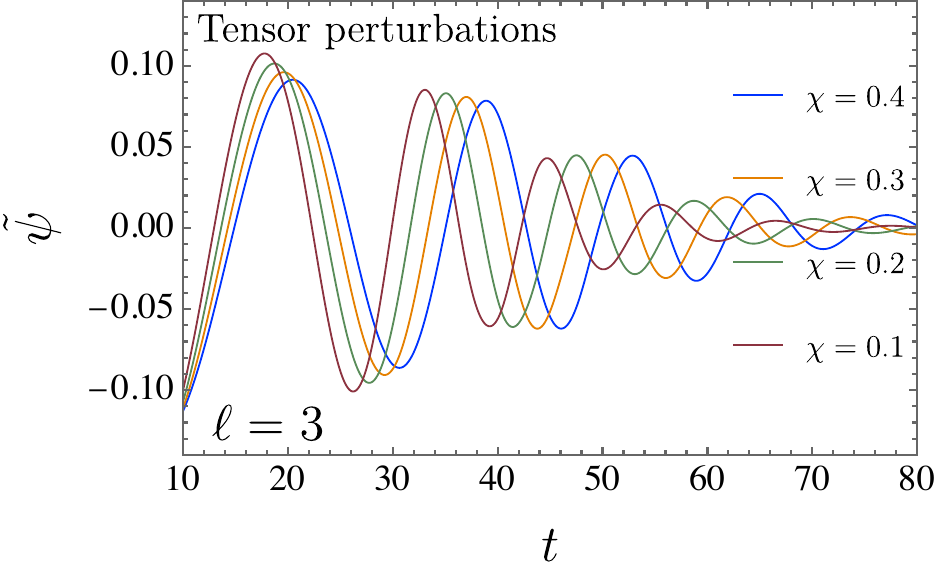}
     \includegraphics[scale=0.51]{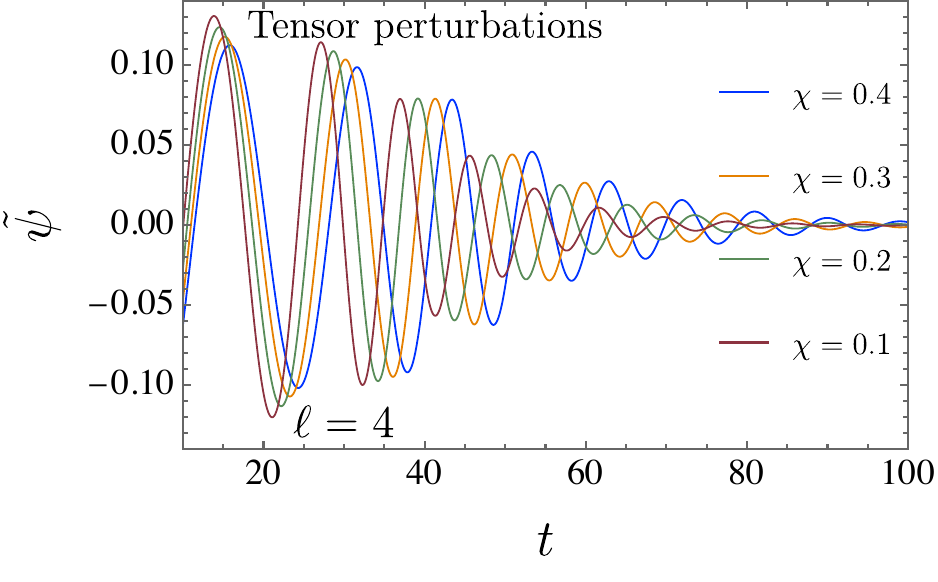}
    \caption{Time profiles of the tensor perturbation $\tilde{\psi}$ for $M=1$ and $\chi = 0.1, 0.2, 0.3,$ and $0.4$. The plots for $l=2$ (top left), $l=3$ (top right), and $l=4$ (bottom) illustrate how the waveform evolves as $\chi$ increases. }
    \label{timedomaintensor2}
\end{figure}

\begin{figure}
    \centering
    \includegraphics[scale=0.51]{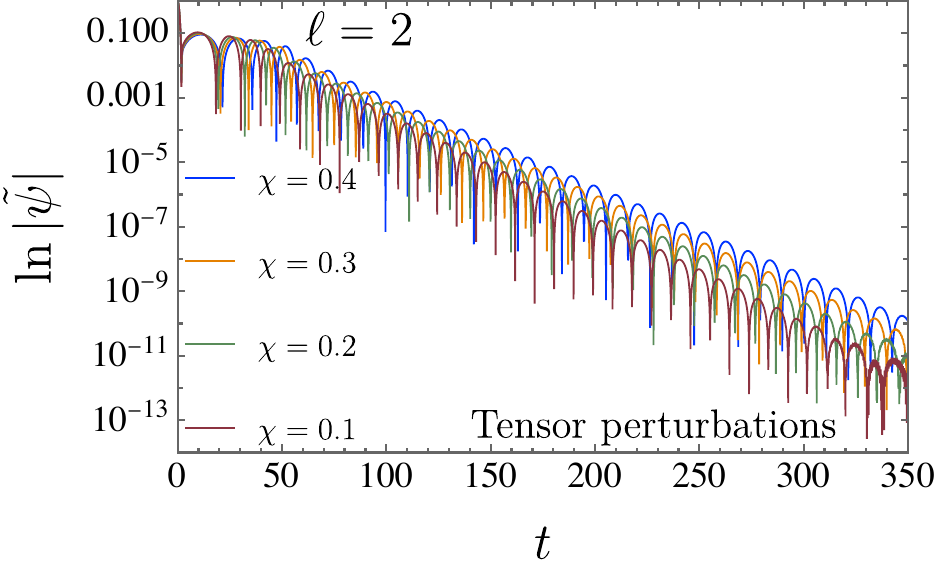}
    \includegraphics[scale=0.51]{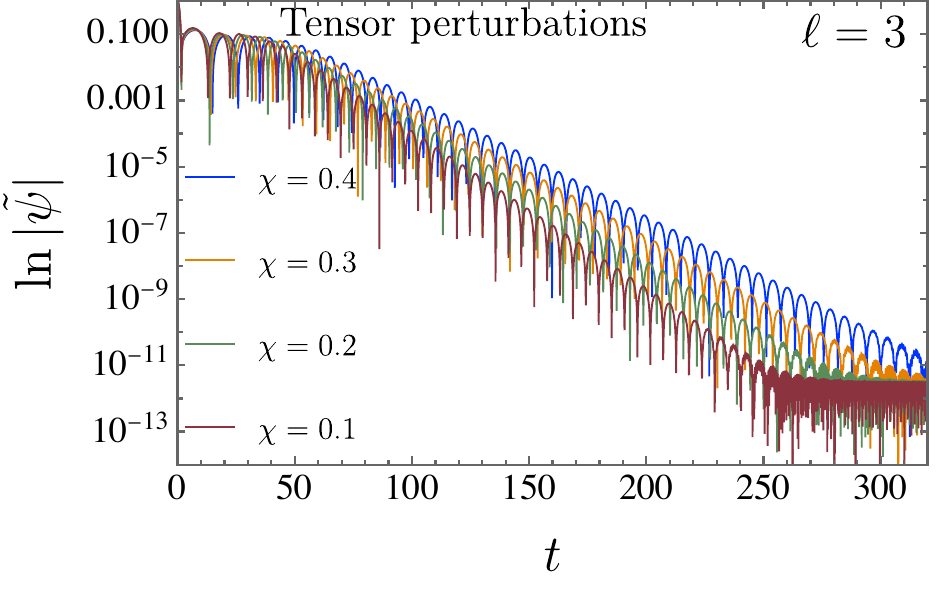}
     \includegraphics[scale=0.51]{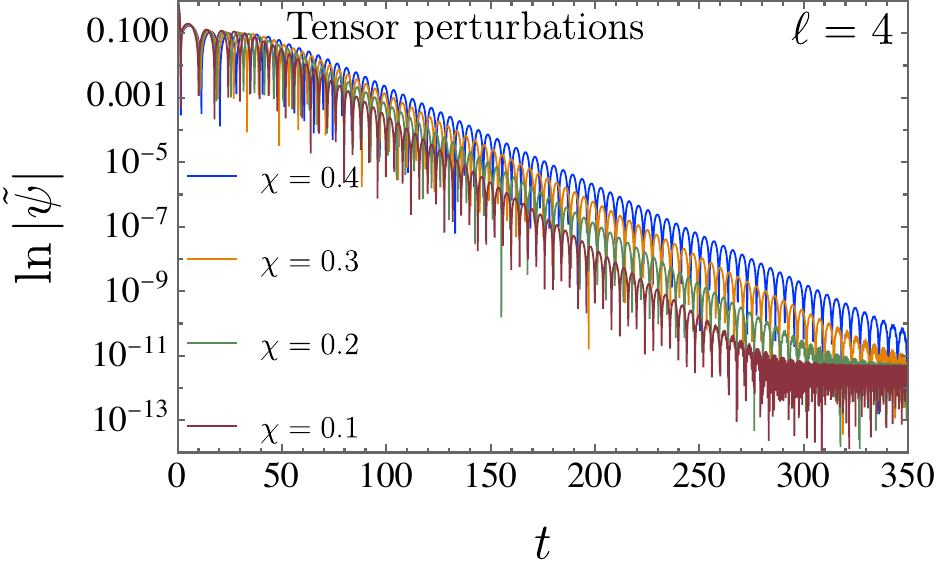}
    \caption{Time evolution of $\ln|\tilde{\psi}|$ for tensor perturbations with $M=1$ and $\chi = 0.1, 0.2, 0.3,$ and $0.4$. The subplots for $l=2$ (upper left), $l=3$ (upper right), and $l=4$ (lower) display the decay pattern of each mode as $\chi$ increases. }
    \label{timedomaintensor3}
\end{figure}

\begin{figure}
    \centering
    \includegraphics[scale=0.51]{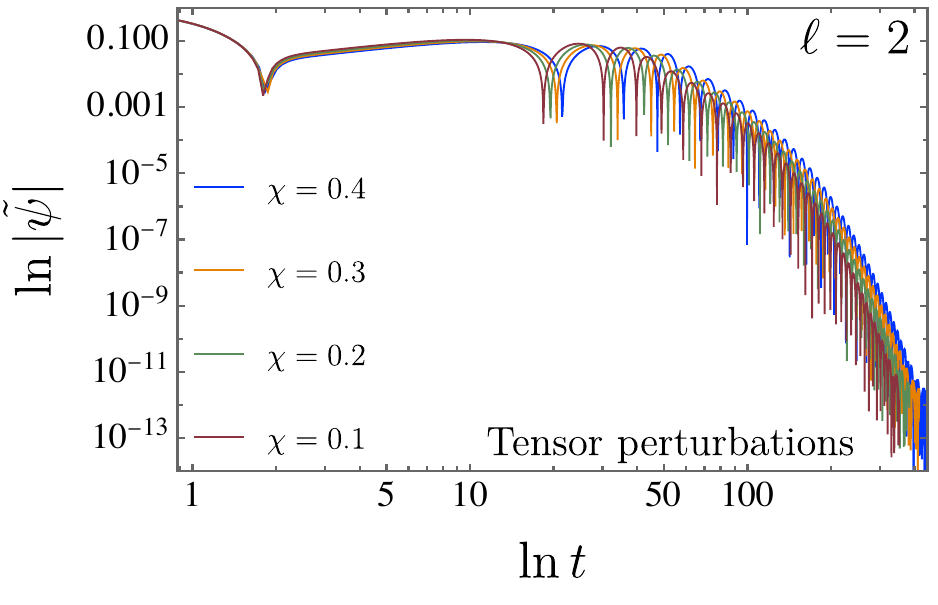}
    \includegraphics[scale=0.51]{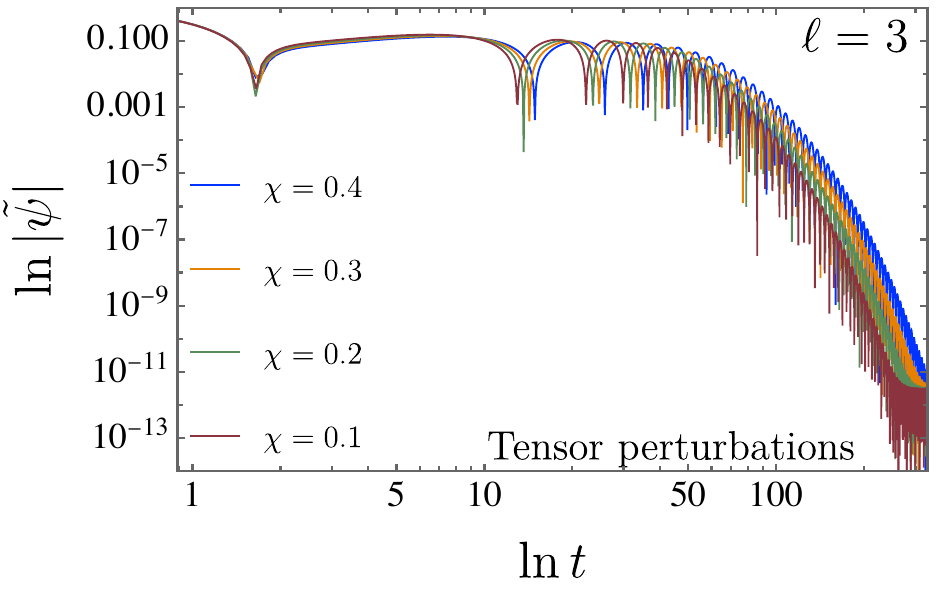}
     \includegraphics[scale=0.51]{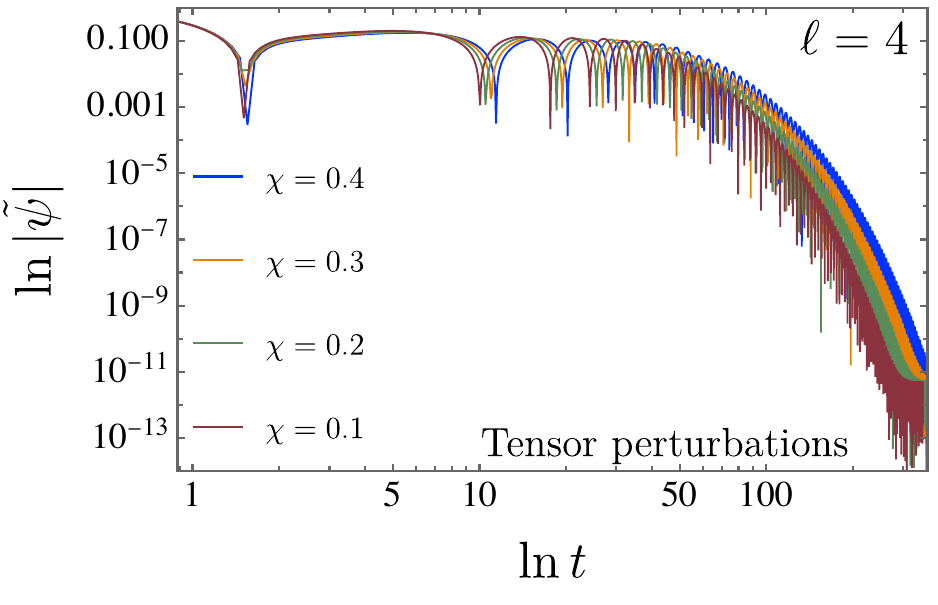}
    \caption{Late--time behavior of the tensor perturbation $\tilde{\psi}$ shown on a double--logarithmic scale, where $\ln|\tilde{\psi}|$ is plotted against $\ln t$ for $M=1$ and $\chi = 0.1, 0.2, 0.3,$ and $0.4$. The subplots for $l=2$ (upper left), $l=3$ (upper right), and $l=4$ (lower) emphasize the power--law decay characteristic of the tail stage of the waveform. }
    \label{timedomaintensor4}
\end{figure}


\subsection{Spinor perturbations }

The temporal behavior of spinor perturbations in the geometry described by Eq.~(\ref{maaaaianametric}) is now investigated for the sake of completing our analysis. Figure~\ref{timedomainspinor12} presents the evolution of the field $\tilde{\psi}$ for a black hole of unit mass ($M=1$) with Lorentz--violating parameters $\chi = 0.1, 0.2, 0.3,$ and $0.4$. The upper panels correspond to $l=1/2$ and $l=3/2$, while the lower one depicts $l=5/2$. The resulting signals exhibit the expected damped oscillations typical of the quasinormal ringing phase. In comparison with the scalar, vector, and tensor sectors, the spinor modes oscillate with slightly higher real frequencies and experience a faster decay than the vector and tensor perturbations, though their damping remains slower than that of the scalar modes. As $\chi$ increases, the oscillations persist for a longer time, confirming that Lorentz violation reduces the overall damping rate.

The attenuation process becomes clearer in Fig.~\ref{timedomainspinor32}, where $\ln|\tilde{\psi}|$ is plotted against time for the same set of parameters. The nearly linear segments correspond to the exponential decay regime of the quasinormal ringing, which gradually transitions into the power--law tail at late times. The slopes of these lines diminish as $\chi$ grows, reflecting a slower decay. When compared with other spin configurations, the spinor perturbations occupy an intermediate position between the rapidly damped scalar modes and the more persistent vector and tensor ones.

Finally, Fig.~\ref{timedomainspinor52} shows the late--time behavior in a double--logarithmic representation, keeping the same layout for direct comparison. The crossover from exponential damping to the power--law tail is clearly visible. Although the overall structure parallels that found for bosonic perturbations, the spinor modes exhibit a distinct intermediate configuration. Among all sectors, tensor perturbations remain the most enduring, followed by the vector, spinor, and scalar modes, in full agreement with the hierarchy obtained from the quasinormal spectra.

In Fig.~\ref{comtimeall}, the time--domain evolution is presented for all spin configurations. As expected from the comparison of the effective potentials in Fig.~\ref{comptortoise}, a similar hierarchy appears in the temporal profiles, with higher frequencies following the sequence $\mathrm{V}_{\psi} > \mathrm{V}_{s} > \mathrm{V}_{v} > \mathrm{V}_{t}$, particularly at early times. The results are calculated through $\chi=0.1$ and $M=1$.

\begin{figure}
    \centering
    \includegraphics[scale=0.51]{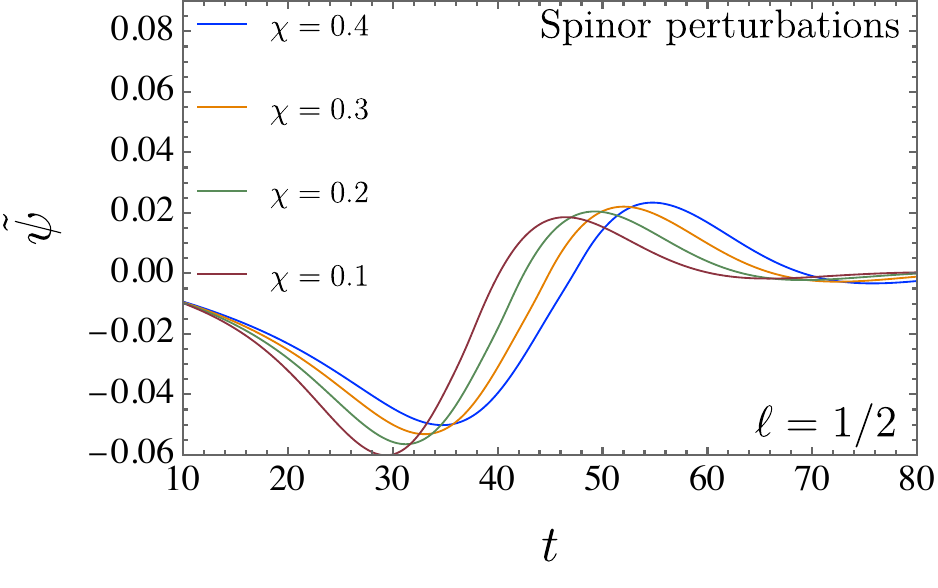}
    \includegraphics[scale=0.51]{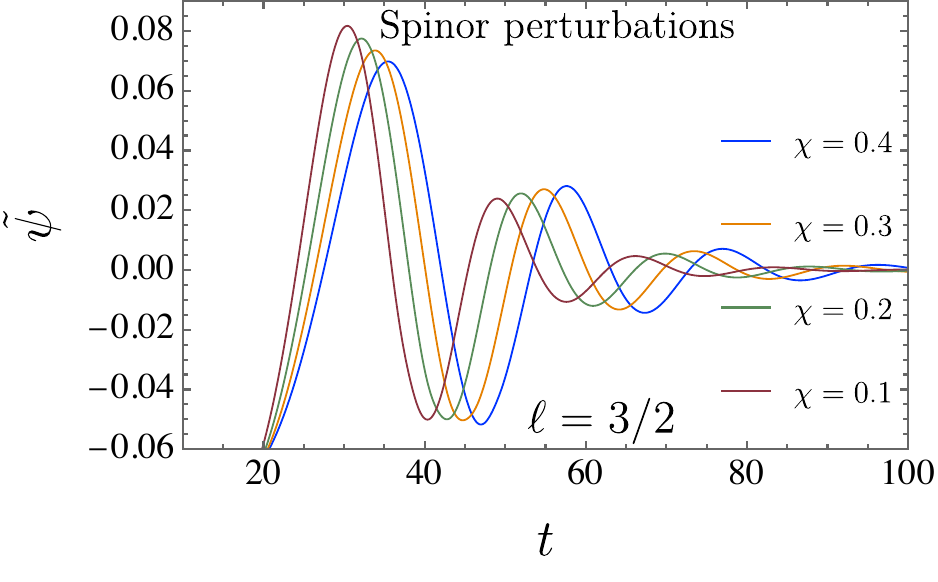}
     \includegraphics[scale=0.51]{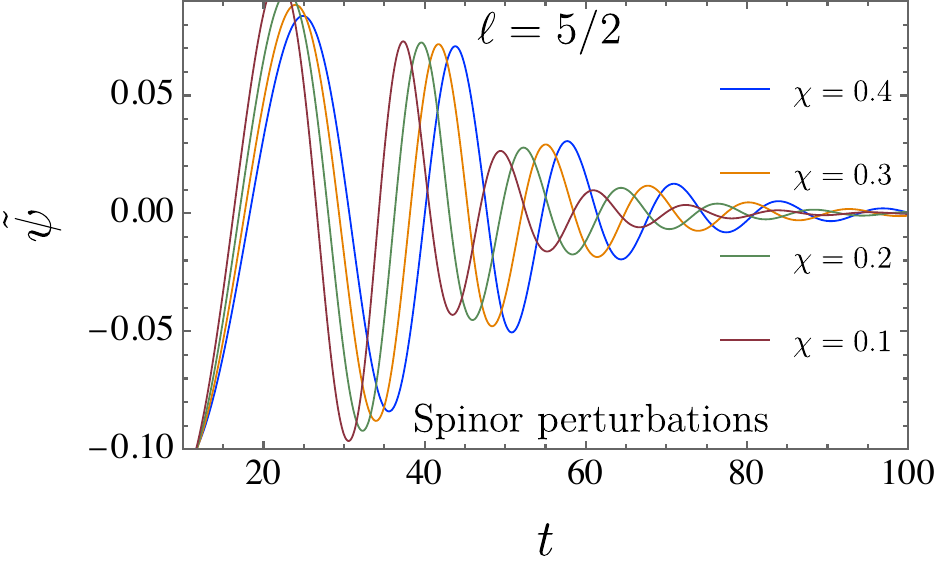}
    \caption{Time evolution of the spinor field $\tilde{\psi}$ for $M=1$ and $\chi = 0.1, 0.2, 0.3,$ and $0.4$. Panels correspond to $l=1/2$ (upper left), $l=3/2$ (upper right), and $l=5/2$ (bottom), showing how the waveform varies with increasing $\chi$. }
    \label{timedomainspinor12}
\end{figure}

\begin{figure}
    \centering
    \includegraphics[scale=0.51]{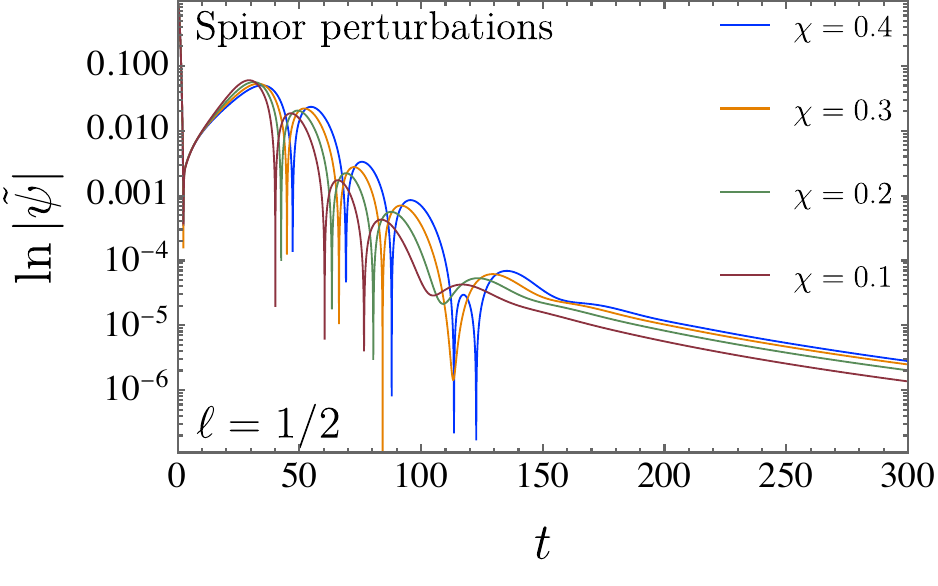}
    \includegraphics[scale=0.51]{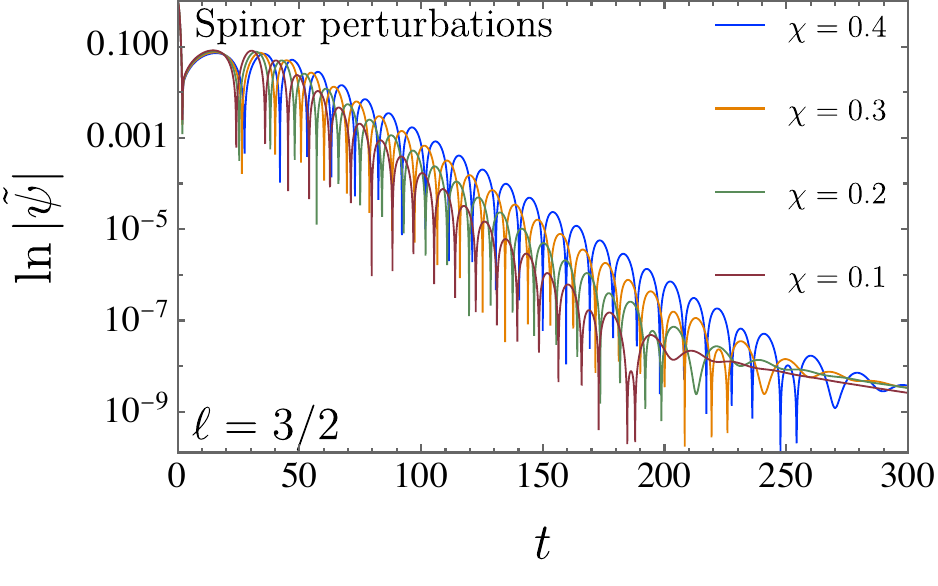}
     \includegraphics[scale=0.51]{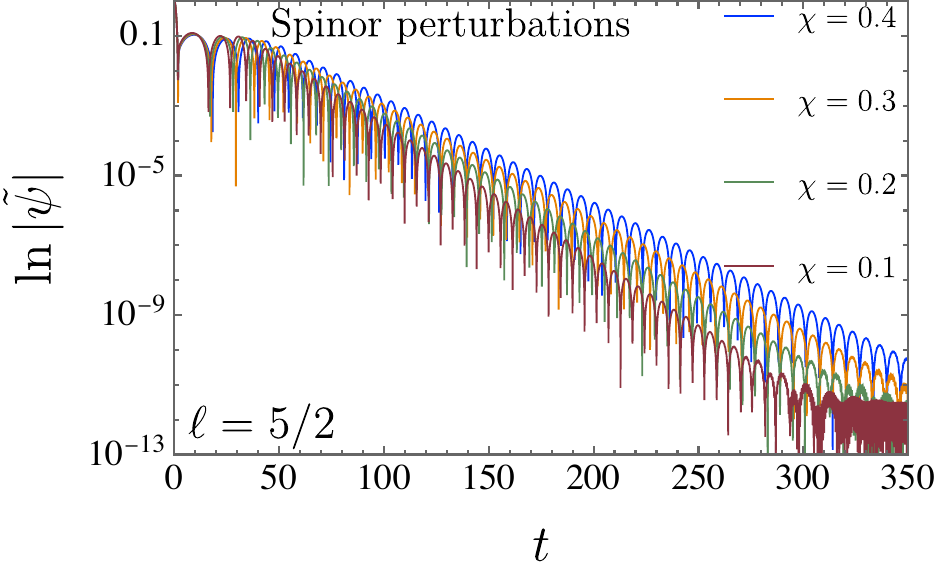}
    \caption{Time evolution of $\ln|\tilde{\psi}|$ for spinor perturbations with $M=1$ and $\chi = 0.1, 0.2, 0.3,$ and $0.4$. Panels correspond to $l=1/2$ (top left), $l=3/2$ (top right), and $l=5/2$ (bottom), illustrating the damping behavior for each mode.}
    \label{timedomainspinor32}
\end{figure}

\begin{figure}
    \centering
    \includegraphics[scale=0.51]{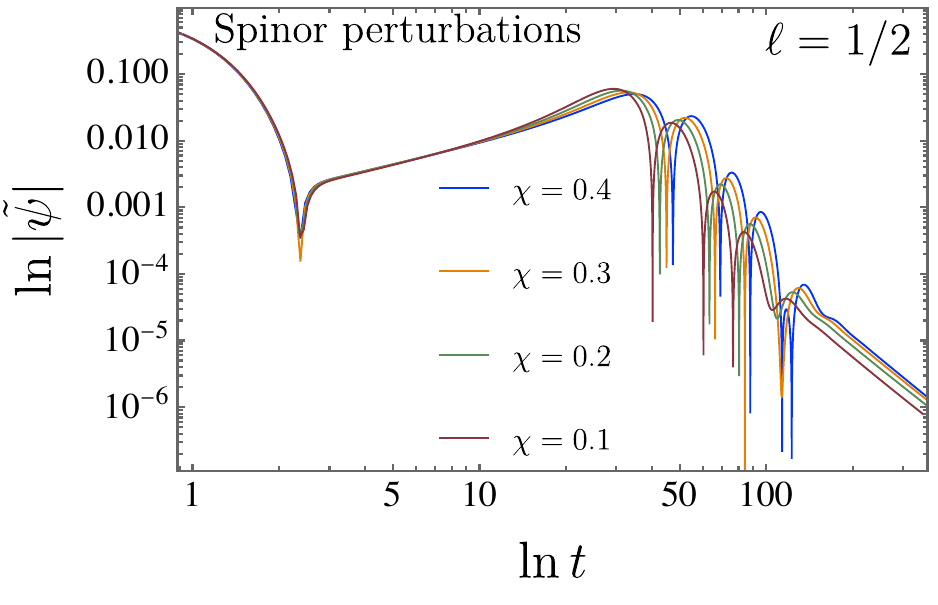}
    \includegraphics[scale=0.51]{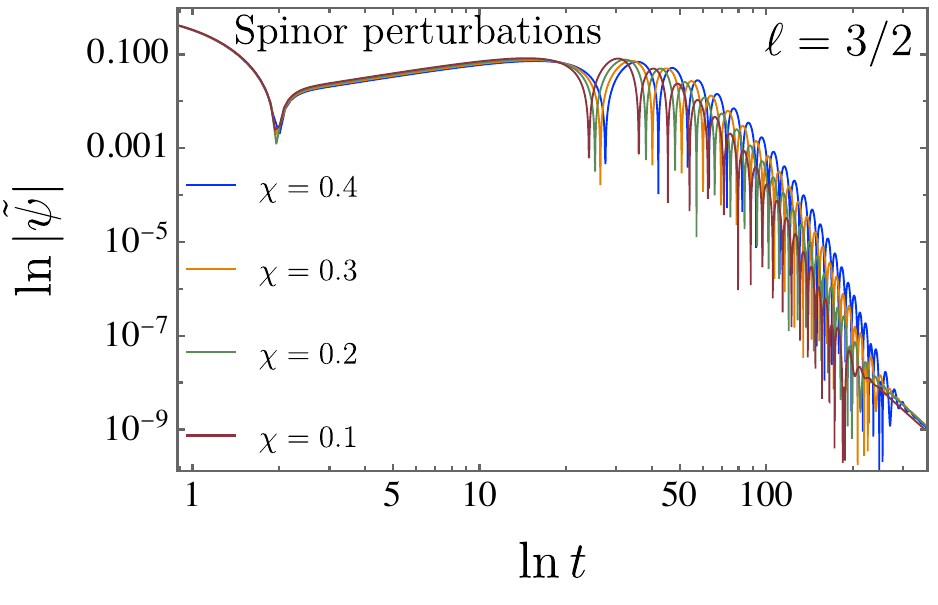}
     \includegraphics[scale=0.51]{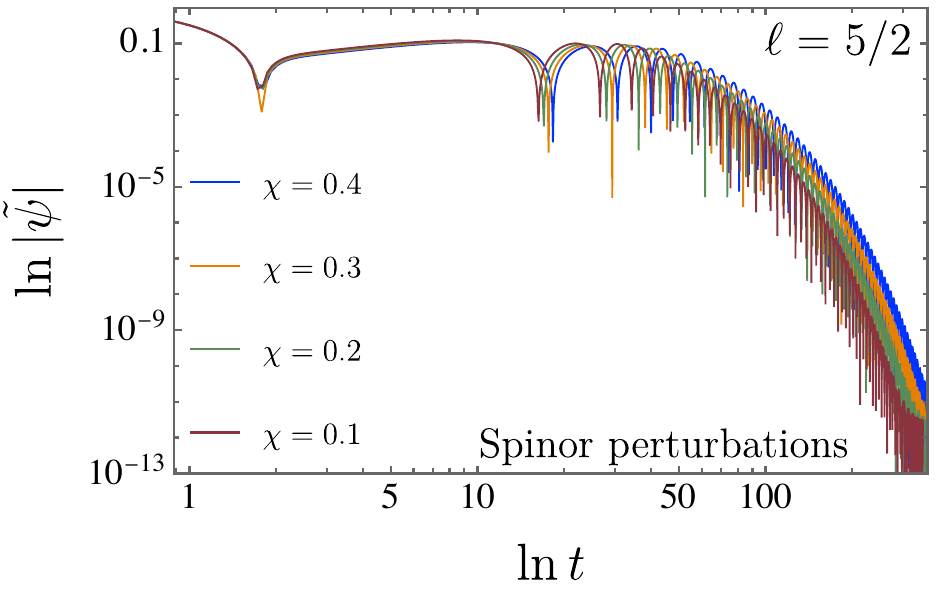}
    \caption{Late--time evolution of the spinor perturbation $\tilde{\psi}$ on a double--logarithmic scale, with $\ln|\tilde{\psi}|$ plotted versus $\ln t$ for $M=1$ and $\chi = 0.1, 0.2, 0.3,$ and $0.4$. Subplots correspond to $l=1/2$ (upper left), $l=3/2$ (upper right), and $l=5/2$ (lower), highlighting the power--law decay that marks the tail phase of the signal. }
    \label{timedomainspinor52}
\end{figure}

\begin{figure}
    \centering
    \includegraphics[scale=0.51]{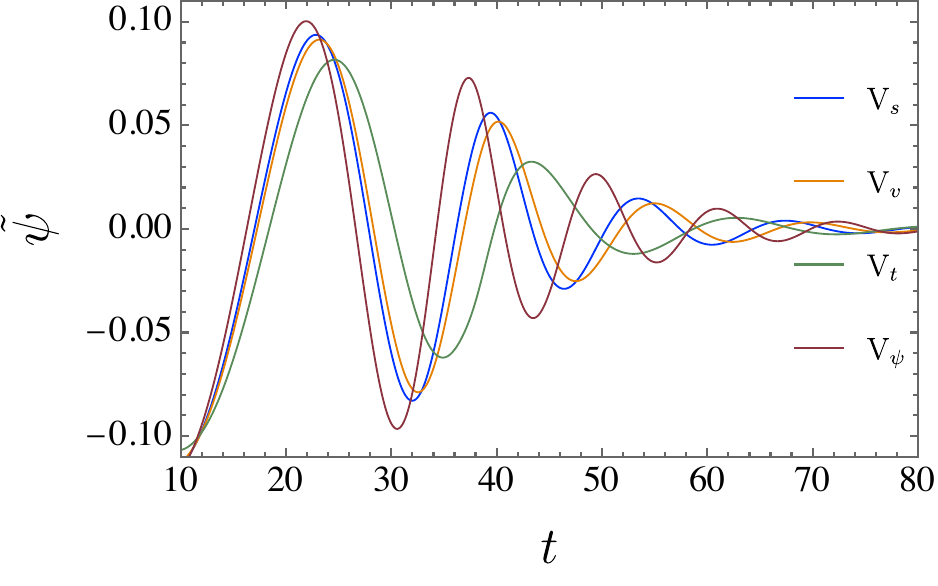}
    \includegraphics[scale=0.51]{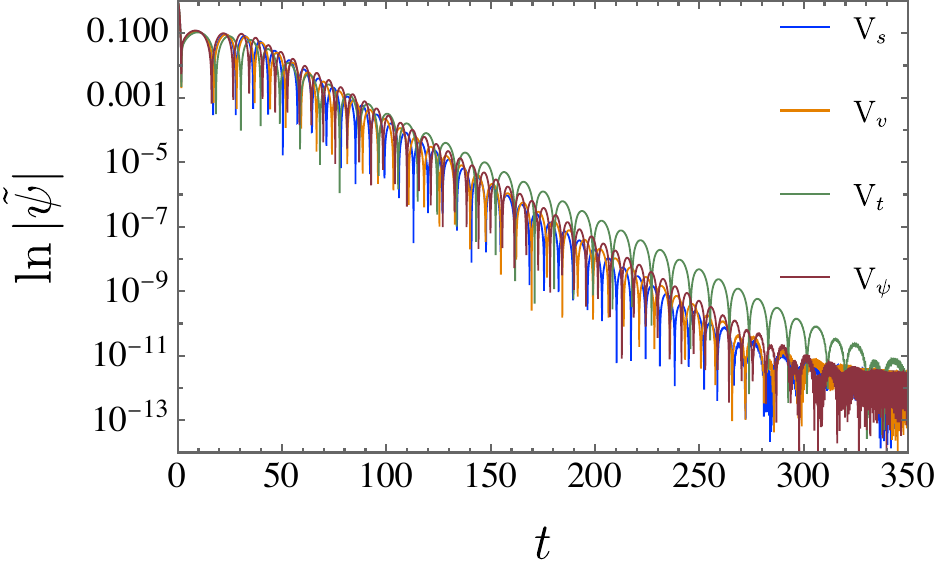}
     \includegraphics[scale=0.51]{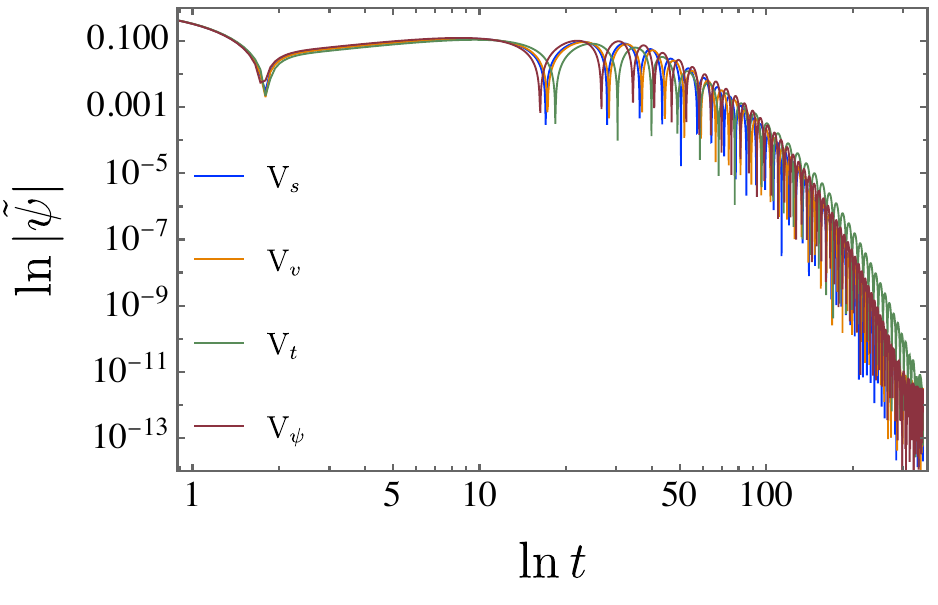}
    \caption{Comparison of the time--domain profiles for all spin configurations, showing $\tilde{\psi}$ and $\ln|\tilde{\psi}|$, with parameters $\chi=0.1$ and $M=1$.}
    \label{comtimeall}
\end{figure}

Finally, for all types of perturbations, as confirmed by the time--domain analysis including spins $s = 0, 1/2, 1,$ and $2$, no echoes were observed. This outcome is consistent with the behavior of the effective potentials plotted as functions of the tortoise coordinate $r^{*}$, where each multipole moment $l$ exhibits a single peak, indicating the absence of potential wells capable of producing echo signals.


\section{Optical Geometry and Weak-Deflection Analysis }

The discussion now turns to the weak--field lensing behavior of the spacetime. Instead of starting from the light trajectories themselves, the procedure adopts the Gauss--Bonnet framework \cite{Gibbons:2008rj}, which links the deflection angle to geometric properties of the associated optical manifold. However, the geometry defined in Eq. (\ref{maaaaianametric}) does not approach flat spacetime at large distances. After suitable redefinitions, the asymptotic region reveals the structure of a globally conical background, similar to that generated by a global monopole. Because of this non-standard asymptotic behavior, the usual weak–field lensing formulas require specific adjustments. Such corrections follow the treatment outlined in Ref. \cite{Jusufi:2017lsl}, where spacetimes with a conical infinity were analyzed in detail.

In this manner, before inserting this method into the computation, it is necessary to understand the behavior of the circular null trajectories. For that purpose, the Gaussian curvature of the optical geometry is evaluated, since it encodes how these orbits respond to small perturbations. The resulting sign of the curvature is essential for our forthcoming analysis: regions where it becomes positive are associated with stable circular photon paths, whereas a negative curvature signals that the corresponding orbits cannot remain bound under radial disturbances.

\subsection{Geometric stability of critical photon orbits }

Circular photon motion around a black hole is best understood through the geometry of the corresponding optical space. Instead of viewing these orbits only as solutions of the geodesic equations, one may reinterpret them as features of the curved two--dimensional manifold that governs the propagation of light. In this picture, the question of whether a photon can keep circling the black hole under small radial disturbances is tied to how this manifold bends. When the geometry forces nearby light rays to separate, a photon initially placed on a circular track cannot remain there: its path either spirals inward toward the horizon or drifts outward until it escapes. If, on the contrary, the geometry confines neighboring null trajectories, the photon can hover near the circular path and execute multiple revolutions within a bounded region \cite{qiao2022geometric,Heidari:2025iiv,qiao2022curvatures,araujo2025impact}.

This behavior can be framed in terms of curvature properties intrinsic to the optical manifold. The essential quantity is the Gaussian curvature $\mathcal{K}(r)$, which determines whether bundles of nearby null rays converge or diverge. According to the Cartan--Hadamard theorem, a domain with non--positive curvature, $\mathcal{K}(r)\le 0$, cannot host conjugate points; consequently, circular photon paths in such regions are necessarily unstable. Conversely, when the curvature becomes positive, $\mathcal{K}(r)>0$, conjugate points may form, allowing for the existence of confined photon rings that behave as stable critical orbits \cite{qiao2024existence}. Under these geometric considerations, the null condition $\mathrm{d}s^{2}=0$ may be recast into an equivalent optical form, written as \cite{AraujoFilho:2024xhm}:
\ie
\mathrm{d}t^2 = \Tilde{\gamma}_{ij}\mathrm{d}x^i \mathrm{d}x^j = \frac{1}{A(r,\chi) \, B(r,\chi)}\mathrm{d}r^2  +\frac{\Bar{D}(r,\chi)}{A(r,\chi)}\mathrm{d}\varphi^2.
\fe

To analyze the optical geometry, it is convenient to restrict attention to the spatial sector of the null metric. Only the coordinates associated with spatial directions are involved, so the indices $i$ and $j$ span the range $1$ to $3$. The optical manifold is characterized by the induced spatial metric $\tilde{\gamma}_{ij}$, which governs the propagation of light rays once the null condition has been imposed. When focusing on the equatorial plane, one isolates the relevant metric function by evaluating it at $\theta=\pi/2$. This leads to the reduced expression $\bar{D}(r,\chi)$, which encapsulates the radial dependence of the optical geometry in that slice and plays the role of an effective metric coefficient in the two--dimensional optical surface. After these ingredients are identified, the entire intrinsic geometry of the optical space is encoded in its Gaussian curvature. For the class of optical manifolds associated with a generic spacetime, the curvature reduces to a single radial function, whose general form was derived in \cite{qiao2024existence}:
\ie
\mathcal{K}(r,\chi) = \frac{R}{2} =  -\frac{ A(r,\chi) \sqrt{B(r,\chi)}}{\sqrt{ \,  \Bar{D}(r,\chi)}}  \frac{\partial}{\partial r} \left[  \frac{A(r,\chi) \sqrt{B(r,\chi)}}{2 \sqrt{  \Bar{D}(r,\chi) }}   \frac{\partial}{\partial r} \left(   \frac{\Bar{D}(r,\chi)}{A(r,\chi)}    \right)    \right].
\fe
In this setting, $R$ denotes the Ricci scalar of the two--dimensional optical geometry. The calculations that follow use the modified line element given in Eq. (\ref{redefinedmetric}), which defines the effective metric employed in the curvature analysis:
\ie
\nonumber
\mathrm{d}s^{2}=-\left(1-\frac{2\Tilde{M}}{\Tilde{r}}\right)\mathrm{d}\Tilde{t}^{\,2}+\frac{1}{1-\frac{2\Tilde{M}}{\Tilde{r}}}\mathrm{d}\Tilde{r}^{\,2}+\Tilde{K}^{2}\Tilde{r}^{2}\mathrm{d}\Omega^{2}.
\fe
To make explicit the conical structure of the geometry, we introduce the rescaled mass parameter $\tilde{M}=\sqrt{1+\chi}\,M$. Because the spacetime is not asymptotically flat—the large-distance region exhibits a global monopole–type configuration—the usual Gauss–Bonnet prescription for weak–field lensing \cite{Gibbons:2008rj} cannot be applied in its standard form. In such geometries, the asymptotic boundary contributes an additional term to the deflection angle, as shown in \cite{Jusufi:2017lsl}. Our treatment follows that approach. The analysis in \cite{Jusufi:2017lsl} includes rotation, but by taking the limit $a\to 0$ and adopting the coordinate redefinitions introduced in Sec. II, their formalism adapts directly to the metric in Eq. (\ref{redefinedmetric}). Thereby, we can properly write the Gaussian curvature as follows
\ie
\begin{split}
\label{gbbhhh}
& \mathcal{K}(r,\chi) = \frac{3 \Tilde{M}^2}{r^4}-\frac{2 \Tilde{M}}{r^3} = \frac{3 M^2 (\chi +1)}{r^4}-\frac{2 M \sqrt{\chi +1}}{r^3}. 
\end{split}
\fe

A series of recent analyses \cite{Heidari:2025iiv,araujo2025impact,qiao2024existence,qiao2022geometric,AraujoFilho:2025huk,qiao2022curvatures,AraujoFilho:2025vgb} has shown that the fate of circular photon trajectories is encoded directly in the sign of the Gaussian curvature $\mathcal{K}(r,\chi)$ of the optical manifold. Rather than relying solely on the effective potential, these works demonstrate that the geometric criterion is decisive: regions with $\mathcal{K}(r,\chi)>0$ promote the convergence of neighboring null rays and may sustain closed photon loops, whereas $\mathcal{K}(r,\chi)<0$ forces nearby geodesics to separate, a behavior incompatible with stability.

This geometric picture is illustrated in Fig.~\ref{gahhssgs}, where $\mathcal{K}(r,\chi)$ is displayed for $M=1$ and $\chi=0.1$. To emphasize the impact of Lorentz violation, the full (non--expanded) curvature expression is used exclusively for this figure. The profile reveals two distinct radial intervals: a sector with positive curvature, shaded light orange, and another where the curvature becomes negative, shaded light purple. The transition between them occurs at $r\simeq 1.57$ (as highlighted by the wine circle), where $\mathcal{K}$ vanishes. Because the photon sphere is located beyond this turning point (as depicted by the wine dot), it lies entirely inside the negative--curvature region, implying that every circular photon orbit in this spacetime is unstable, as happens in the bumblebee case and in the Schwarzschild geometry as well.

\begin{figure}
    \centering
    \includegraphics[scale=0.71]{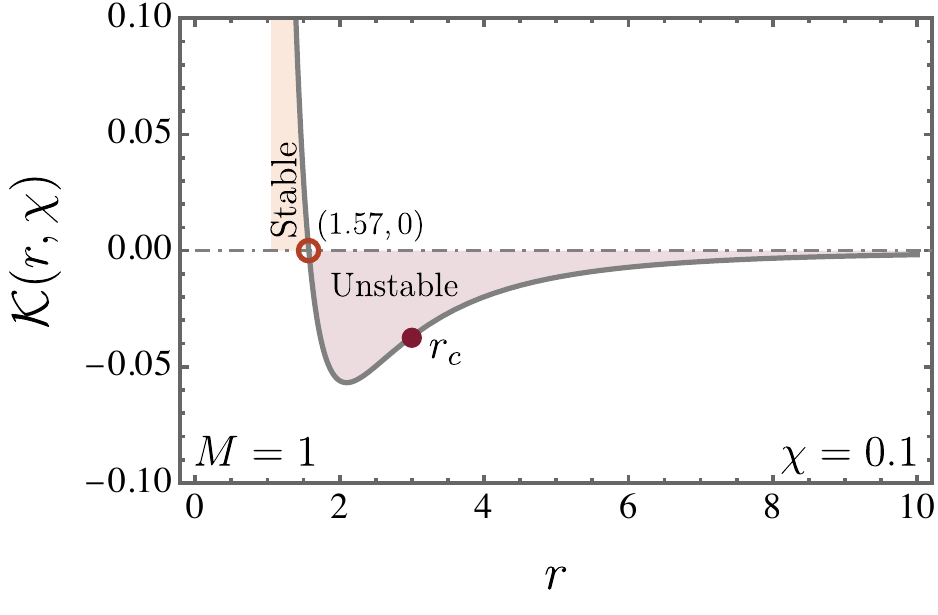}
    \caption{The Gaussian curvature $\mathcal{K}(r,\chi)$ for $M=1$ and $\chi=0.1$ is shown. The wine--colored circle marks the point where $\mathcal{K}$ changes sign, dividing the positive- and negative-curvature domains. The wine--colored dotted marker shows the photon--sphere radius $r_{c}$, situated within the negative--curvature (light purple) region.}
    \label{gahhssgs}
\end{figure}


\subsection{Weak-deflection angle via the Gauss-Bonnet method }

The computation of the weak--deflection angle is carried out through the Gauss--Bonnet framework \cite{Jusufi:2017lsl}, using the curvature term introduced in Eq.~(\ref{gbbhhh}) as the starting point. Since null geodesics relevant for lensing can be confined to the plane $\theta=\pi/2$, the optical manifold effectively reduces to a two--dimensional surface. Within this reduced geometry, the associated surface element takes the form:
\ie
\mathrm{d}S = \sqrt{\Tilde{\gamma}} \, \mathrm{d} r \mathrm{d}\varphi = \sqrt{\frac{1}{A(r,\chi)} \frac{1}{B(r,\chi)} \frac{D(r,\chi)}{A(r,\chi)} } \, \mathrm{d} r \mathrm{d}\varphi.
\fe
For the evaluation of the integral that yields the deflection angle, the Lorentz–violating parameter is handled in the same perturbative manner adopted throughout the paper. The treatment assumes a large impact parameter, $b\gg 2M$, matching the weak--field regime in which the Gauss--Bonnet method was originally implemented in \cite{Gibbons:2008rj}. Under these conditions, the mass term is expanded consistently up to second order.

Starting from the reduced curvature expression under these approximations, the resulting deflection angle can be written in the form derived in \cite{Jusufi:2017lsl}:
\ie
\begin{split}
\label{vfffdddd}
& \alpha (b,\chi) =  \pi \left( \frac{1}{\Tilde{K}} -1 \right) - \frac{1}{\Tilde{K}}\int^{\pi}_{0} \int^{\infty}_{\frac{b}{\sin \varphi}} \mathcal{K} \mathrm{d}S \\
& \approx  \, \, \frac{\pi \chi}{2}  + \left( 1 + \frac{\chi}{2} \right)  \left[ \frac{4 \Tilde{M}}{b} +\frac{3 \pi  \Tilde{M}^2}{4 b^2} -\frac{3 \pi  \Tilde{M}^2 \chi }{8 b^2}-\frac{2 \Tilde{M} \chi }{b} \right] \\ 
& \approx   \, \, \frac{4 M}{b} +\frac{3 \pi  M^2}{4 b^2} +\frac{\pi  \chi }{2}+ \frac{3 \pi  M^2 \chi }{4 b^2}+\frac{2 M \chi }{b}.
\end{split}
\fe
In the above expression, the expansion is carried out to first order in $\chi$. Since the Lorentz–violating parameter is assumed to be small, one may further approximate $\pi(\tilde{K}^{-1}-1)\simeq \pi\chi/2$ and $1/\tilde{K}\simeq 1+\chi/2$. The first two contributions in $\alpha(b,\chi)$ reproduce the standard Schwarzschild result; the next term arises from the conical modification associated with the global monopole; the remaining two terms represent the corrections induced by the bumblebee sector.

Figure~\ref{alphachi} presents the variation of $\alpha(b,\chi)$ for different choices of the Lorentz--violating parameter. For a fixed impact parameter $b=0.5$, the deflection grows as $\chi$ increases. In this manner, the bumblebee contribution enhances the bending of null trajectories and strengthens the lensing effect. This trend is consistent with the behavior observed earlier in the geodesic analysis (see Fig.~\ref{geodesicsmassless}), where larger values of $\chi$ produced more tightly curved photon paths.

\begin{figure}
    \centering
    \includegraphics[scale=0.71]{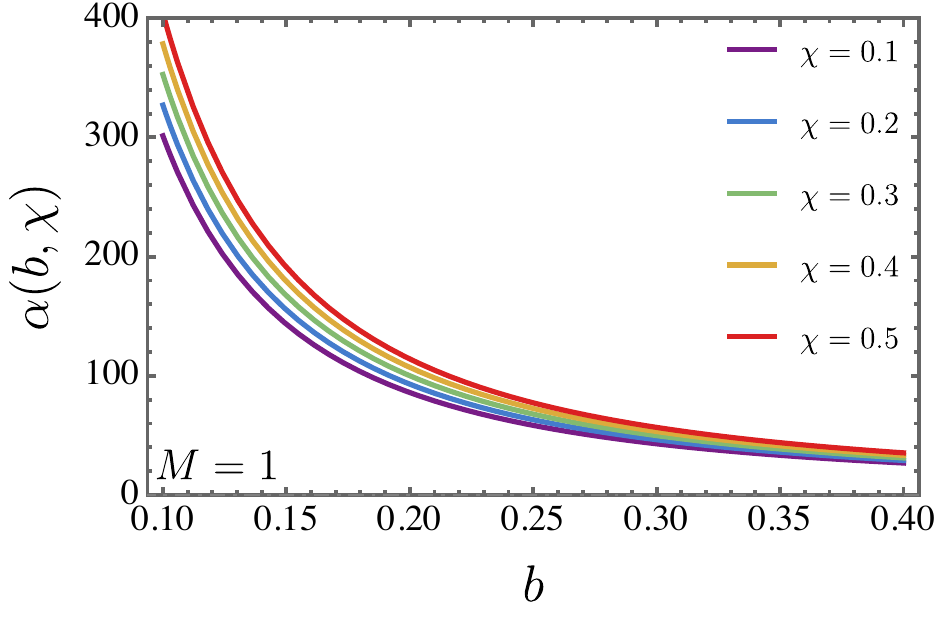}
    \caption{Deflection angle $\alpha(b,\chi)$ for $M=1$ shown for different values of $\chi$. }
    \label{alphachi}
\end{figure}


\section{Optical Geometry and Strong-Field Deflection Analysis }

This part of the analysis deals with the strong--field light deflection. Instead of keeping the notation used in the previous sections, the metric is rewritten following the conventions adopted in Ref. \cite{tsukamoto2017deflection}. The reason for this change is purely technical. With this notation in place, the bending angle can be computed by isolating the divergent contribution generated near the photon sphere and separating it from the regular terms, exactly as prescribed in \cite{tsukamoto2017deflection}
\ie
\label{metricofgl}
\mathrm{d}s^{2} = - A(r,\chi) \mathrm{d}t^{2} + B(r,\chi) \mathrm{d}r^{2} + C(r,\chi)(\mathrm{d}\theta^2 + \sin^{2}\theta\mathrm{d}\phi^2).
\fe

To handle the strong--field contribution to the bending angle, the radial sector is first recast through a new function $\Tilde{D}(r,\chi)$. Rather than confronting the divergence that appears directly at the photon sphere, this auxiliary quantity reorganizes the radial dependence so that the troublesome behavior is absorbed into a regular form. Once this substitution is made, the expression governing the photon’s path becomes manageable, and the analysis proceeds without the singular terms obscuring the leading behavior of the deflection angle
\ie
\Tilde{D}(r,\chi) \equiv \frac{C(r,\chi)^{\prime}}{C(r,\chi)} - \frac{A(r,\chi)^{\prime}}{A(r,\chi)}.
\fe
In this framework, notice that the differentiation with respect to the radial coordinate is marked by a prime. The construction of the function $\Tilde{D}(r,\chi)$ is arranged so that it admits real, positive zeros. Among these, the outermost root plays the role of the critical radius associated with the photon sphere and is labeled $r_{r_{c}}$. For the strong--field method to apply consistently, the metric coefficients $A(r,\chi)$, $B(r,\chi)$, and $C(r,\chi)$ are required to remain regular and positive on the entire interval extending from this critical radius to spatial infinity.

Time--translation symmetry and axial symmetry of the geometry guarantee that a photon moving in this spacetime carries two conserved quantities. One arises from the Killing field associated with $t$, giving a constant proportional to $A(r,\chi)\dot t$, while the other follows from the rotational symmetry, yielding $C(r,\chi)\dot\phi$. Whenever these conserved quantities do not vanish simultaneously, their quotient characterizes the trajectory and is interpreted as the impact parameter
\ie
b \equiv \frac{L}{E} = \frac{C(r,\chi)\Dot{\phi}}{A(r,\chi)\Dot{t}}.
\fe

The rotational symmetry of the spacetime permits the trajectory to be confined to a plane. Choosing $\theta=\pi/2$ fixes this plane without imposing any loss of generality. Under this choice, the null geodesic equations reduce to a single radial relation governing the evolution of the photon’s orbit, which can be expressed as:
\ie
\Dot{r}^{2} = V(r).
\fe
At this stage, it is convenient to introduce an effective potential that encapsulates the radial dynamics of the photon. We define it as
$$
V(r)=\frac{L^{2}\,R(r,\chi)}{B(r,\chi)\,C(r,\chi)}, \qquad \text{and}
\qquad 
R(r,\chi)=\frac{C(r,\chi)}{A(r,\chi)b^{2}}-1.
$$
The quantity above governs how the radial coordinate evolves for a massless particle, functioning as an effective potential. Only the region where this function remains nonnegative is available to photon motion. Since at large distances, the potential tends to a constant value as $r\to\infty$, it indicates that a photon reaching sufficiently large $r$ can propagate outward without any problems. Moreover, we require the function $R(r,\chi)$ to possess at least one positive zero; this root identifies the radial location at which the photon reverses direction along its path.

For lensing applications, the photon path relevant to the strong–field analysis starts far from the central object, moves inward until it reaches a point of closest approach, and then returns to infinity. This closest-approach radius, denoted $r_{\text{o}}$, must sit outside the critical radius associated with the photon sphere, $r_{r_{c}}$, otherwise the trajectory would correspond to a circular null orbit. The value of $r_{\text{o}}$ is selected as the outermost real solution of $R(r,\chi)=0$, under the assumption that the metric functions $B(r,\chi)$ and $C(r,\chi)$ remain regular and positive at that point. At this turning point the condition $V(r_{\text{o}})=0$ holds, so $\mathcal{R}(r_{\text{o}})=0$ becomes the defining relation for identifying the minimum radial distance along the photon’s path
\ie
A_{\text{o}}(r,\chi)\Dot{t}^{2}_{\text{o}} = C_{\text{o}}(r,\chi)\Dot{\phi}^{2}_{\text{o}}.
\fe

From this point on, any symbol marked with the subscript “$\text{o}$” refers to its evaluation at the radius of closest approach. Since reversing the sign of the impact parameter only changes the orientation of the motion, the study of a single trajectory may be restricted to $b>0$ without loss of generality. This constant is preserved along the entire null orbit, and its value can be written in the form
\ie
\label{impcss}
b(r_{\text{o}}) = \frac{L}{E} = \frac{C_{\text{o}}(r,\chi)\Dot{\phi}_{\text{o}}}{A_{\text{o}}(r,\chi)\Dot{t}_{\text{o}}} = \sqrt{\frac{C_{\text{o}}(r,\chi)}{A_{\text{o}}(r,\chi)}}.
\fe
An alternative and often more convenient representation of the function $R(r,\chi)$ can be obtained by rewriting it in the form shown below:
\ie
R(r,\chi)= \frac{A_{\text{o}}(r,\chi)C(r,\chi)}{A(r,\chi)C_{\text{o}}(r,\chi)} - 1.
\fe

The existence of a circular photon orbit can be characterized through the method outlined in Ref.~\cite{hasse2002gravitational}. In that treatment, the motion of a massless particle is encoded in the following relation, which provides the criterion for identifying such null circular paths:
\ie
\frac{B(r,\chi)\,C(r,\chi)\, \Dot{r}^{2}}{E^{2}} + b^{2} = \frac{C(r,\chi)}{A(r,\chi)},
\fe
which allows the expression to be recast in the form
\ie
\ddot{r} + \frac{1}{2}\left( \frac{B(r,\chi)^{\prime}}{B(r,\chi)} + \frac{C(r,\chi)^{\prime}}{C(r,\chi)} \Dot{r}^{2} \right) = \frac{E^{2}\Tilde{D}(r)}{A(r,\chi)B(r,\chi)}. 
\fe

In this context, for radii extending beyond the circular null orbit, the metric functions $A(r,\chi)$, $B(r,\chi)$, and $C(r,\chi)$ must remain positive and differentiable. Once a positive photon energy is assumed, the equation $\tilde{D}(r,\chi)=0$ becomes the condition selecting the radius of a circular light path. At this critical point, the radial function $R(r,\chi)$ must also satisfy a stationarity condition. Evaluating its derivative at the photon sphere gives
$$
R'_{r*{c}}
=
\frac{\tilde{D}_{r_{c}}, C_{r_{c}}(r,\chi), A_{r_{c}}(r,\chi)}{b^{2}}
= 0 ,
$$
where every quantity carrying the subscript $r_{c}$ is taken at $r=r_{r_{c}}$, the location of the circular photon orbit.

The analysis then turns to the limiting value of the impact parameter, here labeled $b_{c}$. This quantity marks the boundary between two distinct classes of null trajectories: photons with $b>b_{c}$ eventually return to infinity, while those with $b=b_{c}$ asymptotically wind around the photon sphere.
\ie
b_{c}(r_{r_{c}}) \equiv \lim_{r_{\text{o}} \to r_{r_{c}}} \sqrt{\frac{C_{\text{o}}(r,\chi)}{A_{\text{o}}(r,\chi)}}.
\fe

This interval of radii corresponds to the region where the bending angle grows without bound, characterizing the strong--deflection sector. In this setting, the criterion for locating the critical orbit arises from differentiating the effective potential with respect to $r$, which imposes the condition
\ie
V^{\prime}(r) = \frac{L^{2}}{B(r,\chi)C(r,\chi)} \left[ R(r,\chi)^{\prime} + \left( \frac{C(r,\chi)^{\prime}(r)}{C(r,\chi)} - \frac{B(r,\chi)^{\prime}(r)}{B(r,\chi)}   \right)   R(r,\chi)  \right].
\fe
In this limit, where the turning point $r_{\text{o}}$ moves ever closer to the photon–sphere radius $r_{r_{c}}$, both the effective potential and its radial derivative simultaneously approach zero. When this occurs, the null–geodesic relation reduces to the simpler form
\ie
\left(  \frac{\mathrm{d}r}{\mathrm{d}\phi}     \right)^{2} = \frac{R(r,\chi)C(r,\chi)}{B(r,\chi)}.
\fe

As a result, the deflection experienced by a photon whose orbit reaches the radius $r_{\text{o}}$ may be expressed through the integral representation
\ie
\alpha(r_{\text{o}}) = I(r_{\text{o}}) - \pi,
\fe
where the quantity $I(r_{\text{o}})$ is written as
\ie
I(r_{\text{o}}) \equiv 2 \int^{\infty}_{r_{\text{o}}} \frac{\mathrm{d}r}{\sqrt{\frac{R(r,\chi)C(r,\chi)}{B(r,\chi)}}}.
\fe

The computation starts with the integral expression governing the bending of light, a step that is notoriously involved from an analytical perspective, as pointed out by Tsukamoto \cite{tsukamoto2017deflection}. To streamline the procedure, it is useful to adopt the auxiliary function introduced in that reference, which reorganizes the integrand in a more manageable form:
\ie
z \equiv 1 - \frac{r_{\text{o}}}{r}.
\fe

With this substitution in place, the deflection integral can be rearranged and expressed in the form
\ie
I(r_{\text{o}}) = \int^{1}_{0} f(z,r_{\text{o}}) \mathrm{d}z,
\fe
where, said differently, we have 
\ie
f(z,z_{0}) \equiv \frac{2r_{\text{o}}}{\sqrt{G(z,r_{\text{o}})}}, \,\,\,\,\,\,\,\, \text{and} \,\,\,\,\,\,\,\,  G(z,r_{\text{o}}) \equiv R(r,\chi) \frac{C(r,\chi)}{B(r,\chi)}(1-z)^{4}.
\fe

Expressing the radial coordinate through the variable $z$ leads to a different representation of the function $R(r,\chi)$. In terms of this new variable, it can be recast as
\ie
R(r,\chi) = \Tilde{{\Bar{D}}}_{\text{o}} \, r_{\text{o}} z + \left[ \frac{r_{\text{o}}}{2}\left( \frac{C(r,\chi)^{\prime\prime}_{\text{o}}}{C_{\text{o}}(r,\chi)} - \frac{A_{\text{o}}(r,\chi)^{\prime\prime}}{A_{\text{o}}(r,\chi)}  \right) + \left( 1 - \frac{A_{\text{o}}(r,\chi)^{\prime}r_{\text{o}}}{A_{\text{o}}(r,\chi)}  \right) \Tilde{{\Bar{D}}}_{\text{o}}  \right] r_{\text{o}} z^{2} + \mathcal{O}(z^{3})+ ...    \,\,\,\,.
\fe

By performing a Taylor expansion of the function $G(z,r_{\text{o}})$ around the point $z=0$, its behavior near this limit can be expressed through the series form given below:
\ie
G(z,r_{\text{o}}) = \sum^{\infty}_{n=1} c_{n}(r_{\text{o}})z^{n}.
\fe
In this expansion, the coefficients accompanying the linear and quadratic powers of $z$ can be singled out. These functions, denoted $c_{1}(r)$ and $c_{2}(r)$, take the form
\ie
c_{1}(r_{\text{o}}) = \frac{C_{\text{o}}(r,\chi)\Tilde{D}_{\text{o}}(r,\chi)r_{\text{o}}}{B(r,\chi)_{\text{o}}},
\fe
and
\ie
c_{2}(r_{\text{o}}) = \frac{C_{\text{o}}(r,\chi)r_{\text{o}}}{B(r,\chi)_{\text{o}}} \left\{ \Tilde{D}_{\text{o}} \left[ \left( \Tilde{D}_{\text{o}} - \frac{B(r,\chi)^{\prime}_{\text{o}}}{B(r,\chi)_{\text{o}}}  \right)r_{\text{o}} -3       \right] + \frac{r_{\text{o}}}{2} \left(  \frac{C(r,\chi)^{\prime\prime}_{\text{o}}}{C_{\text{o}}(r,\chi)} - \frac{A(r,\chi)^{\prime\prime}_{\text{o}}}{A_{\text{o}}(r,\chi)}  \right)                 \right\}.
\fe

In addition, once the strong–deflection limit is invoked, the expression simplifies and one arrives at the relation
\ie
c_{1}(r_{r_{c}}) = 0, \,\,\,\,\,\, \text{and} \,\,\,\,\,\, c_{2}(r_{r_{c}}) =  \frac{C_{r_{c}}(r,\chi)r^{2}_{r_{c}}}{2 B_{r_{c}}(r,\chi)}\Tilde{D}^{\prime}_{r_{c}}(r,\chi), \,\,\,\,\,\,\, \text{with} \,\,\,\,\, \Tilde{D}^{\prime}_{r_{c}}(r,\chi) = \frac{C(r,\chi)^{\prime\prime}}{C_{r_{c}}(r,\chi)} - \frac{A(r,\chi)^{\prime\prime}}{A_{r_{c}}(r,\chi)}.
\fe
Within the strong–deflection limit, the expansion obtained above reduces considerably. Under this approximation, the expression takes the form
\ie
G_{r_{c}}(z) = c_{2}(r_{r_{c}})z^{2} + \mathcal{O}(z^{3}).
\fe

As the turning point $r_{\text{o}}$ approaches the critical radius $r_{r_{c}}$, the quantity $f(z,r_{\text{o}})$ acquires a pole at $z=0$, with its dominant contribution scaling as $1/z$. This singular behavior is responsible for the logarithmic blow-up of the integral $I(r_{\text{o}})$. To deal with this divergence in a controlled way, the integral is separated into two distinct components: a part that contains the singular structure, denoted $I_{\text{Div}}(r_{\text{o}})$, and another, $I_{\text{Reg}}(r_{\text{o}})$, that remains finite in the limit. The piece capturing the divergent contribution can thus be expressed as
\ie
I_{_{\text{Div}}}(r_{\text{o}}) \equiv \int^{1}_{0} f_{_{\text{Div}}}(z,r_{\text{o}}) \mathrm{d}z, \,\,\,\,\,\,\, \text{with} \,\,\,\,\,\,f_{_{\text{Div}}}(z,r_{\text{o}}) \equiv \frac{2 r_{\text{o}}}{\sqrt{c_{1}(r_{\text{o}})z + c_{2}(r_{\text{o}})z^{2}}}.
\fe
Performing the integration associated with the divergent sector leads to the compact result
\ie
I_{_{\text{Div}}} (r_{\text{o}}) = \frac{4 r_{\text{o}}}{\sqrt{c_{2}(r_{\text{o}})}} \ln \left[  \frac{\sqrt{c_{2}(r_{\text{o}})} + \sqrt{c_{1}(r_{\text{o}}) + c_{2}(r_{\text{o}})     }  }{\sqrt{c_{1}(r_{\text{o}})}}  \right].
\fe

In addition, expanding the functions $c_{1}(r_{\text{o}})$ and $b(r_{\text{o}})$ around the critical radius $r_{r_{c}}$ produces the series
\ie
c_{1}(r_{\text{o}}) = \frac{C_{r_{c}}(r,\chi)r_{r_{c}}\Tilde{D}(r,\chi)^{\prime}_{r_{c}}}{B_{r_{c}}(r,\chi)} (r_{\text{o}}-r_{r_{c}}) + \mathcal{O}((r_{\text{o}}-r_{r_{c}})^{2}),
\fe
and
\ie
b(r_{\text{o}}) = b_{c}(r_{r_{c}}) + \frac{1}{4} \sqrt{\frac{C_{r_{c}}(r,\chi)}{A_{r_{c}}(r,\chi)}}  \Tilde{D}(r,\chi)^{\prime}_{r_{c}}(r_{\text{o}}-r_{r_{c}})^{2} + \mathcal{O}((r_{\text{o}}-r_{r_{c}})^{3}),
\fe
so that we get
\ie
\lim_{r_{\text{o}} \to r_{r_{c}}} c_{1}(r_{\text{o}})  =  \lim_{b \to b_{c}} \frac{2 C_{r_{c}}(r,\chi) r_{r_{c}} \sqrt{\Tilde{{D}}^{\prime}_{r_{c}}(r,\chi)}}{B_{r_{c}}(r,\chi)} \left(  \frac{b}{b_{c}} -1  \right)^{1/2}.
\fe

Substituting these series into the expression for the singular sector of the integral leads to the following representation for $I_{\text{Div}}(b)$:
\ie
I_{_{\text{Div}}}(b) = - \frac{r_{r_{c}}}{\sqrt{c_{2}(r_{r_{c}})}} \ln\left[ \frac{b}{b_{c}} - 1 \right] + \frac{r_{r_{c}}}{\sqrt{c_{2}(r_{r_{c}})}}\ln \left[ r^{2}\Tilde{D}^{\prime}_{r_{c}}(r,\chi)\right] + \mathcal{O}[(b-b_{c})\ln(b-b_{c})].
\fe

The remaining finite contribution, which stays well behaved as $r_{\text{o}}$ approaches the photon sphere, is introduced through the definition
\ie
I_{_{\text{Reg}}}(b) = \int^{0}_{1} f_{\text{Reg}}(z,b_{c})\mathrm{d}z + \mathcal{O}[(b-b_{c})\ln(b-b_{c})].
\fe
To isolate the nonsingular behavior, one defines the function ($f_{\text{Reg}}$) by removing from the full integrand the term responsible for the divergence. In other words, ($f_{\text{Reg}}$) is introduced through the subtraction
$$
f_{\text{Reg}} = f(z,r_{\text{o}}) - f_{\text{Div}}(z,r_{\text{o}}).
$$
Once the integrand has been regularized and the singular term removed, the remaining contribution can be handled directly. Employing this smooth part of the integrand and restricting the analysis to the strong–deflection limit, the bending angle reduces to the following expression
\ie
\label{strgffddd}
a(b) = - \Tilde{a} \ln \left[ \frac{b}{b_{c}}-1    \right] + \Tilde{b} + \mathcal{O}[(b-b_{c})\ln(b-b_{c})],
\fe
where we have considered
\ie
\Tilde{a} = \sqrt{\frac{2 \,B_{r_{c}}(r,\chi)A_{r_{c}}(r,\chi)}{C(r,\chi)^{\prime\prime}_{r_{c}}A_{r_{c}}(r,\chi) - C_{r_{c}}(r,\chi)A(r,\chi)^{\prime\prime}_{r_{c}}}},
\fe
and
\ie
\Tilde{b} = \Tilde{a} \ln\left[ r^{2}_{r_{c}}\left( \frac{C_{r_{c}}(r,\chi)^{\prime\prime}}{C_{r_{c}}(r,\chi)}  -  \frac{A(r,\chi)^{\prime\prime}_{r_{c}}}{C_{r_{c}}(r,\chi)} \right)   \right] + I_{_{\text{Reg}}}(r_{r_{c}}) - \pi.
\fe

The next subsection applies the general strong–deflection framework to the particular spacetime introduced in Eq.~(\ref{maaaaianametric}), allowing the coefficients of the formalism to be evaluated explicitly for this black hole geometry.


\subsection{Deflection of light in the new bumblebee black hole spacetime }

Having established the general framework, attention can now be redirected to the particular geometry introduced in Eq. (\ref{maaaaianametric}). Once this line element is inserted into the definition given in Eq. (\ref{impcss}), the corresponding form of the impact parameter follows directly, now written explicitly for the bumblebee configuration
\ie
b_{c} \approx \,\,  3 \sqrt{3} M + \frac{3}{2} \sqrt{3} M \chi.
\fe
It is important to emphasize that the behavior found here departs from what typically occurs in asymptotically flat geometries. In spacetimes with a global conical structure, the critical impact parameter is no longer identical to the shadow radius, and the two quantities must be treated separately. In obtaining this result, the expression was handled through a perturbative expansion in the Lorentz–violating parameter, retaining only the contributions linear in $\chi$. Within this approximation, the coefficients $\tilde{a}$ and $\tilde{b}$ take the following explicit forms:
\ie
\Tilde{a} \approx \, 1 + \frac{\chi }{2},
\fe
and restricting the expansion to terms linear in $\chi$, one is allowed to reorganize the expression accordingly. Under this approximation, the relation takes the form shown below:
\ie
\begin{split}
\Tilde{b} = & \left(1 +\frac{\chi}{2} \right) \ln[6]  + I_{_{\text{Reg}}}(r_{r_{c}}) - \pi.
\end{split}
\fe

In contrast with the Schwarzschild case—where $\tilde{a}$ assumes a much simpler structure—the coefficient here is substantially influenced by the presence of the Lorentz–violating parameter $\chi$, which governs most of its modification.
In addition, the regular contribution of the integral, once evaluated at $r = r_{r_{c}}$, can be presented in the following form:
\ie
\begin{split}
 & I_{_{\text{Reg}}}(r_{r_{c}}) \approx \,    \int_{0}^{1} \mathrm{d}z \left\{  \frac{(\chi +2) \left(\sqrt{3}-\sqrt{3-2 z}\right)}{\sqrt{3-2 z} z}     \right\}  =  \, \, (2 + \chi) \ln \left[6 \left(2-\sqrt{3}\right)\right].
\end{split}
\fe

The steps outlined above allow one to express the result in a closed analytic form. It is also worth noting that, once isolated, the regular part of the integral $I_{\text{Reg}}(r_{r_{c}})$ reproduces the Schwarzschild contribution and acquires additional corrections proportional to $\chi$, as expected from the structure of the metric. Substituting this expression into Eq.~(\ref{strgffddd}) leads directly to the strong–deflection formula for the bending angle, which reads:
\ie
\begin{split}
& a (b,\chi)  =  -\frac{1}{2} (\chi +2) \left\{\ln [6]-\ln \left[\frac{2 b}{\sqrt{3} (3 M \chi +6 M)}-1\right]\right\} +  (2 + \chi) \ln \left[6 \left(2-\sqrt{3}\right)\right] -\pi.
\end{split}
\fe

Figure \ref{alphachistrong} illustrates the behavior of the strong--field bending angle $a(b,\chi)$ for $M=1$ across several choices of the parameter $\chi$. The curves make clear that larger values of $\chi$ lead to a stronger deviation of light. A similar trend had already appeared in the weak--deflection analysis displayed in Fig. \ref{alphachi}, and the same qualitative behavior emerged in the geodesic trajectories discussed earlier. Taken together, these results indicate a consistent pattern: as the Lorentz--violating parameter increases, the bending of light becomes more pronounced, even in the strong--deflection regime. { The smooth deformation of the Schwarzschild geometry, governed by the same parameter in the weak and strong field regimes is a phenomenological useful pattern to be tested through precision measurements.}

\begin{figure}
    \centering
    \includegraphics[scale=0.61]{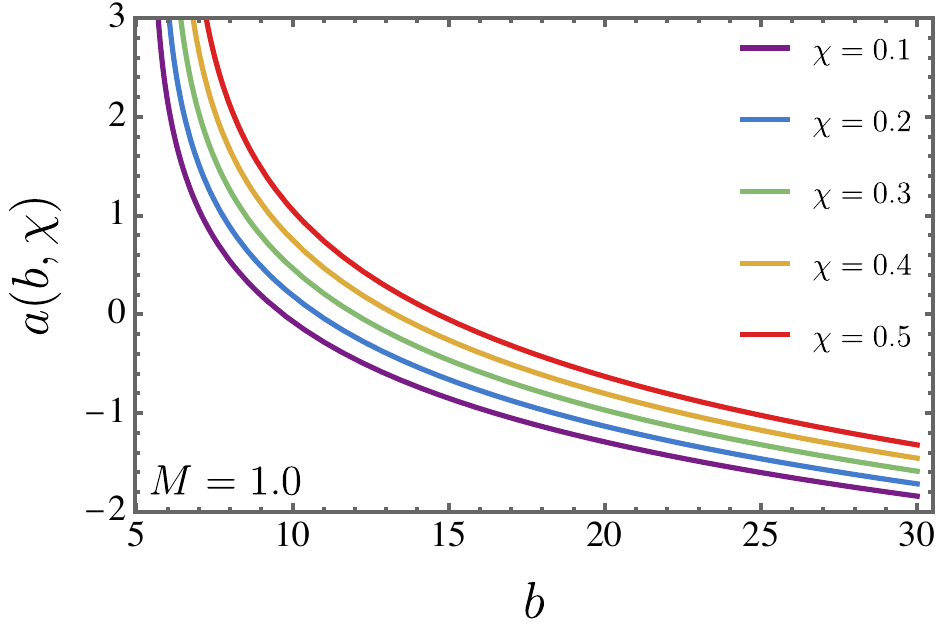}
    \caption{Strong–field deflection angle $a(b,\chi)$ for $M=1$ plotted for several choices of the Lorentz–violating parameter $\chi$. }
    \label{alphachistrong}
\end{figure}


\section{Time delay of light }

For starting off our forthcoming analysis in this section, let us rewrite Eq. (\ref{maaaaianametric}) as follows
\ie
\label{frmetric2}
\mathrm{d}s^{2} = - \frac{1}{1+\chi}\left(1 - \frac{2M}{r}         \right)\mathrm{d}t^{2} + \frac{1}{\frac{1}{1+\chi} \left(1 - \frac{2M}{r} \right)} \mathrm{d}r^{2} + r^{2}\mathrm{d}\Omega^{2},
\fe
or, in a more compact form, we have
\ie
\label{frmetric1}
\mathrm{d}s^{2} = - \tilde{f}(r,\chi)\mathrm{d}t^{2} + \frac{1}{\tilde{f}(r,\chi)} \mathrm{d}r^{2} + r^{2}\mathrm{d}\Omega^{2},
\fe
where $\tilde{f}(r,\chi) \equiv \, \frac{1}{1+\chi} \left(1 - \frac{2M}{r} \right)$. Before proceeding, it is worth mentioning that the discussion in this section follows the treatment presented in Ref.~\cite{Qiao:2024ehj}. In this manner, this section lays out the geometric setting used throughout the analysis. Instead of starting from the time–delay formula itself, we first specify how light rays propagate in a generic static and spherically symmetric background. For such a geometry, the trajectory of a photon follows a null curve, and its path is determined by the associated geodesic equations. Along a null worldline parameterized by $\lambda$, the spacetime symmetries guarantee the existence of conserved quantities. The Killing vector associated with time translations provides a constant of motion identified with the photon energy,
\ie
E = \tilde f(r,\chi)\,\frac{\mathrm{d}t}{\mathrm{d}\lambda},
\fe
while spherical symmetry ensures conservation of angular momentum. For motion in the equatorial plane, this takes the form
\ie
L = r^{2}\sin^{2}\theta\,\frac{\mathrm{d}\phi}{\mathrm{d}\lambda}.
\fe

The remaining condition defining the trajectory arises from imposing the null character of the four–velocity, encoded in the Lagrangian $\mathcal{L}$ constructed from the metric. Written in terms of the radial coordinate and the constants $(E,L)$, this constraint closes the system of equations that is later used to evaluate the gravitational time delay experienced by the photon. Now, by using the Lagrangian approach, we have
\ie
\mathcal{L} = g_{\mu\nu}\mathrm{d}x^{\mu}\mathrm{d}x^{\nu} = \Tilde{f}(r,\chi) \bigg( \frac{\mathrm{d}t}{\mathrm{d}\lambda} \bigg)^2
		- \frac{1}{\Tilde{f}(r,\chi)} \bigg( \frac{\mathrm{d}r}{\mathrm{d}\lambda} \bigg)^{2}
		- r^{2} \bigg( \frac{\mathrm{d}\theta}{\mathrm{d}\lambda} \bigg)^{2} 
		- r^{2}\sin^{2}\theta \bigg( \frac{\mathrm{d}\phi}{\mathrm{d}\lambda} \bigg)^{2}.
\fe

Stationarity and spherical symmetry imply two constants of motion: the energy $E$ associated with time translations and the angular momentum $L$ related to rotational invariance. Since any geodesic in a spherically symmetric space can be confined to a plane, we select the equatorial slice $\theta=\pi/2$. Under this restriction, the geodesic equations reduce to a simpler set involving only $r$, $\phi$, and the constants $(E,L)$, which govern the evolution of the photon path
\ie
		\frac{1}{2} \bigg( \frac{\mathrm{d}r}{\mathrm{d}\lambda} \bigg)^{2} + \frac{1}{2} \Tilde{f}(r,\chi) \bigg[ \frac{L^{2}}{r^{2}} + \mathcal{L} \bigg]
		= \frac{1}{2} \bigg( \frac{\mathrm{d}r}{\mathrm{d}\lambda} \bigg)^{2} + V(r)
		= \frac{1}{2}E^{2}.
\fe

For radial motion in a static and spherically symmetric background, the geodesic equations can be reorganized so that the dynamics resembles a one-dimensional problem. After expressing the trajectory in terms of the conserved quantities, the radial equation introduces a single function that controls whether the motion admits turning points or continues monotonically. This function is interpreted as the effective potential, which for this spacetime reads as follows:
\ie
V(r) = \frac{\Tilde{f}(r,\chi)}{2} \left[ \frac{L^2}{r^2} + \mathcal{L} \right].
\fe 
The ratio between the conserved angular momentum and the conserved energy defines a characteristic length scale for the trajectory, usually written as the impact parameter $b = |L/E|$. For null rays, the geodesic constraint eliminates the contribution from the Lagrangian term $\mathcal{L}$, since the four–velocity has zero norm. When this condition is imposed and the motion is restricted to photons, the radial equation reduces to the simplified form:  
\ie
\frac{\mathrm{d}r}{\mathrm{d}t} = \frac{\mathrm{d}r}{\mathrm{d}\lambda}  \frac{\mathrm{d}\lambda}{\mathrm{d}t}
= \pm \Tilde{f}(r,\chi) \sqrt{1 - b^{2}\frac{\Tilde{f}(r,\chi)}{r^{2}}}.
\fe
When dealing with null trajectories, the condition $\mathcal{L}=0$ removes the norm term from the geodesic equations, and the conserved quantity associated with time translations enters through
\ie
E=\tilde f(r,\chi)\,\frac{\mathrm{d}t}{\mathrm{d}\lambda}.
\fe
The radial equation then admits two branches, encoded by the $\pm$ sign, corresponding to segments of the path in which the photon either moves inward or recedes from the gravitational source. Along the inbound segment, the coordinate $r$ decreases monotonically until the trajectory reaches its minimum radius $r_{0}$, defining the point of closest approach. Beyond this point, the sign of the radial derivative flips, and the photon follows the outward branch of the same equation. Interpreting the motion in this way, one obtains the pair of relations:
\ie
\frac{\mathrm{d}r}{\mathrm{d}t} = - \Tilde{f}(r,\chi) \sqrt{1 - b^{2} \frac{\Tilde{f}(r,\chi)}{r^{2}}} < 0.
\fe

The radial evolution of the null trajectory naturally splits into two segments. Starting on the incoming branch, the coordinate value $r$ decreases as the photon moves through the spacetime toward the point where the radial motion momentarily halts. This occurs at a particular radius $r_{0}$, which marks the closest distance attained during the passage. Beyond this radius, the solution switches to the outgoing branch of the geodesic equation, and the radial coordinate grows again as the photon departs from the gravitational source. In this two-stage description, the geodesic equations governing the radial path take the form:
\ie
\frac{\mathrm{d}r}{\mathrm{d}t} = \Tilde{f}(r,\chi) \sqrt{1 - b^{2} \frac{\Tilde{f}(r,\chi)}{r^{2}}} > 0,  
\fe
Once the photon has passed the radius of closest approach $r_{0}$, the radial coordinate grows along the outward branch of the null trajectory until it reaches the observer at $r=r_{\text{O}}$. In a typical lensing configuration, the emission point lies at $r=r_{S}$ and the detection point at $r=r_{\text{O}}$. The total coordinate time accumulated along this two–stage path encodes the gravitational time delay produced by the curved geometry. Following the general prescription outlined in \cite{Qiao:2024ehj}, the delay can be written as:
\ie
\begin{split}
\label{asdasdddd}
 \Delta T & = T - T_{0} \\
 & = -\int_{r_{S}}^{r_{0}} \frac{\mathrm{d}r}{\Tilde{f}(r,\chi)\sqrt{1-\frac{b^{2} \Tilde{f}(r,\chi)}{r^{2}}}}
	      + \int_{r_{0}}^{r_{\text{O}}} \frac{\mathrm{d}r}{\Tilde{f}(r,\chi)\sqrt{1-\frac{b^{2} \Tilde{f}(r,\chi)}{r^{2}}}}
	      - T_{0}
	      \\
	& = \int_{r_{0}}^{r_{S}} \frac{\mathrm{d}r}{\Tilde{f}(r,\chi)\sqrt{1-\frac{b^{2} \Tilde{f}(r,\chi)}{r^{2}}}}
          + \int_{r_{0}}^{r_{\text{O}}} \frac{\mathrm{d}r}{\Tilde{f}(r,\chi)\sqrt{1-\frac{b^{2} \Tilde{f}(r,\chi)}{r^{2}}}}
          - \sqrt{r_{S}^{2}-r_{0}^{2}} - \sqrt{r_{\text{O}}^{2}-r_{0}^{2}}.
\end{split}
\fe
If one ignores gravity, the travel time of a light signal connecting the source at $r_{S}$ to the observer at $r_{\text{O}}$—while reaching a minimum radius $r_{0}$ along the way—reduces to the flat–space expression
\ie
T_{0}= \sqrt{r_{S}^{2}-r_{0}^{2}} \; + \; \sqrt{r_{\text{O}}^{2}-r_{0}^{2}} .
\fe

Once spacetime curvature is taken into account, the total coordinate time accumulates an additional contribution, interpreted as the gravitational time delay. This extra term becomes larger as either the emission point or the detection point is placed farther from the lens. For trajectories characterized by small values of $\chi$ and of the impact parameter $b$, the expression obtained earlier simplifies considerably. In this limit, Eq.~(\ref{asdasdddd}) reduces to the following approximate form:
\ie
\begin{split}
\Delta T = &\,  +\frac{1}{2} b^2 \left(\frac{2}{r_{0}}-\frac{r_{O}  +r_{S}}{r_{O} r_{S}}\right) + (\chi +1) (-(2 r_{0}-r_{O}-r_{S}))\\
& + \frac{b^4 (\chi -1) \left(-3 M r_{0}^4 \left(r_{O}^4+r_{S}^4\right)+6 M r_{O}^4 r_{S}^4+2 r_{0}^4 r_{O} r_{S} \left(r_{O}^3+r_{S}^3\right)-4 r_{0} r_{O}^4 r_{S}^4\right)}{16 r_{0}^4 r_{O}^4 r_{S}^4} \\
& +2 M (\chi +1) \Big[-2 \ln (r_{0}-2 M)+\ln (r_{O}-2 M)+\ln (r_{S}-2 M)\Big]  \\
& - \sqrt{r_{S}^{2}-r_{0}^{2}} - \sqrt{r_{\text{O}}^{2}-r_{0}^{2}}.
\end{split}
\fe

Figure \ref{deallllaa} illustrates how the time delay behaves under different parameter choices. In the upper panel, $\Delta T$ is plotted for a fixed set of values, namely $\chi=0.1$, $M=1$, $r_{0}=3$, $r_{\text{O}}=10$, $b=0.1$, and a source located at $r_{S}=6$. The lower panel keeps the same configuration except that both $\chi$ and $r_{S}$ are allowed to vary. Within the range of parameters considered, the dependence of $\Delta T$ on $\chi$ remains essentially linear for modest values of the Lorentz--violating parameter, while the variation with respect to $r_{S}$ is displayed across multiple choices of $\chi$. In this second case, increasing the Lorentz--violating parameter $\chi$ leads to a corresponding growth in the time delay $\Delta T$.

\begin{figure}
    \centering
     \includegraphics[scale=0.7]{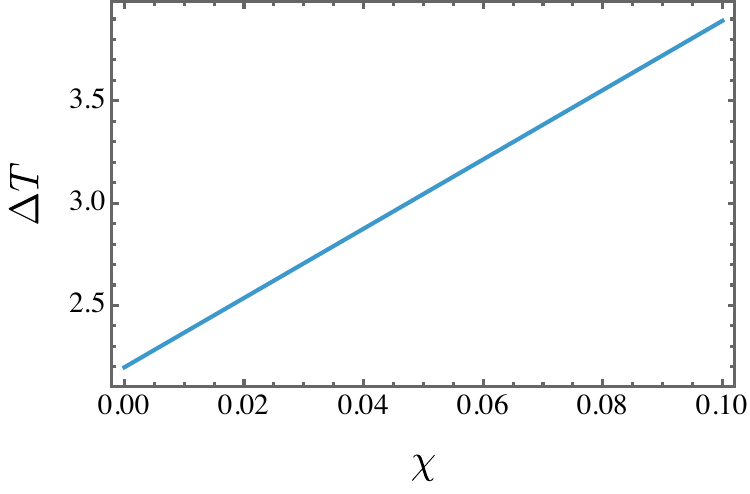}
     \includegraphics[scale=0.7]{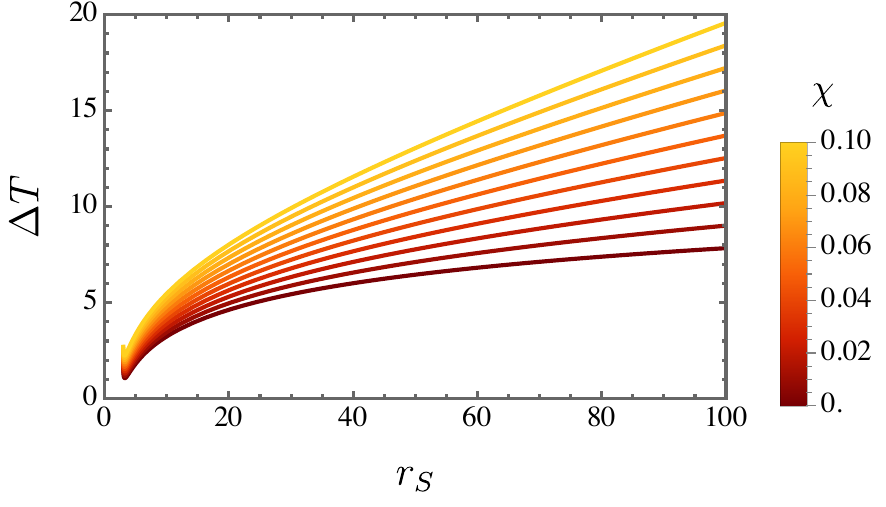}
    \caption{Time delay $\Delta T$ evaluated with $M=1$, $r_{0}=3$, $r_{\text{O}}=10$, and $b=0.1$. The top panel corresponds to the choice $\chi=0.1$ and a fixed source at $r_{S}=6$. In contrast, the lower panel varies both $\chi$ and $r_{S}$ while keeping the remaining quantities unchanged, highlighting the near--linear trend of $\Delta T$ for small $\chi$ and its dependence on the source position for several values of $\chi$. }
    \label{deallllaa}
\end{figure}


\section{Bounds based on the Solar system tests }

Before closing this section, it is important to note that the entire analysis was carried out using the generic metric specified in Eq. (\ref{metricofgl}). The analysis of solar system tests of general relativity is based on the study of massive and massless geodesics of the exterior solution of the static and spherically symmetric metric. Besides that, to study such motion, we consider the simplest case where $\theta = \pi/2$, which means that the test particles are confined in the equatorial plane. The Lagrangian ${\cal L}$ that governs this motion is given by
\begin{equation}\label{eq:lagr1}
-2{\cal L}(x,\dot{x})=A(r,\chi) \, {{\dot t}^2} - B(r,\chi)\,{{\dot r}^2} - r^2\,{\dot \varphi }^2 = \eta\, ,
\end{equation}
where $\eta=0$ describes massless particles, while, on the other hand, $\eta = 1$ describes to massive particles.

It is also interesting to identify the conserved quantities related to the symmetry under time translations and rotations, which are the energy $E$ and the angular momentum $L$:

\begin{equation}\label{constant2}
E = A(r,\chi) \,\dot t \quad\mathrm{and}\quad L = r^2\,\dot \varphi.
\end{equation}

From this, we can write the equation \eqref{eq:lagr1} as

\begin{equation}\label{massive}
    \left[\frac{\mathrm{d}}{\mathrm{d}\varphi}\left(\frac{1}{r}\right)\right]^2\frac{E^2}{A(r,\chi)B(r,\chi)L^2}-\frac{1}{B(r,\chi)L^2}\left(\eta+\frac{L^2}{r^2}\right)\, .
\end{equation}

Differentiating Eq.~\eqref{massive} with respect to $\varphi$ and defining the function $u = L^2/(M r)$, the perturbed geodesic equation in first order in $\chi=\alpha\ell$ is
\begin{align}\label{eq:u-both}
&\frac{\mathrm{d}^2 u}{\mathrm{d}\varphi^2} = \left(1-\chi\right)\left(\eta - u + \frac{3 M^2 u^2}{L^2}\right).
\end{align}

Notice that we have a contribution from the Lorentz violating parameter even in the absence of the mass $M$. This means that we have a non-trivial Minkowski solution, which is a property that is also present in others Lorentz violating approaches in quantum gravity and opens appealing phenomenological opportunities \cite{Amelino-Camelia:2008aez,Addazi:2021xuf,AlvesBatista:2023wqm}. { As we will verify, the conformal correction shown in Eq.\eqref{eq:u-both} will provide stringent constraints on the parameter $\chi$.}


\subsection{Mercury's perihelion shift }

The advance of Mercury's perihelion is one of the most important predictions of general relativity and we can set constraints on the Lorentz violating parameters by analyzing it. It is based on the assumption that Mercury acts as a massive test particle in the vicinity of the Sun. This way, we set $\eta=1$ in \eqref{eq:u-both}. This gives an equation that can be straightforwardly treated perturbatively. We can write down the solution as $u=u_0+\chi u_{\chi}+\frac{M^2}{L^2}u_{M}$, where $u_0$ is the Newtonian correction, $u_{\chi}$ is the contribution proportional to $\chi$ and we have the first relativistic correction $u_M$. A direct substitution of this perturbation in the differential equation \eqref{eq:u-both}, gives a series of corrections, however the only ones that accumulate over revolutions of the planet are
\begin{equation}
    u\approx 1-\chi+e\cos(\varphi)+3\frac{\widetilde{M}^2(\chi)}{L^2}e\varphi \sin(\varphi)\approx 1-\chi+e\cos\left[\left(1-\frac{3\widetilde{M}^2}{L^2}\right)\varphi\right],
\end{equation}
where $e$ is the eccentricity of the orbit, $\widetilde{M}^2(\chi)=M^2(1+\chi L^2/(6M^2)$ and we used the fact that the corrections are tiny in a trigonometric identity for the cosine function. From this, we verify that the dimensionless correction of the perihelion advance of Mercury is given by 
\begin{equation}
    \delta_{\chi,\text{Per}}=\chi \frac{L^2}{6M^2}.
\end{equation}

Using the relation between the angular momentum, and energy of the test particle as function of the eccentricity $e$ and the semi-major axis $a$ of the orbit as $L^2=Ma(1-e^2)$ and $E=-M/(2a)$, from celestial mechanics \cite{Goldstein:2002}, we can use data from \cite{ParticleDataGroup:2024cfk,Casana:2017jkc} to derive deviations from general relativity. In fact, using $e=0.2056$, $a=3.583\times 10^{45}$ and $M=9.138\times 10^{37}$ (in natural units), and the experimental result $\Delta\Phi_{\text{Exp}}=(42.9794\pm 0.0030)''/\text{century}$ \cite{Casana:2017jkc,Yang:2023wtu}, we must have
\begin{equation}
    -1.817 \times 10^{-11} \leq \chi \leq 3.634 \times 10^{-12}.
\end{equation}


\subsection{Light bending }

Concerning the light bending, the situation is somewhat similar. The main difference is that we must consider the massless case, in which $\eta=0$ in \eqref{eq:u-both}. We consider that light emitted from a star passes near the surface of the Sun, such that its path is bended by the curvature of spacetime. This modifies the apparent position of the star in comparison with its position when the Sun is not between the star and the detector. 

To deal with this situation, we define $u$ as $u=1/r$ instead. The perturbed solution is also given by the Newtonian term plus a correction of the form
\begin{equation}
    u\approx \frac{1}{b}\sin\left(\left(1-\frac{\chi}{2}\right)\varphi\right)+\frac{{\bar M}}{b^2(1-\chi)}\left[1+\cos^2\left(\left(1-\frac{\chi}{2}\right)\varphi\right)\right],
\end{equation}
where $b\doteq L/E$ is the impact parameter \cite{Wang:2024fiz} and $\bar{M}=(1-\chi)M$. The asymptotic behavior of the light ray is such that $u\rightarrow 0$, when $r\rightarrow \infty$, which determines the bended angle. By imposing this condition, we find that the external scattering angle is $\delta_{\chi,\text{ex}}=-2M/b(1+\chi/2)$, which gives a total angle of $\delta{\chi}=2\delta_{\chi,\text{ex}}$. This means that the dimensionless correction is given by $\chi/2$. Again, using experimental inputs of \cite{dsasdas}, we can estimate deviations of general relativity by comparing the GR result $\delta_{\text{GR}}=4M/b=1.7516687''$ and  the experimental result $\delta_{\text{Exp}}=\frac{1}{2}(1+\gamma)\delta_{\text{GR}}$ with $\delta_{\chi}$. This gives a rough bound
\begin{equation}
    -1 \times 10^{-4} \leq \chi \leq 2 \times 10^{-5},
\end{equation}
which far less accurate than the Mercury's perihelion result. 


\subsection{Shapiro delay }

This effect is produced when a light ray emitted by a source passes near the surface of a massive object, then reaches a receiver that reflects back the signal to the original source. The delay in time produced due to the curvature of spacetime, in comparison to the flat case, is called Shapiro delay \cite{Shapiro:1964uw}.

To consider this effect, we analyze light-like geodesics from ${\cal L}=0$, which can be seen from \eqref{eq:lagr1}
\begin{equation}
  \left(  \frac{\mathrm{d}r}{\mathrm{d}t}\right)^2=\frac{A(r,\chi)\,r^2-\frac{L^2}{E^2}A(r,\chi)}{B(r,\chi)r^2}\, .
\end{equation}

The point of closest approach $r_{\text{min}}$ can be found from $\dot{r}=0$, which allows us to find a relation between $L$ and $E$ as $L^2/E^2 = r_{\text{min}}^2/A(r_{\text{min}},\chi)$ (which is also the square of the impact parameter). From this, we can write a dependence on time and radius that will be integrated as

\begin{equation}\label{eq:shapiro_main}
    \mathrm{d} t=\pm \frac{1}{A(r,\chi)}\frac{1}{\sqrt{\frac{1}{A(r,\chi)\,B(r,\chi)}-\frac{r_{\text{min}}^2/A(r_{\text{min}},\chi)}{B(r,\chi)\,D(r,\chi)}}}\, .
\end{equation}

Integrating this expression leads us to the time from one of the observers to the massive object
\begin{align}
   t(r)=\left(1+\chi\right)\left[\sqrt{r^2-r_{\text{min}}^2}+M\left(\sqrt{\frac{r-r_{\text{min}}}{r+r_{\text{min}}}}+2\ln\left(\frac{r+\sqrt{r^2-r_{\text{min}}^2}}{r_{\text{min}}}\right)\right)\right]\, ,
\end{align}
which is the relativistic result corrected by a conformal factor $1+\chi$. Let us consider that the impact parameter is much smaller than the typical length scale of the problem, which is reasonable considering the solar system, we have $r_{min}\ll r$. Let us also analyze the round trip of the light ray from the emitter (located in $r_E$) to the massive object, then from the massive object to the receiver (located in $r_R$), than back from the receiver to the massive object and, finally, back to the original emitter. This give the time $T=2t(r_E)+2t(r_R)$, which in our case of a $\chi$-deformation is simply 

\begin{equation} \label{eq:new-shapiro}
T_{\chi}=\left(1+\chi\right)\left\{2(r_E+r_R)+4M\left[1+\ln\left(\frac{4r_Rr_E}{r_{\text{min}}^2}\right)\right]\right\}=T_{\text{Mink}}+\delta T_{M,\chi}\, ,
\end{equation}
where $T_{\text{Mink}}=2(r_E+r_R)$ is the Minkowskian term and $\delta T_{M,\chi}$ is the contribution from GR and the Bumblebee parameter.

From the parametrized post-Newtonian formalism, this extra time is expressed as
\begin{equation}\label{eq:ppn-shapiro}
    \delta T = 4M\left(1+\frac{1+\gamma}{2}\ln \left(\frac{4r_Rr_E}{r_{\text{min}}^2}\right)\right)\, ,
\end{equation}
where $\gamma$ is the parameter that is observationally constrained. From the Cassini mission \cite{Bertotti:2003rm,Will:2014kxa} a tight constraint on it is given by $|\gamma-1|< 2.3\times 10^{-5}$. We compare \eqref{eq:ppn-shapiro} with \eqref{eq:new-shapiro} and the Cassini constraint to estimate the allowed magnitude of the parameter $\chi$. Using astronomical units, we have with $r_E = 1\,  \text{AU} = 2.457 \times 10^{45}$, $r_R = 8.46\, \text{AU}$, and $r_{\text{min}} = 1.6 R_{\odot}$ where $R_{\odot} = 4.305 \times 10^{43}$ is the solar radius. Considering $M$ as the mass of the Sun, we can set the following constraint 
\begin{equation}
-1.140 \times 10^{-5}\leq \chi \leq 1.140 \times 10^{-5},
\end{equation}
which is also of the same order of magnitude of the light bending. We summarize our results in table \ref{tab:constr}.

\begin{table}[ht!]
\centering
\caption{Bounds for $\chi$  derived from Solar System tests.}
\label{tab:constr}
\begin{tabular}{lc}
\hline\hline
\textbf{Solar System Experiments} & Bounds  \\
\hline
{\bf{Mercury's perihelion shift}}   & \makecell{$-1.817 \times 10^{-11} \leq \chi \leq 3.634 \times 10^{-12}$} \\
{\bf{Light bending}}     & \makecell{$-1 \times 10^{-4} \leq \chi \leq 2 \times 10^{-5}$}  \\
{\bf{Shapiro delay}}   & \makecell{$-1.140 \times 10^{-5}\leq \chi \leq 1.140 \times 10^{-5}$}  \\
\hline\hline
\end{tabular}
\end{table}


\section{Conclusion}\label{Sec:Conclusion}

This work investigated the gravitational consequences introduced by the Lorentz--violating parameter $\chi$ in a recently proposed black hole solution formulated within bumblebee gravity \cite{Zhu:2025fiy}. A preliminary change of coordinates was implemented to make explicit the globally canonical character of the spacetime, a feature that later proved important for the lensing analysis.

The study then turned to the motion of test particles. Both massless and massive trajectories were integrated, and the deformation governed by $\chi$ produced a systematic ``contraction'' of the light paths as $\chi$ increased. The circular null geodesics were examined separately, showing that neither the critical radius $r_{c}$ nor the shadow radius $R_{\text{sh}}$ was modified by the Lorentz–violating term. A comparison was also made with several Lorentz--violating geometries (black holes) discussed in the literature, covering vector (bumblebee with or without charge or cosmological constant) and tensor (Kalb--Ramond with similar extensions) configurations. The topological aspects of the new solution were analyzed as well.

The behavior of massive particles was explored through the effective radial force and the innermost stable circular orbit. In this case, the specific energy at the ISCO acquired a dependence on $\chi$, given by $E_{\text{ISCO}}=\frac{2\sqrt{2},M}{3\sqrt{1+\chi}}$. Even though the parameter $\chi$ reshaped the effective dynamics governing particle motion, the underlying geometric structure remained unchanged: the radii of the circular null and timelike orbits, including the ISCO, were left unaffected.

After separating the perturbation equations for all perturbations, the corresponding effective potentials were obtained, allowing the computation of the quasinormal frequencies and the time--domain evolution for all spin sectors. In particular, the quasinormal spectra were computed with the 6th--order WKB approximation for the scalar, vector, and tensor sectors, while the spinor modes were evaluated using the 3rd--order WKB method. In addition, the time--domain evolution was examined for each of these cases. In general lines, when the fermionic sector was contrasted with the scalar, vector, and tensor cases, the spinor frequencies turned out to be slightly larger in their real parts and moderately higher in their imaginary components (with the exception of the comparison with the scalar field). This showed that spinor perturbations decayed more rapidly than vector and tensor modes, although at a slower rate than the scalar ones. This raised the question of whether the same hierarchy would remain if the scalar quasinormal modes were also computed with the third--order WKB method used for the spinor sector. After carrying out this check, the same ordering was confirmed. In every spin channel, the introduction of Lorentz violation produced a uniform trend: the effective potential barrier became lower, the oscillation frequencies shifted to smaller real values, and the damping weakened as $\chi$ increased. As in the other analyses performed in this work, the fundamental mode governed the late--time signal, whereas the higher overtones decayed much faster.

For the gravitational lensing sector, two distinct approaches were employed. First, the weak–deflection regime was treated through the Gauss–Bonnet method applied to canonical global spacetimes, following Ref. \cite{Jusufi:2017lsl}. This procedure yielded the expression
$\alpha(b,\chi)=\frac{4M}{b}+\frac{3\pi M^{2}}{4 b^{2}}+\frac{\pi\chi}{2}+\frac{3\pi M^{2}\chi}{4 b^{2}}+\frac{2M\chi}{b}$,
and the deflection angle increased as the Lorentz–violating parameter $\chi$ grew. The second method relied on the regularization technique at the photon sphere, in the spirit of Ref. \cite{tsukamoto2017deflection}, through which the strong–field deflection angle
$a (b,\chi)  =  -\frac{1}{2} (\chi +2) \left\{\ln [6]-\ln \left[\frac{2 b}{\sqrt{3} (3 M \chi +6 M)}-1\right]\right\} +  (2 + \chi) \ln \left[6 \left(2-\sqrt{3}\right)\right] -\pi$
was obtained. In this regime as well, the quantity $a(b,\chi)$ became larger for increasing $\chi$. The behavior observed in both lensing treatments matched the trends identified previously in the analysis of null geodesics.

Finally, bounds on the Lorentz--violating parameter $\chi$ were obtained from the classical Solar System tests. The constraints derived from each observable were the following: perihelion shift of Mercury, $-1.817\times 10^{-11} \leq \chi \leq 3.634\times 10^{-12}$; light bending, $-1\times 10^{-4} \leq \chi \leq 2\times 10^{-5}$; and Shapiro time delay, $-1.140\times 10^{-5} \leq \chi \leq 1.140\times 10^{-5}$. { The bumblebee parameter was tightly constrained by Solar System tests. This constituted an important result, as it set the order of magnitude that should be adopted for meaningful comparisons with observations. This represented a distinct contribution of the bumblebee field to black hole solutions, showing that Lorentz--violating effects can be incorporated in a consistent way at the level of the spacetime geometry.}

As a continuation of this work, we intend to extend the analysis to all spin sectors ($0, 1, 2, 1/2$), focusing on the computation of thermodynamics, particle creation, greybody factors, absorption cross sections, correspondence of QNMs and greybody factors, and evaporation lifetimes, following the framework discussed in Ref.~\cite{AraujoFilho:2025zzf,AraujoFilho:2025hkm,Heidari:2024bvd}. We also plan to explore the formation of black hole shadows in the presence of a thin accretion disk, as in Refs.~\cite{Rosa:2023hfm,Shi:2025oek}, and to examine the consequences for neutrino oscillations, inspired by Refs.~\cite{Shi:2025rfq,Shi:2025plr,Shi:2025ywa}. All these analyses are nearing completion and are currently under final revision by the authors, with submission to the arXiv expected in the coming days.

In addition, it would be worthwhile to examine the Unruh effect for accelerated detectors in this background, following the line of investigation carried out in Refs. \cite{Barros:2023kor,Barros:2024ftf,Barros:2024wdv,Barros:2025arh,Barros:2025din}. Extending that framework to the newly proposed bumblebee black hole could reveal how the Lorentz--violating deformation modifies the response of uniformly accelerated observers.


\section*{Acknowledgments}
\hspace{0.5cm} A.A.A.F. is supported by Conselho Nacional de Desenvolvimento Cient\'{\i}fico e Tecnol\'{o}gico (CNPq) and Fundação de Apoio à Pesquisa do Estado da Paraíba (FAPESQ), project numbers 150223/2025-0 and 1951/2025. V. B. Bezerra is partially supported by the Conselho Nacional de Desenvolvimento Científico e Tecnológico (CNPq) grant number 307211/2020-7. I. P. L. was partially supported by the National Council for Scientific and Technological Development - CNPq, grant 312547/2023-4.

\section*{Data Availability Statement}

Data Availability Statement: No Data associated with the manuscript

\bibliographystyle{ieeetr}
\bibliography{main}

\begin{thebibliography}{100}

\bibitem{Liu:2025oho}
J.-Z. Liu, S.-P. Wu, S.-W. Wei, and Y.-X. Liu, ``{Exact Black Hole Solutions in
  Bumblebee Gravity with Lightlike or Spacelike VEVS},'' 10 2025.

\bibitem{Zhu:2025fiy}
J.~Zhu and H.~Li, ``{Full Classification of Static Spherical Vacuum Solutions
  to Bumblebee Gravity with General VEVs},'' 11 2025.

\bibitem{colladay1997cpt}
D.~Colladay and V.~A. Kosteleck{\`y}, ``Cpt violation and the standard model,''
  {\em Physical Review D}, vol.~55, no.~11, p.~6760, 1997.

\bibitem{kostelecky1989spontaneous}
V.~A. Kosteleck{\`y} and S.~Samuel, ``Spontaneous breaking of lorentz symmetry
  in string theory,'' {\em Physical Review D}, vol.~39, no.~2, p.~683, 1989.

\bibitem{kostelecky2004gravity}
V.~A. Kosteleck{\`y}, ``Gravity, lorentz violation, and the standard model,''
  {\em Physical Review D}, vol.~69, no.~10, p.~105009, 2004.

\bibitem{kostelecky2011data}
V.~A. Kosteleck{\`y} and N.~Russell, ``Data tables for lorentz and cpt
  violation,'' {\em Reviews of Modern Physics}, vol.~83, no.~1, pp.~11--31,
  2011.

\bibitem{kostelecky1999constraints}
V.~A. Kosteleck{\`y} and C.~D. Lane, ``Constraints on lorentz violation from
  clock-comparison experiments,'' {\em Physical Review D}, vol.~60, no.~11,
  p.~116010, 1999.

\bibitem{bluhm2005spontaneous}
R.~Bluhm and V.~A. Kosteleck{\`y}, ``Spontaneous lorentz violation,
  nambu-goldstone modes, and gravity,'' {\em Physical Review D—Particles,
  Fields, Gravitation, and Cosmology}, vol.~71, no.~6, p.~065008, 2005.

\bibitem{Bluhm:2023kph}
R.~Bluhm and Y.~Zhi, ``{Spontaneous and Explicit Spacetime Symmetry Breaking in
  Einstein{\textendash}Cartan Theory with Background Fields},'' {\em Symmetry},
  vol.~16, no.~1, p.~25, 2024.

\bibitem{Bluhm:2019ato}
R.~Bluhm, H.~Bossi, and Y.~Wen, ``{Gravity with explicit spacetime symmetry
  breaking and the Standard-Model Extension},'' {\em Phys. Rev. D}, vol.~100,
  no.~8, p.~084022, 2019.

\bibitem{bluhm2008spontaneous}
R.~Bluhm, S.-H. Fung, and V.~A. Kosteleck{\`y}, ``Spontaneous lorentz and
  diffeomorphism violation, massive modes, and gravity,'' {\em Physical Review
  D—Particles, Fields, Gravitation, and Cosmology}, vol.~77, no.~6,
  p.~065020, 2008.

\bibitem{Maluf:2013nva}
R.~V. Maluf, V.~Santos, W.~T. Cruz, and C.~A.~S. Almeida, ``{Matter-gravity
  scattering in the presence of spontaneous Lorentz violation},'' {\em Phys.
  Rev. D}, vol.~88, no.~2, p.~025005, 2013.

\bibitem{Maluf:2014dpa}
R.~V. Maluf, C.~A.~S. Almeida, R.~Casana, and M.~M. Ferreira, Jr.,
  ``{Einstein-Hilbert graviton modes modified by the Lorentz-violating
  bumblebee Field},'' {\em Phys. Rev. D}, vol.~90, no.~2, p.~025007, 2014.

\bibitem{kostelecky1991photon}
V.~A. Kosteleck{\`y} and S.~Samuel, ``Photon and graviton masses in string
  theories,'' {\em Physical Review Letters}, vol.~66, no.~14, p.~1811, 1991.

\bibitem{jacobson2004einstein}
T.~Jacobson and D.~Mattingly, ``Einstein-aether waves,'' {\em Physical Review
  D}, vol.~70, no.~2, p.~024003, 2004.

\bibitem{Bertolami:2005bh}
O.~Bertolami and J.~Paramos, ``{The Flight of the bumblebee: Vacuum solutions
  of a gravity model with vector-induced spontaneous Lorentz symmetry
  breaking},'' {\em Phys. Rev. D}, vol.~72, p.~044001, 2005.

\bibitem{Casana:2017jkc}
R.~Casana, A.~Cavalcante, F.~P. Poulis, and E.~B. Santos, ``{Exact
  Schwarzschild-like solution in a bumblebee gravity model},'' {\em Phys. Rev.
  D}, vol.~97, no.~10, p.~104001, 2018.

\bibitem{Liu:2024wpa}
W.~Liu, C.~Wen, and J.~Wang, ``{Lorentz violation alleviates gravitationally
  induced entanglement degradation},'' {\em JHEP}, vol.~01, p.~184, 2025.

\bibitem{AraujoFilho:2025hkm}
A.~A. Ara{\'u}jo~Filho, ``{How does non-metricity affect particle creation and
  evaporation in bumblebee gravity?},'' {\em JCAP}, vol.~06, p.~026, 2025.

\bibitem{Liang:2022hxd}
D.~Liang, R.~Xu, X.~Lu, and L.~Shao, ``{Polarizations of gravitational waves in
  the bumblebee gravity model},'' {\em Phys. Rev. D}, vol.~106, no.~12,
  p.~124019, 2022.

\bibitem{Oliveira:2021abg}
R.~Oliveira, D.~M. Dantas, and C.~A.~S. Almeida, ``{Quasinormal frequencies for
  a black hole in a bumblebee gravity},'' {\em EPL}, vol.~135, no.~1, p.~10003,
  2021.

\bibitem{Gullu:2020qzu}
I.~G\"ull\"u and A.~\"Ovg\"un, ``{Schwarzschild-like black hole with a
  topological defect in bumblebee gravity},'' {\em Annals Phys.}, vol.~436,
  p.~168721, 2022.

\bibitem{KumarJha:2020ivj}
S.~Kumar~Jha, H.~Barman, and A.~Rahaman, ``{Bumblebee gravity and particle
  motion in Snyder noncommutative spacetime structures},'' {\em JCAP}, vol.~04,
  p.~036, 2021.

\bibitem{Maluf:2020kgf}
R.~V. Maluf and J.~C.~S. Neves, ``{Black holes with a cosmological constant in
  bumblebee gravity},'' {\em Phys. Rev. D}, vol.~103, no.~4, p.~044002, 2021.

\bibitem{Ding:2019mal}
C.~Ding, C.~Liu, R.~Casana, and A.~Cavalcante, ``{Exact Kerr-like solution and
  its shadow in a gravity model with spontaneous Lorentz symmetry breaking},''
  {\em Eur. Phys. J. C}, vol.~80, no.~3, p.~178, 2020.

\bibitem{Liu:2019mls}
C.~Liu, C.~Ding, and J.~Jing, ``{Thin accretion disk around a rotating
  Kerr-like black hole in Einstein-bumblebee gravity model},'' 10 2019.

\bibitem{Ovgun:2018ran}
A.~Ovg\"un, K.~Jusufi, and I.~Sakalli, ``{Gravitational lensing under the
  effect of Weyl and bumblebee gravities: Applications of
  Gauss\textendash{}Bonnet theorem},'' {\em Annals Phys.}, vol.~399,
  pp.~193--203, 2018.

\bibitem{Li:2020wvn}
Z.~Li, G.~Zhang, and A.~\"Ovg\"un, ``{Circular Orbit of a Particle and Weak
  Gravitational Lensing},'' {\em Phys. Rev. D}, vol.~101, no.~12, p.~124058,
  2020.

\bibitem{asdasdSekhmani:2025zen}
Y.~Sekhmani, W.~Liu, W.~Deng, and K.~Boshkayev, ``{Quasinormal Modes of Massive
  Scalar Perturbations in Slow-Rotation Bumblebee Black Holes with Traceless
  Conformal Electrodynamics},'' 10 2025.

\bibitem{asdasd1Deng:2025uvp}
W.~Deng, W.~Liu, F.~Long, K.~Xiao, and J.~Jing, ``{Quasinormal Modes of a
  Massive Scalar Field in Slowly Rotating Einstein-Bumblebee Black Holes},'' 7
  2025.

\bibitem{Kumar:2025bim}
A.~Kumar, S.~U. Islam, and S.~G. Ghosh, ``{Probing Lorentz Symmetry Violation
  through Lensing Observables of Rotating Black Holes},'' 8 2025.

\bibitem{Filho:2022yrk}
A.~A.~A. Filho, J.~R. Nascimento, A.~Y. Petrov, and P.~J. Porf{\'\i}rio,
  ``{Vacuum solution within a metric-affine bumblebee gravity},'' {\em Phys.
  Rev. D}, vol.~108, no.~8, p.~085010, 2023.

\bibitem{AraujoFilho:2024ykw}
A.~A. Ara{\'u}jo~Filho, J.~R. Nascimento, A.~Y. Petrov, and P.~J.
  Porf{\'\i}rio, ``{An exact stationary axisymmetric vacuum solution within a
  metric-affine bumblebee gravity},'' {\em JCAP}, vol.~07, p.~004, 2024.

\bibitem{AraujoFilho:2025rvn}
A.~A. Ara{\'u}jo~Filho, N.~Heidari, I.~P. Lobo, Y.~Shi, and F.~S.~N. Lobo,
  ``{The Flight of the Bumblebee in a Non-Commutative Geometry: A New Black
  Hole Solution},'' 9 2025.

\bibitem{Neves:2022qyb}
J.~C.~S. Neves, ``{Kasner cosmology in bumblebee gravity},'' {\em Annals
  Phys.}, vol.~454, p.~169338, 2023.

\bibitem{Jesus:2019nwi}
W.~D.~R. Jesus and A.~F. Santos, ``{Ricci dark energy in bumblebee gravity
  model},'' {\em Mod. Phys. Lett. A}, vol.~34, no.~22, p.~1950171, 2019.

\bibitem{Neves:2024ggn}
J.~C.~S. Neves and F.~G. Gardim, ``{Stars and quark stars in bumblebee
  gravity},'' {\em Annals Phys.}, vol.~475, p.~169950, 2025.

\bibitem{Magalhaes:2025nql}
R.~B. Magalh{\~a}es, L.~A. Lessa, and M.~M. Ferreira, ``{Wormholes in
  Lorentz-violating gravity},'' 5 2025.

\bibitem{AraujoFilho:2024iox}
A.~A. Ara{\'u}jo~Filho, J.~A. A.~S. Reis, and A.~{\"O}vg{\"u}n, ``{Modified
  particle dynamics and thermodynamics in a traversable wormhole in bumblebee
  gravity},'' {\em Eur. Phys. J. C}, vol.~85, no.~1, p.~83, 2025.

\bibitem{Magalhaes:2025lti}
R.~B. Magalh{\~a}es, L.~A. Lessa, and R.~Casana, ``{Lorentz-violating
  wormholes: The role of the matter coupled to Lorentz-violating fields},'' 7
  2025.

\bibitem{Ovgun:2018xys}
A.~{\"O}vg{\"u}n, K.~Jusufi, and {\.I}.~Sakall{\i}, ``{Exact traversable
  wormhole solution in bumblebee gravity},'' {\em Phys. Rev. D}, vol.~99,
  no.~2, p.~024042, 2019.

\bibitem{Pereira:2025xnw}
C.~F.~S. Pereira, M.~V. d.~S. Silva, H.~Belich, D.~C.~Rodrigues, J.~C. Fabris,
  and M.~E. Rodrigues, ``{Black-bounce solutions in a k-essence theory under
  the effects of bumblebee gravity},'' {\em Phys. Rev. D}, vol.~111, no.~12,
  p.~124005, 2025.

\bibitem{Gomes:2018oyd}
D.~A. Gomes, R.~V. Maluf, and C.~A.~S. Almeida, ``{Thermodynamics of
  Schwarzschild-like black holes in modified gravity models},'' {\em Annals
  Phys.}, vol.~418, p.~168198, 2020.

\bibitem{Ovgun:2019ygw}
A.~\"Ovg\"un and I.~Sakall\i{}, ``{Hawking Radiation via Gauss-Bonnet
  Theorem},'' {\em Annals Phys.}, vol.~413, p.~168071, 2020.

\bibitem{Kanzi:2019gtu}
S.~Kanzi and I.~Sakall\i{}, ``{GUP Modified Hawking Radiation in Bumblebee
  Gravity},'' {\em Nucl. Phys. B}, vol.~946, p.~114703, 2019.

\bibitem{Yang:2018zef}
R.-J. Yang, H.~Gao, Y.~Zheng, and Q.~Wu, ``{Effects of Lorentz breaking on the
  accretion onto a Schwarzschild-like black hole},'' {\em Commun. Theor.
  Phys.}, vol.~71, no.~5, pp.~568--572, 2019.

\bibitem{Yang:2025whw}
C.-Y. Yang, H.~Ye, and X.-X. Zeng, ``{Shadow and Polarization Images of
  Rotating Black Holes in Kalb-Ramond Gravity Illuminated by Several Thick
  Accretion Disks},'' 10 2025.

\bibitem{Shi:2025plr}
Y.~Shi and A.~A. Ara{\'u}jo~Filho, ``{Effects of bumblebee gravity on neutrino
  motion},'' 5 2025.

\bibitem{Shi:2025ywa}
Y.~Shi and A.~A. Ara{\'u}jo~Filho, ``{The role of non-metricity on neutrino
  behavior in bumblebee gravity},'' 5 2025.

\bibitem{Shi:2025rfq}
Y.~Shi and A.~A. Ara{\'u}jo~Filho, ``{Influence of a Kalb-Ramond black hole on
  neutrino behavior},'' {\em JHEP}, vol.~08, p.~028, 2025.

\bibitem{LIGOScientific:2016aoc}
e.~a. Abbott, Benjamin~P., ``Observation of gravitational waves from a binary
  black hole merger,'' {\em Physical review letters}, vol.~116, no.~6,
  p.~061102, 2016.

\bibitem{Akiyama2022}
K.~Akiyama, A.~Alberdi, W.~Alef, J.~C. Algaba, R.~Anantua, K.~Asada, R.~Azulay,
  U.~Bach, A.-K. Baczko, D.~Ball, {\em et~al.}, ``First sagittarius a$^*$ event
  horizon telescope results. i. the shadow of the supermassive black hole in
  the center of the milky way,'' {\em The Astrophysical Journal Letters},
  vol.~930, no.~2, p.~L12, 2022.

\bibitem{Akiyama2019}
T.~E. H.~T. Collaboration, ``First {M87$^*$} event horizon telescope results.
  i. the shadow of the supermassive black hole,'' {\em The Astrophysical
  Journal Letters}, vol.~875, no.~1, p.~L1, 2019.

\bibitem{021}
C.~G. Darwin, ``The gravity field of a particle,'' {\em Proceedings of the
  Royal Society of London. Series A. Mathematical and Physical Sciences},
  vol.~249, no.~1257, pp.~180--194, 1959.

\bibitem{019}
O.~Contigiani, ``Lensing efficiency for gravitational wave mergers,'' {\em
  Monthly Notices of the Royal Astronomical Society}, vol.~492, no.~3,
  pp.~3359--3363, 2020.

\bibitem{020}
S.~Mukherjee, B.~D. Wandelt, and J.~Silk, ``Probing the theory of gravity with
  gravitational lensing of gravitational waves and galaxy surveys,'' {\em
  Monthly Notices of the Royal Astronomical Society}, vol.~494, no.~2,
  pp.~1956--1970, 2020.

\bibitem{022}
R.~d. Atkinson, ``On light tracks near a very massive star,'' {\em Astronomical
  Journal, Vol. 70, p. 517}, vol.~70, p.~517, 1965.

\bibitem{mohan2025strong}
G.~Mohan, R.~Karmakar, R.~J. Borah, and U.~D. Goswami, ``Strong lensing effect
  and quasinormal modes of oscillations of black holes in {$f(R,T)$} gravity
  theory,'' {\em arXiv preprint arXiv:2503.08402}, 2025.

\bibitem{araujo2025gasdsadravitational}
A.~A. Ara{\'u}jo~Filho, N.~Heidari, I.~P. Lobo, and V.~Bezerra, ``Gravitational
  signatures of a nonlinear electrodynamics in f (r, t) gravity,'' {\em Journal
  of Cosmology and Astroparticle Physics}, vol.~2025, no.~09, p.~015, 2025.

\bibitem{Cunningham}
C.~T. Cunningham and J.~M. Bardeen, ``{The Optical Appearance of a Star
  Orbiting an Extreme Kerr Black Hole},'' {\em The Astrophysical Journal},
  vol.~183, pp.~237--264, 1973.

\bibitem{Falcke:1999pj}
H.~Falcke, F.~Melia, and E.~Agol, ``Viewing the shadow of the black hole at the
  galacticcenter,'' {\em The Astrophysical Journal}, vol.~528, no.~1, p.~L13,
  1999.

\bibitem{Afrin:2024khy}
M.~Afrin, S.~G. Ghosh, and A.~Wang, ``{Testing EGB gravity coupled to bumblebee
  field and black hole parameter estimation with EHT observations},'' {\em
  Physics of the Dark Universe}, vol.~46, p.~101642, 2024.

\bibitem{Khodadi:2024ubi}
M.~Khodadi, S.~Vagnozzi, and J.~T. Firouzjaee, ``{Event Horizon Telescope
  observations exclude compact objects in baseline mimetic gravity},'' {\em
  Scientific Reports}, vol.~14, no.~1, p.~26932, 2024.

\bibitem{Allahyari:2019jqz}
A.~Allahyari, M.~Khodadi, S.~Vagnozzi, and D.~F. Mota, ``{Magnetically charged
  black holes from non-linear electrodynamics and the Event Horizon
  Telescope},'' {\em Journal of Cosmology and Astroparticle Physics,}, vol.~02,
  p.~003, 2020.

\bibitem{Afrin:2021wlj}
M.~Afrin and S.~G. Ghosh, ``Testing horndeski gravity from eht observational
  results for rotating black holes,'' {\em The Astrophysical Journal},
  vol.~932, no.~1, p.~51, 2022.

\bibitem{Nojiri:2024txy}
S.~Nojiri and S.~D. Odintsov, ``{Improving mimetic gravity with non-trivial
  scalar potential: Cosmology, black holes, shadow and photon sphere},'' {\em
  Physics of the Dark Universe}, vol.~46, p.~101669, 2024.

\bibitem{Afrin:2021imp}
M.~Afrin, R.~Kumar, and S.~G. Ghosh, ``Parameter estimation of hairy kerr black
  holes from its shadow and constraints from {M87$^*$},'' {\em Monthly Notices
  of the Royal Astronomical Society}, vol.~504, no.~4, pp.~5927--5940, 2021.

\bibitem{Nojiri:2024qgx}
S.~Nojiri and S.~D. Odintsov, ``{Black holes and their shadows in {$F(R)$}
  gravity},'' {\em Physics of the Dark Universe}, vol.~47, p.~101785, 2025.

\bibitem{Bambi:2019tjh}
C.~Bambi, K.~Freese, S.~Vagnozzi, and L.~Visinelli, ``{Testing the rotational
  nature of the supermassive object {M87$^*$} from the circularity and size of
  its first image},'' {\em Physical Review D}, vol.~100, no.~4, p.~044057,
  2019.

\bibitem{Khodadi:2021gbc}
M.~Khodadi, G.~Lambiase, and D.~F. Mota, ``{No-hair theorem in the wake of
  Event Horizon Telescope},'' {\em Journal of Cosmology and Astroparticle
  Physics,}, vol.~09, p.~028, 2021.

\bibitem{Liu:2024lve}
W.~Liu, D.~Wu, and J.~Wang, ``{Shadow of slowly rotating Kalb-Ramond black
  holes},'' {\em Journal of Cosmology and Astroparticle Physics,}, vol.~05,
  p.~017, 2025.

\bibitem{Khodadi:2022pqh}
M.~Khodadi and G.~Lambiase, ``{Probing Lorentz symmetry violation using the
  first image of Sagittarius A*: Constraints on standard-model extension
  coefficients},'' {\em Physical Review D}, vol.~106, no.~10, p.~104050, 2022.

\bibitem{Kumar:2020hgm}
R.~Kumar, S.~G. Ghosh, and A.~Wang, ``{Gravitational deflection of light and
  shadow cast by rotating Kalb-Ramond black holes},'' {\em Physical Review D},
  vol.~101, no.~10, p.~104001, 2020.

\bibitem{Afrin:2022ztr}
M.~Afrin, S.~Vagnozzi, and S.~G. Ghosh, ``Tests of loop quantum gravity from
  the event horizon telescope results of {Sgr A$^*$},'' {\em The Astrophysical
  Journal}, vol.~944, no.~2, p.~149, 2023.

\bibitem{Afrin:2023uzo}
M.~Afrin and S.~G. Ghosh, ``Eht observables as a tool to estimate parameters of
  supermassive black holes,'' {\em Monthly Notices of the Royal Astronomical
  Society}, vol.~524, no.~3, pp.~3683--3691, 2023.

\bibitem{Ghosh:2022kit}
S.~G. Ghosh and M.~Afrin, ``An upper limit on the charge of the black hole {Sgr
  A$^*$} from eht observations,'' {\em The Astrophysical Journal}, vol.~944,
  no.~2, p.~174, 2023.

\bibitem{Vagnozzi:2019apd}
S.~Vagnozzi and L.~Visinelli, ``{Hunting for extra dimensions in the shadow of
  {M87$^*$}},'' {\em Physical Review D}, vol.~100, no.~2, p.~024020, 2019.

\bibitem{Nojiri:2024nlx}
S.~Nojiri and S.~D. Odintsov, ``{Black holes, photon sphere, and cosmology in
  ghost-free {$f(G)$} gravity},'' {\em Physics of the Dark Universe}, vol.~46,
  p.~101702, 2024.

\bibitem{Fu:2021fxn}
Q.-M. Fu and X.~Zhang, ``Gravitational lensing by a black hole in effective
  loop quantum gravity,'' {\em Physical Review D}, vol.~105, no.~6, p.~064020,
  2022.

\bibitem{031}
V.~Perlick, ``Theoretical gravitational lensing--beyond the weak-field
  small-angle approximation,'' in {\em The Eleventh Marcel Grossmann Meeting:
  On Recent Developments in Theoretical and Experimental General Relativity,
  Gravitation and Relativistic Field Theories (In 3 Volumes)}, pp.~680--699,
  World Scientific, 2008.

\bibitem{virbhadra2000schwarzschild}
K.~S. Virbhadra and G.~F. Ellis, ``Schwarzschild black hole lensing,'' {\em
  Physical Review D}, vol.~62, no.~8, p.~084003, 2000.

\bibitem{033}
V.~Bozza, S.~Capozziello, G.~Iovane, and G.~Scarpetta, ``Strong field limit of
  black hole gravitational lensing,'' {\em General Relativity and Gravitation},
  vol.~33, pp.~1535--1548, 2001.

\bibitem{032}
S.~Frittelli, T.~P. Kling, and E.~T. Newman, ``Spacetime perspective of
  schwarzschild lensing,'' {\em Physical Review D}, vol.~61, no.~6, p.~064021,
  2000.

\bibitem{034}
V.~Bozza, ``Gravitational lensing in the strong field limit,'' {\em Physical
  Review D}, vol.~66, no.~10, p.~103001, 2002.

\bibitem{heidari2023gravitational}
N.~Heidari, H.~Hassanabadi, A.~A. Ara{\'u}jo~Filho, J.~Kriz, S.~Zare, and
  P.~Porf{\'\i}rio, ``Gravitational signatures of a non--commutative stable
  black hole,'' {\em Physics of the Dark Universe}, p.~101382, 2023.

\bibitem{araujo2025antisymmetric}
A.~A. Ara{\'u}jo~Filho, ``Antisymmetric tensor influence on charged black hole
  lensing phenomena and time delay,'' {\em Journal of High Energy
  Astrophysics}, p.~100401, 2025.

\bibitem{nascimento2024gravitational}
A.~A. Araújo~Filho, J.~R. Nascimento, A.~Y. Petrov, and P.~J. Porf{\'\i}rio,
  ``Gravitational lensing by a lorentz-violating black hole,'' {\em arXiv
  preprint arXiv:2404.04176}, 2024.

\bibitem{chakraborty2017strong}
S.~Chakraborty and S.~SenGupta, ``Strong gravitational lensing—a probe for
  extra dimensions and kalb-ramond field,'' {\em Journal of Cosmology and
  Astroparticle Physics}, vol.~2017, no.~07, p.~045, 2017.

\bibitem{40}
R.~Shaikh and S.~Kar, ``Gravitational lensing by scalar-tensor wormholes and
  the energy conditions,'' {\em Physical Review D}, vol.~96, no.~4, p.~044037,
  2017.

\bibitem{Lobo:2020jfl}
I.~P. Lobo, M.~G. Richarte, J.~P. Morais~Gra\c{c}a, and H.~Moradpour,
  ``{Thin-shell wormholes in Rastall gravity},'' {\em Eur. Phys. J. Plus},
  vol.~135, no.~7, p.~550, 2020.

\bibitem{38.5}
R.~Shaikh, P.~Banerjee, S.~Paul, and T.~Sarkar, ``Strong gravitational lensing
  by wormholes,'' {\em Journal of Cosmology and Astroparticle Physics,},
  vol.~2019, no.~07, p.~028, 2019.

\bibitem{38.2}
G.~W. Gibbons and M.~Vyska, ``The application of weierstrass elliptic functions
  to schwarzschild null geodesics,'' {\em Classical and Quantum Gravity},
  vol.~29, no.~6, p.~065016, 2012.

\bibitem{38.1}
N.~Tsukamoto, T.~Harada, and K.~Yajima, ``Can we distinguish between black
  holes and wormholes by their einstein-ring systems?,'' {\em Physical Review
  D}, vol.~86, no.~10, p.~104062, 2012.

\bibitem{38.3}
N.~Tsukamoto, ``Strong deflection limit analysis and gravitational lensing of
  an ellis wormhole,'' {\em Physical Review D}, vol.~94, no.~12, p.~124001,
  2016.

\bibitem{38.4}
N.~Tsukamoto, ``Retrolensing by a wormhole at deflection angles $\pi$ and 3
  $\pi$,'' {\em Physical Review D}, vol.~95, no.~8, p.~084021, 2017.

\bibitem{virbhadra2024conservation}
K.~Virbhadra, ``Conservation of distortion of gravitationally lensed images,''
  {\em Physical Review D}, vol.~109, no.~12, p.~124004, 2024.

\bibitem{Konoplya:2019hlu}
R.~A. Konoplya, A.~Zhidenko, and A.~F. Zinhailo, ``{Higher order WKB formula
  for quasinormal modes and grey-body factors: recipes for quick and accurate
  calculations},'' {\em Class. Quant. Grav.}, vol.~36, p.~155002, 2019.

\bibitem{Konoplya:2013rxa}
R.~A. Konoplya and A.~Zhidenko, ``{Massive charged scalar field in the
  Kerr-Newman background I: quasinormal modes, late-time tails and
  stability},'' {\em Physical Review D}, vol.~88, p.~024054, 2013.

\bibitem{karmakar2024quasinormal}
R.~Karmakar and U.~D. Goswami, ``{Quasinormal modes, thermodynamics and shadow
  of black holes in Hu--Sawicki {$f(R)$} gravity theory},'' {\em The European
  Physical Journal C}, vol.~84, no.~9, p.~969, 2024.

\bibitem{Konoplya:2007zx}
R.~A. Konoplya and A.~Zhidenko, ``{Decay of a charged scalar and Dirac fields
  in the Kerr-Newman-de Sitter background},'' {\em Physical Review D}, vol.~76,
  no.~8, p.~084018, 2007.
\newblock [Erratum: Phys.Rev.D 90, 029901 (2014)].

\bibitem{Kokkotas:2010zd}
K.~D. Kokkotas, R.~A. Konoplya, and A.~Zhidenko, ``{Quasinormal modes,
  scattering and Hawking radiation of Kerr-Newman black holes in a magnetic
  field},'' {\em Physical Review D}, vol.~83, p.~024031, 2011.

\bibitem{karmakar2022quasinormal}
R.~Karmakar, D.~J. Gogoi, and U.~D. Goswami, ``Quasinormal modes and
  thermodynamic properties of gup-corrected schwarzschild black hole surrounded
  by quintessence,'' {\em International Journal of Modern Physics A}, vol.~37,
  no.~28n29, p.~2250180, 2022.

\bibitem{Konoplya:2011qq}
R.~A. Konoplya and A.~Zhidenko, ``Quasinormal modes of black holes: From
  astrophysics to string theory,'' {\em Reviews of Modern Physics}, vol.~83,
  no.~3, pp.~793--836, 2011.

\bibitem{Jusufi:2020dhz}
K.~Jusufi, ``{Connection Between the Shadow Radius and Quasinormal Modes in
  Rotating Spacetimes},'' {\em Physical Review D}, vol.~101, no.~12, p.~124063,
  2020.

\bibitem{Konoplya:2024lir}
R.~A. Konoplya and A.~Zhidenko, ``{Correspondence between grey-body factors and
  quasinormal modes},'' {\em Journal of Cosmology and Astroparticle Physics,},
  vol.~09, p.~068, 2024.

\bibitem{Konoplya:2024vuj}
R.~Konoplya and A.~Zhidenko, ``Correspondence between grey-body factors and
  quasinormal frequencies for rotating black holes,'' {\em Physics Letters B},
  vol.~861, p.~139288, 2025.

\bibitem{Franchini:2023eda}
N.~Franchini and S.~H. V{\"o}lkel, {\em Testing general relativity with black
  hole quasi-normal modes}, pp.~361--416.
\newblock Springer, 2024.

\bibitem{Hindmarsh:1994re}
M.~B. Hindmarsh and T.~W.~B. Kibble, ``{Cosmic strings},'' {\em Rept. Prog.
  Phys.}, vol.~58, pp.~477--562, 1995.

\bibitem{Liu:2024oas}
W.~Liu, D.~Wu, and J.~Wang, ``{Static neutral black holes in Kalb-Ramond
  gravity},'' {\em JCAP}, vol.~09, p.~017, 2024.

\bibitem{Filho:2023etf}
A.~A.~A. Filho, H.~Hassanabadi, N.~Heidari, J.~Kriz, and S.~Zare,
  ``{Gravitational traces of bumblebee gravity in metric{\textendash}affine
  formalism},'' {\em Class. Quant. Grav.}, vol.~41, no.~5, p.~055003, 2024.

\bibitem{Filho:2023ycx}
A.~A.~A. Filho, J.~A. A.~S. Reis, and H.~Hassanabadi, ``{Exploring
  antisymmetric tensor effects on black hole shadows and quasinormal
  frequencies},'' {\em JCAP}, vol.~05, p.~029, 2024.

\bibitem{AraujoFilho:2024ctw}
A.~A. Ara{\'u}jo~Filho, ``{Particle creation and evaporation in Kalb-Ramond
  gravity},'' {\em JCAP}, vol.~04, p.~076, 2025.

\bibitem{Liu:2024axg}
J.-Z. Liu, W.-D. Guo, S.-W. Wei, and Y.-X. Liu, ``{Charged spherically
  symmetric and slowly rotating charged black hole solutions in bumblebee
  gravity},'' {\em Eur. Phys. J. C}, vol.~85, no.~2, p.~145, 2025.

\bibitem{Yang:2023wtu}
K.~Yang, Y.-Z. Chen, Z.-Q. Duan, and J.-Y. Zhao, ``{Static and spherically
  symmetric black holes in gravity with a background Kalb-Ramond field},'' {\em
  Phys. Rev. D}, vol.~108, no.~12, p.~124004, 2023.

\bibitem{Duan:2023gng}
Z.-Q. Duan, J.-Y. Zhao, and K.~Yang, ``{Electrically charged black holes in
  gravity with a background Kalb{\textendash}Ramond field},'' {\em Eur. Phys.
  J. C}, vol.~84, no.~8, p.~798, 2024.

\bibitem{Wei2020}
S.-W. Wei, ``Topological charge and black hole photon spheres,'' {\em Physical
  Review D}, vol.~102, no.~6, p.~064039, 2020.

\bibitem{Cunha2020}
P.~V. Cunha and C.~A. Herdeiro, ``Stationary black holes and light rings,''
  {\em Physical Review Letters}, vol.~124, no.~18, p.~181101, 2020.

\bibitem{Sadeghi2024}
J.~Sadeghi and M.~A.~S. Afshar, ``The role of topological photon spheres in
  constraining the parameters of black holes,'' {\em Astroparticle Physics},
  vol.~162, p.~102994, 2024.

\bibitem{BahrozBrzo2025}
A.~B. Brzo, S.~N. Gashti, B.~Pourhassan, and S.~Beikpour, ``Thermodynamic
  topology of ads black holes within non-commutative geometry and barrow
  entropy,'' {\em Nuclear Physics B}, vol.~1012, 3 2025.

\bibitem{alipour2024weak}
M.~R. Alipour, M.~A.~S. Afshar, S.~N. Gashti, and J.~Sadeghi, ``Weak gravity
  conjecture validation with photon spheres of quantum corrected
  ads-reissner-nordstrom black holes in kiselev spacetime,'' {\em arXiv
  preprint arXiv:2410.14352}, 2024.

\bibitem{Heidari:2025sku}
N.~Heidari, ``{Imprints of Non-commutativity on Charged Black Holes},'' {\em
  Classical and quantum Gravity}, vol.~43, p.~035004, 2026.

\bibitem{Duane1984}
Y.~Duane, ``The structure of the topological current*,'' tech. rep., 1984.

\bibitem{Heidari:2024bkm}
N.~Heidari, A.~A. Ara{\'u}jo~Filho, R.~C. Pantig, and A.~{\"O}vg{\"u}n,
  ``{Absorption, scattering, geodesics, shadows and lensing phenomena of black
  holes in effective quantum gravity},'' {\em Phys. Dark Univ.}, vol.~47,
  p.~101815, 2025.

\bibitem{AraujoFilho:2024xhm}
A.~A. Ara{\'u}jo~Filho, ``{Analysis of a nonlinear electromagnetic
  generalization of the Reissner{\textendash}Nordstr{\"o}m black hole},'' {\em
  Eur. Phys. J. C}, vol.~85, no.~4, p.~454, 2025.

\bibitem{AraujoFilho:2024lsi}
A.~A. Ara{\'u}jo~Filho, ``{Remarks on a nonlinear electromagnetic extension in
  AdS Reissner-Nordstr{\"o}m spacetime},'' {\em JCAP}, vol.~01, p.~072, 2025.

\bibitem{Baruah:2025ifh}
A.~Baruah, Y.~Sekhmani, S.~K. Maurya, A.~Deshamukhya, and M.~K. Jasim,
  ``{Quasinormal modes, greybody factors, and Hawking radiation sparsity of
  black holes influenced by a global monopole charge in Kalb-Ramond gravity},''
  {\em JCAP}, vol.~08, p.~023, 2025.

\bibitem{Filho:2023abd}
A.~A.~A. Filho, K.~Jusufi, B.~Cuadros-Melgar, and G.~Leon, ``{Dark matter
  signatures of black holes with Yukawa potential},'' {\em Phys. Dark Univ.},
  vol.~44, p.~101500, 2024.

\bibitem{Filho:2024ilq}
A.~A.~A. Filho, K.~Jusufi, B.~Cuadros-Melgar, G.~Leon, A.~Jawad, and C.~E.
  Pellicer, ``{Charged black holes with Yukawa potential},'' {\em Phys. Dark
  Univ.}, vol.~46, p.~101711, 2024.

\bibitem{AraujoFilho:2025hnf}
A.~A. Ara{\'u}jo~Filho, N.~Heidari, I.~P. Lobo, and V.~B. Bezerra,
  ``{Gravitational signatures of a nonlinear electrodynamics in f(R,T)
  gravity},'' {\em JCAP}, vol.~09, p.~015, 2025.

\bibitem{AraujoFilho:2025vgb}
A.~A. Ara{\'u}jo~Filho, N.~Heidari, and I.~P. Lobo, ``{Black Hole Gravitational
  Phenomena in Higher-Order Curvature-Scalar Gravity},'' 9 2025.

\bibitem{Chen:2019iuo}
C.-Y. Chen and P.~Chen, ``{Gravitational perturbations of nonsingular black
  holes in conformal gravity},'' {\em Phys. Rev. D}, vol.~99, no.~10,
  p.~104003, 2019.

\bibitem{Bouhmadi-Lopez:2020oia}
M.~Bouhmadi-L{\'o}pez, S.~Brahma, C.-Y. Chen, P.~Chen, and D.-h. Yeom, ``{A
  consistent model of non-singular Schwarzschild black hole in loop quantum
  gravity and its quasinormal modes},'' {\em JCAP}, vol.~07, p.~066, 2020.

\bibitem{albuquerque2023massless}
S.~Albuquerque, I.~P. Lobo, and V.~B. Bezerra, ``Massless dirac perturbations
  in a consistent model of loop quantum gravity black hole: quasinormal modes
  and particle emission rates,'' {\em Classical and Quantum Gravity}, vol.~40,
  no.~17, p.~174001, 2023.

\bibitem{al2024massless}
A.~Al-Badawi and S.~K. Jha, ``Massless dirac perturbations of black holes in f
  (q) gravity: quasinormal modes and a weak deflection angle,'' {\em
  Communications in Theoretical Physics}, vol.~76, no.~9, p.~095403, 2024.

\bibitem{arbey2021hawking}
A.~Arbey, J.~Auffinger, M.~Geiller, E.~R. Livine, and F.~Sartini, ``Hawking
  radiation by spherically-symmetric static black holes for all spins:
  Teukolsky equations and potentials,'' {\em Physical Review D}, vol.~103,
  no.~10, p.~104010, 2021.

\bibitem{devi2020quasinormal}
S.~Devi, R.~Roy, and S.~Chakrabarti, ``Quasinormal modes and greybody factors
  of the novel four dimensional gauss--bonnet black holes in asymptotically de
  sitter space time: scalar, electromagnetic and dirac perturbations,'' {\em
  The European Physical Journal C}, vol.~80, no.~8, p.~760, 2020.

\bibitem{Gundlach:1993tp}
C.~Gundlach, R.~H. Price, and J.~Pullin, ``{Late time behavior of stellar
  collapse and explosions: 1. Linearized perturbations},'' {\em Physical Review
  D}, vol.~49, pp.~883--889, 1994.

\bibitem{Skvortsova:2024wly}
M.~Skvortsova, ``Ringing of extreme regular black holes,'' {\em Gravitation and
  Cosmology}, vol.~30, no.~3, pp.~279--288, 2024.

\bibitem{Bolokhov:2024ixe}
S.~V. Bolokhov, ``{Late time decay of scalar and Dirac fields around an
  asymptotically de Sitter black hole in the Euler\textendash{}Heisenberg
  electrodynamics},'' {\em The European Physical Journal C}, vol.~84, no.~6,
  p.~634, 2024.

\bibitem{Guo:2023nkd}
W.-D. Guo, Q.~Tan, and Y.-X. Liu, ``{Quasinormal modes and greybody factor of a
  Lorentz-violating black hole},'' {\em Journal of Cosmology and Astroparticle
  Physics,}, vol.~07, p.~008, 2024.

\bibitem{Yang:2024rms}
Z.-H. Yang, C.~Xu, X.-M. Kuang, B.~Wang, and R.-H. Yue, ``Echoes of massless
  scalar field induced from hairy schwarzschild black hole,'' {\em Physics
  Letters B}, vol.~853, p.~138688, 2024.

\bibitem{Baruah:2023rhd}
A.~Baruah, A.~\"Ovg\"un, and A.~Deshamukhya, ``{Quasinormal modes and bounding
  greybody factors of GUP-corrected black holes in Kalb\textendash{}Ramond
  gravity},'' {\em Annals of Physics}, vol.~455, p.~169393, 2023.

\bibitem{Shao:2023qlt}
C.-Y. Shao, C.~Zhang, W.~Zhang, and C.-G. Shao, ``{Scalar fields around a loop
  quantum gravity black hole in de Sitter spacetime: Quasinormal modes,
  late-time tails and strong cosmic censorship},'' {\em Physical Review D},
  vol.~109, no.~6, p.~064012, 2024.

\bibitem{Lutfuoglu:2025kqp}
B.~C. L{\"u}tf{\"u}o{\u{g}}lu, ``{Long-lived quasinormal modes, grey-body
  factors and absorption cross-section of the black hole immersed in the
  Hernquist galactic halo},'' 10 2025.

\bibitem{Santos:2025xbk}
A.~C.~L. Santos, L.~A. Lessa, R.~V. Maluf, and G.~J. Olmo, ``{Echoes and
  quasinormal modes of asymmetric black bounces},'' 8 2025.

\bibitem{Gibbons:2008rj}
G.~W. Gibbons and M.~C. Werner, ``{Applications of the Gauss-Bonnet theorem to
  gravitational lensing},'' {\em Classical and Quantum Gravity}, vol.~25,
  p.~235009, 2008.

\bibitem{Jusufi:2017lsl}
K.~Jusufi, M.~C. Werner, A.~Banerjee, and A.~{\"O}vg{\"u}n, ``{Light Deflection
  by a Rotating Global Monopole Spacetime},'' {\em Phys. Rev. D}, vol.~95,
  no.~10, p.~104012, 2017.

\bibitem{qiao2022geometric}
C.-K. Qiao and M.~Li, ``Geometric approach to circular photon orbits and black
  hole shadows,'' {\em Physical Review D}, vol.~106, no.~2, p.~L021501, 2022.

\bibitem{Heidari:2025iiv}
N.~Heidari, A.~A. Ara{\'u}jo~Filho, and I.~P. Lobo, ``{Non-commutativity in
  Hayward spacetime},'' 3 2025.

\bibitem{qiao2022curvatures}
C.-K. Qiao, ``Curvatures, photon spheres, and black hole shadows,'' {\em
  Physical Review D}, vol.~106, no.~8, p.~084060, 2022.

\bibitem{araujo2025impact}
A.~A. Ara{\'u}jo~Filho, N.~Heidari, J.~A. A.~S. Reis, and H.~Hassanabadi, ``The
  impact of an antisymmetric tensor on charged black holes: evaporation
  process, geodesics, deflection angle, scattering effects and quasinormal
  modes,'' {\em Classical and Quantum Gravity}, vol.~42, no.~6, p.~065026,
  2025.

\bibitem{qiao2024existence}
C.-K. Qiao, ``{The Existence and Distribution of Photon Spheres Near
  Spherically Symmetric Black Holes -- A Geometric Analysis},'' {\em Eur. Phys.
  J. C}, vol.~85, p.~191, 2025.

\bibitem{AraujoFilho:2025huk}
A.~A. Ara{\'u}jo~Filho, N.~Heidari, I.~P. Lobo, and Y.~Shi, ``{Optical
  Phenomena in a Non-Commutative Kalb-Ramond Black Hole Spacetime},'' 8 2025.

\bibitem{tsukamoto2017deflection}
N.~Tsukamoto, ``{Deflection angle in the strong deflection limit in a general
  asymptotically flat, static, spherically symmetric spacetime},'' {\em Phys.
  Rev. D}, vol.~95, no.~6, p.~064035, 2017.

\bibitem{hasse2002gravitational}
W.~Hasse and V.~Perlick, ``Gravitational lensing in spherically symmetric
  static spacetimes with centrifugal force reversal,'' {\em Gen. Rel. Grav.},
  vol.~34, pp.~415--433, 2002.

\bibitem{Qiao:2024ehj}
C.-K. Qiao and P.~Su, ``{Time delay of light in the gravitational lensing of
  supermassive black holes in dark matter halos},'' {\em Eur. Phys. J. C},
  vol.~84, no.~10, p.~1032, 2024.

\bibitem{Amelino-Camelia:2008aez}
G.~Amelino-Camelia, ``{Quantum-Spacetime Phenomenology},'' {\em Living Rev.
  Rel.}, vol.~16, p.~5, 2013.

\bibitem{Addazi:2021xuf}
A.~Addazi {\em et~al.}, ``{Quantum gravity phenomenology at the dawn of the
  multi-messenger era\textemdash{}A review},'' {\em Prog. Part. Nucl. Phys.},
  vol.~125, p.~103948, 2022.

\bibitem{AlvesBatista:2023wqm}
R.~Alves~Batista {\em et~al.}, ``{White paper and roadmap for quantum gravity
  phenomenology in the multi-messenger era},'' {\em Class. Quant. Grav.},
  vol.~42, no.~3, p.~032001, 2025.

\bibitem{Goldstein:2002}
H.~Goldstein, C.~Poole, and J.~Safko, {\em {Classical Mechanics}}.
\newblock San Francisco, USA: Addison-Wesley, 3rd~ed., 2002.

\bibitem{ParticleDataGroup:2024cfk}
S.~Navas {\em et~al.}, ``{Review of particle physics},'' {\em Phys. Rev. D},
  vol.~110, no.~3, p.~030001, 2024.

\bibitem{Wang:2024fiz}
R.-B. Wang, S.-J. Ma, J.-B. Deng, and X.-R. Hu, ``{Estimating the strength of
  Lorentzian distribution in noncommutative geometry by solar system tests},''
  {\em Mod. Phys. Lett. A}, vol.~40, no.~13n14, p.~2550034, 2025.

\bibitem{dsasdas}
S.~Lambert and C.~Le~Poncin-Lafitte, ``Improved determination of $\gamma$ by
  vlbi,'' {\em Astronomy \& Astrophysics}, vol.~529, p.~A70, 2011.

\bibitem{Shapiro:1964uw}
I.~I. Shapiro, ``{Fourth Test of General Relativity},'' {\em Phys. Rev. Lett.},
  vol.~13, pp.~789--791, 1964.

\bibitem{Bertotti:2003rm}
B.~Bertotti, L.~Iess, and P.~Tortora, ``{A test of general relativity using
  radio links with the Cassini spacecraft},'' {\em Nature}, vol.~425,
  pp.~374--376, 2003.

\bibitem{Will:2014kxa}
C.~M. Will, ``{The Confrontation between General Relativity and Experiment},''
  {\em Living Rev. Rel.}, vol.~17, p.~4, 2014.

\bibitem{AraujoFilho:2025zzf}
A.~A. Ara{\'u}jo~Filho, N.~Heidari, and F.~S.~N. Lobo, ``{Spin effects on
  particle creation and evaporation in $f(R,T)$ gravity},'' 10 2025.

\bibitem{Heidari:2024bvd}
N.~Heidari, C.~F.~B. Macedo, A.~A.~A. Filho, and H.~Hassanabadi, ``{Scattering
  effects of bumblebee gravity in metric-affine formalism},'' {\em Eur. Phys.
  J. C}, vol.~84, no.~11, p.~1221, 2024.

\bibitem{Rosa:2023hfm}
J.~L. Rosa, ``{Observational properties of relativistic fluid spheres with thin
  accretion disks},'' {\em Phys. Rev. D}, vol.~107, no.~8, p.~084048, 2023.

\bibitem{Shi:2025oek}
Y.~Shi and H.~Cheng, ``{Probing Hernquist dark matter with black hole shadows:
  A comprehensive study of various accretions},'' 9 2025.

\bibitem{Barros:2023kor}
P.~H.~M. Barros, I.~G. da~Paz, O.~P. d.~S. Neto, and H.~A.~S. Costa,
  ``{Robustness of Wave{\textendash}Particle Duality under Unruh Effect},''
  {\em Entropy}, vol.~26, no.~1, p.~1, 2024.

\bibitem{Barros:2024ftf}
P.~H.~M. Barros and H.~A.~S. Costa, ``{Dispersive vacuum as a decoherence
  amplifier of an Unruh{\textendash}DeWitt detector},'' {\em J. Phys. A},
  vol.~57, no.~44, p.~445305, 2024.

\bibitem{Barros:2024wdv}
P.~H.~M. Barros and H.~A.~S. Costa, ``{Detecting gravitational waves via
  coherence degradation induced by the Unruh effect},'' {\em Eur. Phys. J. C},
  vol.~84, no.~12, p.~1261, 2024.

\bibitem{Barros:2025arh}
P.~H.~M. Barros, P.~R.~S. Carvalho, and H.~A.~S. Costa, ``{On the information
  behavior from quadratically coupled accelerated detectors},'' {\em Eur. Phys.
  J. C}, vol.~85, no.~8, p.~868, 2025.

\bibitem{Barros:2025din}
P.~H.~M. Barros, F.~C.~E. Lima, C.~A.~S. Almeida, and H.~A.~S. Costa,
  ``{Mitigating the information degradation in a massive Unruh-DeWitt
  theory},'' {\em JHEP}, vol.~04, p.~165, 2025.

\end{thebibliography}

\end{document}